\documentclass[12pt]{article}
\usepackage[utf8]{inputenc}

\usepackage{CJKutf8}
\usepackage{empheq}

\usepackage{footmisc}
\usepackage{comment}

\usepackage[dvipsnames]{xcolor}
\definecolor{amber(sae/ece)}{rgb}{1.0, 0.49, 0.0}
\definecolor{ballblue}{rgb}{0.13, 0.67, 0.8 }

\usepackage{amsmath, amssymb, amscd, amsthm, amsfonts}
\usepackage[colorlinks = true, linkcolor = ballblue, citecolor = red, final]{hyperref}
\usepackage{physics}
\usepackage{color}
\usepackage{bm}
\usepackage{graphicx}
\usepackage{multicol}
\usepackage{ marvosym }
\usepackage{tensor}
\usepackage[export]{adjustbox}

\newcommand{\cN}{\mathcal{N}}
\newcommand{\bea}{\begin{align}}
\newcommand{\mH}{\mathcal{H}}
\newcommand{\eea}{\end{align}}
\newcommand{\nn}{\nonumber}
\numberwithin{equation}{section}

\newcommand{\be}{\begin{equation}}
\newcommand{\ee}{\end{equation}}

\newcommand{\ii}{\mathrm{i}}
\newcommand{\tPsi}{\tilde{\Psi}}
\newcommand{\Schw}{\text{Sch}}
\allowdisplaybreaks

\begin{document}

\thispagestyle{plain}
\begin{center}
\vspace*{1cm}
\vspace*{1cm}
\LARGE
\textbf{Constructing all BPS black hole microstates \\ 
 from the gravitational path integral}
\vspace*{1cm}
\normalsize
        
\vspace{0.4cm}

\textbf{Jan Boruch${}^1$, Luca V.~Iliesiu${}^{2,3}$, Cynthia Yan${}^2$}
\,\vspace{0.5cm}\\
{${}^1$  Department of Physics, University of Warsaw, ul. Pasteura 5, 02-093 Warsaw, Poland}
\,\vspace{0.5cm}\\
{${}^2$ Stanford Institute for Theoretical Physics, Stanford University, Stanford, CA 94305, USA}
\,\vspace{0.5cm}\\{${}^3$ Department of Physics, University of California,
Berkeley, CA 94720, USA}

\vspace{0.9cm}
\textbf{Abstract}
\end{center}

Understanding how to prepare and count black hole micro-states by using the gravitational path integral is one of the most important problems in quantum gravity. Nevertheless, a state-by-state count of black hole microstates is difficult because the apparent number of degrees of freedom available in the gravitational effective theory can vastly exceed the entropy of the black hole, even in the special case of BPS black holes.  In this paper, we show that we can use the gravitational path integral to prepare a basis for the Hilbert space of all BPS black hole microstates. We find that the dimension of this Hilbert space computed by an explicit state count is in complete agreement with the degeneracy obtained from the Gibbons-Hawking prescription. Specifically, this match includes all non-perturbative corrections in $1/G_N$. Such corrections are, in turn, necessary in order for this degeneracy of BPS states to match the non-perturbative terms in the $1/G_N$ expansion in the string theory count of such microstates. 
\vspace{0.4cm}

\newpage
\tableofcontents

\newpage

\section{Introduction}

One of the most celebrated achievements of string theory has been the count of BPS black hole microstates as a count of string states \cite{Strominger:1996sh}, providing a microscopic explanation for the macroscopic entropy that black hole exhibit. More recently, such counts over black hole microstates have also been reproduced exactly from the Euclidean gravitational path integral, viewing the black hole as a quantum system in which a periodic Euclidean time can be used to compute the partition function of the system \cite{Dabholkar:2010rm, Dabholkar:2010uh, Dabholkar:2011ec, Gupta:2012cy, Dabholkar:2014ema, Gupta:2015gga, Murthy:2015yfa,deWit:2018dix,Jeon:2018kec, Iliesiu:2022kny}. To reproduce the count of BPS black holes, one first fixes the charges of the black hole and imposes that the length of the Euclidean thermal circle is infinite, thus isolating the lowest-energy (BPS) states in a given charged sector.\footnote{\label{indexFootnote} To reproduce the string theory index computation more directly, one instead considers Euclidean solutions, which have a thermal circle of finite size $\beta$ but also have an angular velocity fixed to $\Omega = 2\pi/\beta$ \cite{Cabo-Bizet:2018ehj, Iliesiu:2021are}. By fixing such boundary conditions, one can directly compute the black hole index in a fixed charge sector by using the gravitational path integral. In the case of BPS black holes that are purely bosonic, the index agrees with the zero temperature partition function, and consequently, the two approaches yield the same result.}  Nevertheless, in contrast to conventional quantum systems where performing the path integral with a periodic identification of time can easily be argued to compute a trace over its Hilbert space (and the integral yields the degeneracy of the lowest energy states in the limit where the size of the time circle becomes infinite), for black holes, no direct argument exists.\footnote{Crucially, the argument that is typically applied in the case of conventional quantum systems relies on the fact that the thermal circle is non-contractible, while when computing the thermal partition function of a black hole, the time circle becomes contractible on the Euclidean horizon of the black hole in order for the overall geometry to be smooth. Even in more subtle cases, such as for the path integral computation of the index of black holes mentioned in footnote \footref{indexFootnote}, for which the time circle is non-contractible, there is always a contractible circle involving the time direction, which makes the standard statistical argument inapplicable.  See, however, \cite{Penington:2023dql} for recent progress.}
Instead, the equality between the Euclidean gravitational path integral with the time circle periodically identified, and the trace of a conventional quantum system is one of the basic assumptions in the duality between black holes and conventional quantum systems. We will refer to this assumption as the Gibbons-Hawking prescription \cite{GibbonsHawking77}.

Moreover, while both the string theory description and the Gibbons-Hawking prescription are useful for computing the overall number of BPS black hole microstates, neither method is sufficiently powerful to give a geometric description for individual BPS states.  The goal of this paper is to provide an explicit preparation and consequent count of all BPS black hole microstates, i.e.~without relying on the Gibbons-Hawking prescription, rather using the gravitational path integral to construct a complete basis of BPS states. We will be able to show that this explicit count of these states completely agrees with the result of the Gibbons-Hawking prescription and, moreover, includes precisely the same type of non-perturbative corrections in $1/G_N$ as the result of the string microstate count or the exact Gibbons-Hawking result obtained from localizing the gravitational path integral.\footnote{In contrast to \cite{Penington:2023dql}, where the Gibbons-Hawking prescription was used to obtain an exact match between the string count and the gravitational path integral, in this paper, we do not have sufficient control to obtain an exact match to the string theory result to all orders in $1/G_N$. Instead, we show that the result of the Gibbons-Hawking prescription, the string count, and our explicit count from the gravitational path integral all share the same kind of non-perturbative corrections in $1/G_N$. In the latter case, we do not have control over the perturbative expansion around each saddle that contributes to the count.}

We start by constructing the BPS version of the thermofield double state, the maximally entangled two-sided BPS state in a sector of fixed charge. 
In gravitational language, such a state is dual to the two-sided black hole with a fixed set of charges prepared by the Euclidean gravitational path integral over half-disk geometry, where the renormalized proper length of the asymptotic boundary is taken to infinity.
By adding matter excitations in the preparation, which in gravitational language can be viewed as exciting particles (not necessarily BPS excitations) in the black hole geometry, we argue that we are able to prepare a large number of states which are naively orthogonal to one another.\footnote{As we shall explain, the states are orthogonal if only including the naive leading contribution in $1/G_N$ to the inner product \cite{Penington:2019kki, Stanford:2020wkf,Hsin:2020mfa}. This contribution is
given by the Euclidean black hole geometry in which matter excitations are considered in the preparation of the bra and ket states.  } If such black holes are indeed dual to a conventional quantum system, the infinite Euclidean evolution would imply that all such states should belong to  the Hilbert space $\mH_\text{BPS}^{Q_i}\otimes \mH_\text{BPS}^{Q_i}$, a two-sided BPS Hilbert space in a sector with fixed charges $Q_i$, whose dimension is predicted from string counts or from the Gibbons-Hawking prescription. However, the number of orthogonal states that we can seemingly prepare in this way is unrelated to the dimension of this Hilbert space $\mH_\text{BPS}^{Q_i}\otimes \mH_\text{BPS}^{Q_i}$ and, in particular, the number of states can be much larger than the dimension of this Hilbert space computed from the Gibbons-Hawking prescription or from string theory constructions. If black holes have a quantum dual, this would, of course, be a contradiction, and the puzzle that we thus raise can be viewed as a version of the black hole information paradox that solely involves BPS black holes. More precisely, our formulation of the paradox is similar in spirit to the ``bags of gold'' paradox \cite{Wheeler:1964qna} in which one finds gravitational solutions where the number of degrees of freedom that can be placed behind the horizon of a black hole is seemingly much larger than the entropy of the black hole in question.

The resolution to our paradox comes from the fact that the inner product between states prepared by the gravitational path integral should include non-perturbative corrections. While the gravitational path integral cannot fully determine the exact correction to each inner product, it is now widely believed that it is a tool that is sufficiently powerful to determine the statistics of such inner products \cite{Penington:2019kki, Stanford:2020wkf,Hsin:2020mfa} both for non-BPS as well as BPS states. Specifically, the statistics of a product of $n$ inner products can be computed from gravity by summing over all geometries with $n$ asymptotic boundaries, whose boundary conditions are given by the preparation of each state entering the inner product. Crucially, even when considering the inner product of BPS states, such geometries include spacetime wormholes that connect different asymptotic boundaries.  Exploiting such non-perturbative corrections to the statistics of these inner products, we can find that the maximal rank of any density matrix constructed from the gravitationally prepared states described above completely agrees with the degeneracy of BPS states computed from the Gibbons-Hawking prescription. This thus ensures that the gravitationally prepared states with various matter excitations inserted in the preparation can yield a complete basis of states. To further verify this statement, we check that an arbitrary state in the two-sided BPS Hilbert space can be written as a linear combination of the gravitationally prepared states described above.

Our results should be compared to similar recent counts in toy models of non-supersymmetric black holes \cite{Penington:2019kki, Hsin:2020mfa,Balasubramanian:2022lnw,Balasubramanian:2022gmo}. The count of non-supersymmetric black holes requires us to prescribe what exact set of microstates we would like to count -- for instance, one can try to count the total number of microstates within some energy interval. Nevertheless, the exact number of such microstates depends on the exact energy level of each microstate and, therefore, sensitively depends on the full UV completion of the gravitational theory. At the level of the gravitational path integral, this can be seen by noticing that the count over microstates within a certain energy window has a non-zero variance. This shows that the path integral can solely compute a coarse-grained number of states within an energy window rather than the exact number, or alternatively, it can compute the average number of states within that window in an ensemble of theories.  In contrast, the supergravity path integral has shown that the BPS black hole microstates are isolated from the non-supersymmetric states by a large gap \cite{Stanford:2017thb,Mertens:2017mtv,Heydeman:2020hhw,Boruch:2022tno}; this isolation suggests that an exact count is possible even without relying on the UV completion of the theory. At the level of the supergravity path integral, this can be seen through the fact that our computation of the maximal rank is not only self-averaging, as is also the case when counting non-supersymmetric black hole microstates, but rather has an exactly zero standard deviation (even though individual inner products between BPS states do have a non-zero standard deviation). This shows that we can perform an exact rather than coarse-grained count and that the states we have constructed span the two-sided BPS Hilbert space for each member in a putative ensemble of boundary theories. To further contrast the past non-supersymmetric counts and the supersymmetric count discussed in this paper, in our computation, there are only a limited number of non-perturbative corrections when computing the rank itself rather than its standard deviation. We show that such corrections come solely from wormholes of genus-zero as wormholes of higher genus turn out to yield a vanishing contribution to the gravitational path integral. Consequently, this leads to a convergent sum over geometries in the rank computation for BPS states; the analogous computation for non-supersymmetric states yields an asymptotic sum over geometries which requires the study of doubly non-perturbative corrections whose geometric meaning has so far remained mysterious. The absence of geometries with higher genera in our supersymmetric computation brings yet another advantage. We are able to argue that our results are only minimally affected by loop corrections from matter fields which, for the purposes of this paper, we therefore neglect. In contrast, in the non-supersymmetric case, such loop corrections result in a well-known divergence for wormholes of higher genus whose regularization is still not understood.\footnote{In the absence of matter sources, this divergence appears even for wormholes with genus zero in non-supersymmetric examples. In the supergravity effective theory studied in this paper ($\mathcal N=2$ JT supergravity), wormhole contributions vanish in the absence of matter sources when considering asymptotic boundaries whose time circle has infinite proper length, i.e.~the kind of boundary conditions that, according to the Gibbons-Hawking prescription, compute the ground state degeneracy of the black hole. Thus, in this case, there is again no one-loop matter divergence to be concerned about.  }

The remainder of this paper is organized as follows. In section \ref{subsec:Gibbons-Hawking-presc}, we start by reviewing how to obtain the number of BPS states in a fixed charged sector when assuming the Gibbons-Hawking prescription. Then, in section \ref{subsec:SUSY-bh-paradox}, we contrast this to the naive state count obtained through the state preparation in which we consider the Hartle-Hawking state in the presence of matter excitations. The large difference between the number of states that we can seemingly construct in this way and the number of states obtained from the Gibbons-Hawking prescription or from string theory counts leads to a formulation of the black hole information paradox that solely involves BPS black hole states. In section \ref{subsec:resolution-of-paradoxes-summary}, we present a summary of the resolution of the paradox, explaining why the Hartle-Hawking states with added matter excitations that were naively orthogonal, in fact, have non-perturbatively small overlaps. In section \ref{subsec:basic-ing} and \ref{subsec:basics_correlators}, we present the basic ingredients that allow for the computation of these non-perturbatively small overlaps. Using these results, we are able to compute the actual dimension of the Hilbert space in section \ref{subsec:leadingorder} and now find agreement with the result obtained using the Gibbons-Hawking prescription at leading order in $1/G_N$; this thus resolves the BPS version of the black hole information paradox that we had posed earlier in the paper. Then, in section \ref{subsec:non-perturbative}, we explain why in our calculation of the dimension of the BPS Hilbert space, the sum over geometries with different topologies is under control, and we show that the non-perturbative $1/G_N$ corrections to our count take the same form as those in the Gibbons-Hawking result or from the string theory count. In subsection \ref{subsec:Interpretation-Haar-random-states}, we identify a  model of Gaussian Haar random states dual to the states that we generated by using the gravitational path integral. We explain how this model reproduces all the nonzero contributions to the gravitational path integral computing statistics of the products of inner products.
In subsection \ref{sec:vanishing-standard-deviation}, we argue that the standard deviation of the rank completely vanishes, thus showing that the calculation of the rank can, in principle, be made exact. We explain the possible sources of error in our count in section \ref{subsec:error} and revisit all our assumptions. In section \ref{sec:reconstructing-an-arbitrary-state}, we provide a non-trivial check for our count by showing that an arbitrary state in  $\mH_\text{BPS}^{Q_i}\otimes \mH_\text{BPS}^{Q_i}$ can be constructed by considering a linear combination of the states that we had previously counted thus confirming in an explicit fashion that these states form a complete basis.  We end with a discussion about the boundary interpretation of our results and about the future directions that the present work opens up. For those interested in more technical details about the $\mathcal N=2$ super-Schwarzian, we review the quantization of the theory with a variety of boundary conditions in appendix \ref{app:N=2-super-JT-part-functions}. In appendix \ref{app:length-of-spatial-wormhole}, we compute non-perturbative corrections to the expectation value of the length of a supersymmetric Einstein-Rosen bridge, including the contribution of geometries that have supersymmetric defects. 
Finally, in appendix \ref{app:hyperbolic-geometry}, we discuss the ingredients from hyperbolic geometry that are necessary in order to be able to estimate the size of the throat for the wormholes that contribute to the maximal rank.

\section{A paradox and a summary of its resolution }
\label{sec:problems-and-a-summary-of-its-resolution}

\subsection{The Gibbons-Hawking prescription}
\label{subsec:Gibbons-Hawking-presc}

We begin by first describing the Gibbons-Hawking prescription \cite{GibbonsHawking77} and its unique features when used for computing the degeneracies of BPS black holes. As briefly described in the introduction, the Gibbons-Hawking prescription assumes that a black hole partition function is computed by performing the gravitational path integral with time identified periodically, with the period identified with the inverse-temperature $\beta$, and a set of boundary conditions for all gauge fields and various metric components that determine whether we are working in the canonical (if fixing the field strength or angular momentum) or grand-canonical ensembles (if fixing the gauge field or the angular velocity). Since BPS states are the lowest energy states in a sector with fixed charges, to compute the BPS degeneracy, one should take $\beta \to \infty$ and fix the electromagnetic field strengths such that the overall electric and magnetic fluxes fix the charges of the black hole to the sector of interest.\footnote{Below, we will ignore the contribution of geometries where the charges are divided between multiple    centers. It would be interesting to understand how the contribution of such geometries can also be taken into account. } 

Using the gravitational path integral, one then sums over all bulk geometries satisfying these boundary conditions. Nevertheless, it is oftentimes the case that as one takes the limit $\beta \to \infty$, most of these geometries yield a vanishing contribution.  For instance, for $1/8$-BPS black holes in $4D$ $\cN=8$ supergravity with charges $Q_i$ (labeling both electric and magnetic charges), the sum over geometries in the limit $\beta \to \infty$ takes the following schematic form \cite{ Dabholkar:2014ema, Iliesiu:2022kny, Banerjee:2009af,Murthy:2009dq}, 
\be 
\label{eq:BH-degeneracy-Gibbons-Hawking}
Z_{BH}^{\beta \to \infty}(Q_i) = \underbrace{\includegraphics[valign=c,width=0.3\textwidth]{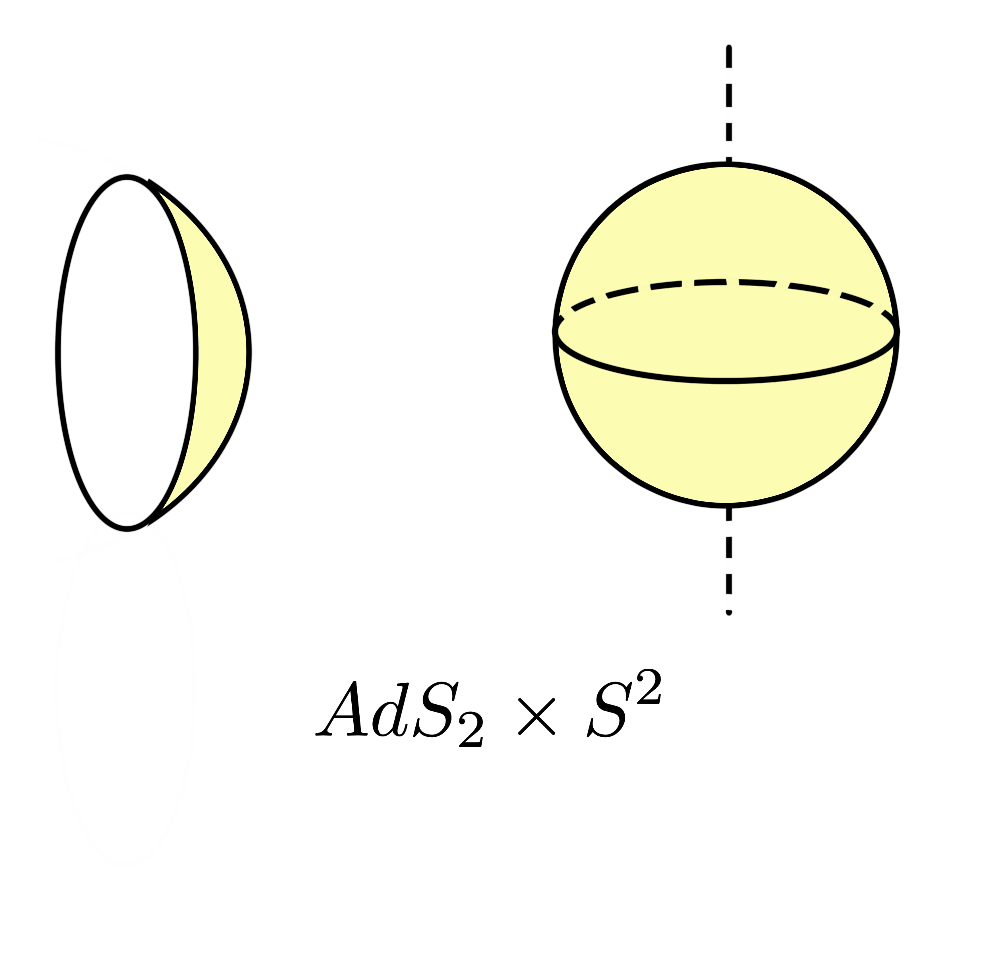}}_{\substack{\text{Extremal BH }\\ \text{contribution}}} \quad + \quad  \underbrace{\includegraphics[valign=c,width=0.3\textwidth]{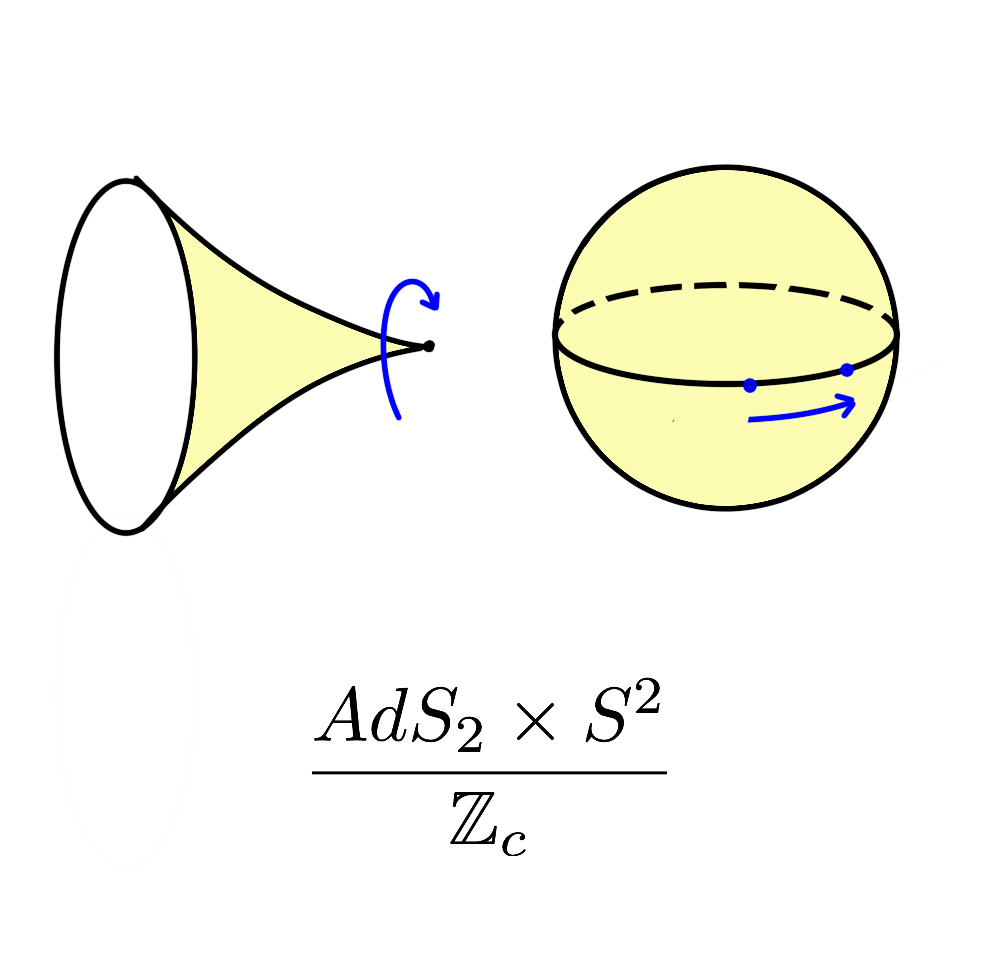}}_{\substack{\text{Orbifolded BH }\\ \text{geometry}}}\,,
\ee
where the first figure represents the extremal black hole geometry with an AdS$_2 \times S^2$ near-horizon region while the second figure represents a $\mathbb Z_c$ quotient of the extremal geometry which includes an $\frac{\text{AdS}_2 \times S^2}{\mathbb Z_c}$ region. At the semi-classical level, by evaluating the action in a saddle-point approximation, the geometry with an AdS$_2 \times S^2$ near-horizon region gives a contribution $\sim e^{\pi \sqrt{\Delta(Q_i)}}$, while the orbifold contributions provide an exponentially subleading  correction $\sim e^{\frac{\pi \sqrt{\Delta(Q_i)}}{c}}$ with $c \in \mathbb Z$. Here, $\Delta(Q_i)$ is a function of all the black hole charges that fixes the leading order in the extremal entropy  and is U-duality invariant.\footnote{To be concrete, for Type IIA supergravity on $\mathbb R^4 \times T^6$ the U-duality invariant $\Delta(Q_i)$ for arbitrary R-R charges and vanishing NS-NS charges is given by, 
\be 
\Delta(Q_I) = 4C_{IJK} Q_m^I Q_m^J Q_m^K \Big(Q^e_0+\frac{1}{12}  Q^e_I C^{IJ} Q^e_J\Big),\label{eq:DeltaRRcharges}
\ee 
where $C_{IJK}$ is the intersection matrix on $T^6$, $C^{IJ}$ is the inverse of the matrix $C_{IJ}\equiv C_{IJK} Q_m^K$ and $Q^I_{m}$ and $Q^e_{I}$ denote the magnetic and electric charges respectively. See \cite{Iliesiu:2022kny} for detailed conventions.
} The exact result, including the entire perturbative series in $1/\Delta(Q_i)$, or equivalently in terms of $1/G_N$, can be obtained by applying the technique of supersymmetric localization to the supergravity path integral. The resulting degeneracy, obtained by only summing over the limited number of geometries in \eqref{eq:BH-degeneracy-Gibbons-Hawking} precisely reproduces the index in a stringy construction of such black holes, $Z_{BH}^{\beta \to \infty}(Q_i) = d_\text{BPS}^{Q_i}$. For such a match to be achieved and for the degeneracy computed by the supergravity path integral to produce an integer, solely the geometries shown in \eqref{eq:BH-degeneracy-Gibbons-Hawking} need to be included. 

The explanation for why only such geometries need to be included comes from studying the quantum fluctuations of the metric that dominate in the near-extremal limit. The modes that control the low-temperature limit of such black holes are given by a set of large super-diffeomorphisms in the near-horizon region and are captured by an effective theory called the $\mathcal N=4$ super-JT gravity theory \cite{Heydeman:2020hhw}. One can quantize these fluctuations exactly around near-horizon geometries with different topologies to obtain the contribution to the density of states near extremality from each such geometry. The result found in \cite{Heydeman:2020hhw} is that only fluctuations around the AdS$_2\times S^2$ and $\frac{\text{AdS}_2 \times S^2}{\mathbb Z_c}$ near-horizon geometries yield a non-vanishing density of states at extremality. Geometries with any other topology have a vanishing one-loop determinant for these fluctuations. In particular, geometries that might be used to connect multiple boundaries in the limit $\beta \to \infty$, also yield a vanishing answer,
\be 
\includegraphics[valign=c,width=0.2\textwidth]{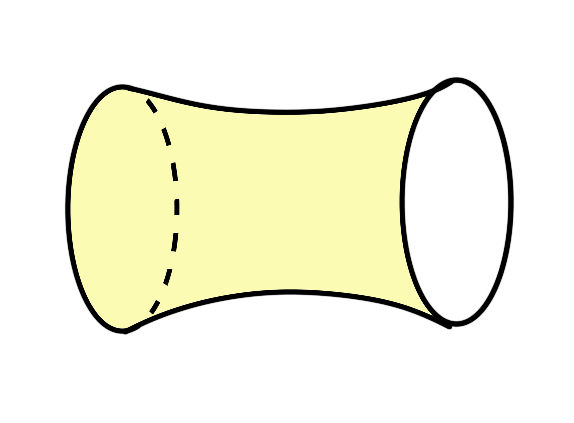} = 0\,,\qquad  \includegraphics[valign=c,width=0.18\textwidth]{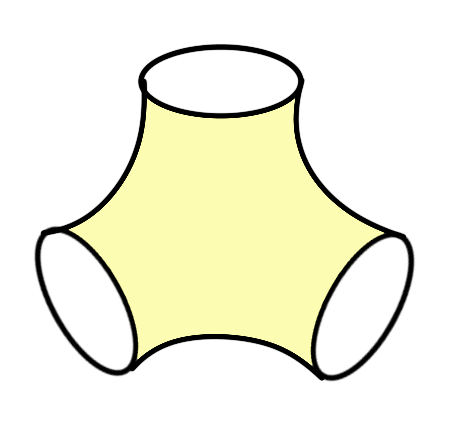} = 0\,.
\label{eq:vanishing-wormhole-contribution}
\ee
More broadly, wormholes are conjectured to compute the statistics of the observables, where the observable is indicated for the Gibbons-Hawking prescription \eqref{eq:vanishing-wormhole-contribution}. Thus, \eqref{eq:vanishing-wormhole-contribution} implies that the dimension of the Hilbert space of BPS states in a sector of fixed charge, i.e.~the degeneracy of such states, has a vanishing variance. This is consistent with the fact that the supergravity path integral is capable of reproducing the integer value of this degeneracy exactly without relying on the full UV completion of the theory. 

The limited number of geometries that contribute to the degeneracy of $1/8$-BPS black holes in  $4d$ flatspace $\cN=8$ supergravity and the vanishing of the variance of this degeneracy is not an isolated example. For instance, the index of $1/16$-BPS AdS$_5$ black holes admits a similar expansion at large $N$ \cite{Aharony:2021zkr},\footnote{There are additional contributions to the large $N$-expansion of the index of the $\mathcal N=4$ super-Yang Mills boundary dual in addition to those schematically shown in \eqref{eq:BH-degeneracy-Gibbons-Hawking}.   },  but this time the effective theory that governs the modes whose quantum effects are large in the low-temperature limit is the $\cN=2$ super-Schwarzian. In this theory, once again, fluctuations around geometries with different topologies yield a non-vanishing one-loop determinant solely in the case of AdS$_2$ near-horizon geometries or specific orbifolds of AdS$_2$. Once again, wormhole contributions are vanishing, which indicates that the degeneracy computed from the Gibbons-Hawking prescription is not an averaged quantity but rather is exact.

Nevertheless, the above computations all rely on the Gibbons-Hawking prescription for which it is still unclear why the gravitational path integral computation actually computes a degeneracy of states. Below, we will proceed by using the gravitational path integral to construct individual states and then we will count the overall number of linearly independent states that we can construct by studying all inner products between such states. This will thus provide a state-by-state count which does not rely on the Gibbons-Hawking prescription. Nevertheless, our count will fully reproduce the terms in the non-perturbative expansion associated with all the geometries in \eqref{eq:BH-degeneracy-Gibbons-Hawking}.

\subsection{A supersymmetric version of the black hole information paradox}
\label{subsec:SUSY-bh-paradox}

To explain how to construct a complete basis of BPS states, it is useful to first consider a thought experiment that can be viewed as a supersymmetric version of the black hole information paradox that is applicable to BPS black holes. Even though the ground state sector does not evolve with time, and therefore such black holes do not evaporate, such black holes still exhibit a puzzle. As we shall describe below, the naive enumeration of orthogonal groundstates in the gravitational effective field theory yields a much larger number than the degeneracy predicted by computing the entropy at extremality using the Gibbons-Hawking prescription.

The paradox that we shall formulate is similar in spirit to the ``bag of gold'' problem that is typically encountered in non-supersymmetric examples.  In the ``bag of gold'' problem \cite{Marolf:2008tx}, one can attach a space whose volume is arbitrarily large behind the horizon of a black hole. Within this volume, one can consider an arbitrary number of orthogonal excitations which would yield different orthogonal states for the black hole. Thus, by making this volume arbitrarily large, the black hole can seemingly have an arbitrarily large number of internal states, contradicting the fact that the Bekenstein-Hawking entropy $S_\text{BH}$ is finite. One resolution to the ``bag of gold'' problem is that the orthogonal matter excitations that we insert behind the horizon of the black hole do not all result in orthogonal black hole microstates -- instead, these microstates can have small corrections to their inner product coming from non-perturbative gravitational effects. 

Let us see now how, naively, we can similarly construct an arbitrarily large number of orthogonal states in the BPS sector by using the gravitational path integral. 
We start by preparing a two-sided BPS black hole using Euclidean path integral by considering a state for which the boundary proper length is infinite and for which the field-strength is fixed. The infinite Euclidean evolution thus projects the state to its groundstate sector and the result can be viewed as a supersymmetric version of the Hartle-Hawking state $\ket{HH}$. From the perspective of a putative boundary dual, since we believe that the Hartle-Hawking state with a period of Euclidean-evolution of $\beta/2$ is equivalent to the thermofield double state whose inverse-temperature is $\beta$, the boundary state that we are constructing by taking $\beta \to \infty$  is the maximally entangled state in a two-sided BPS Hilbert space, $\mathcal H_\text{BPS} \otimes \mathcal H_\text{BPS}$, in a sector of fixed charge (from now on we will suppress the index $Q_i$ indicating the fixed charge sector in which the degeneracy is computed). 

Having constructed a specific two-sided BPS state, we can now construct other states in the two-sided Hilbert space by
adding matter excitations in the Euclidean preparation. For the state to remain in the two-sided BPS Hilbert space we still perform the Euclidean boundary evolution for an infinite amount of boundary proper time to the left and right of the operator insertion,
\be
\ket{HH}=\includegraphics[valign=c,width=0.2\textwidth]{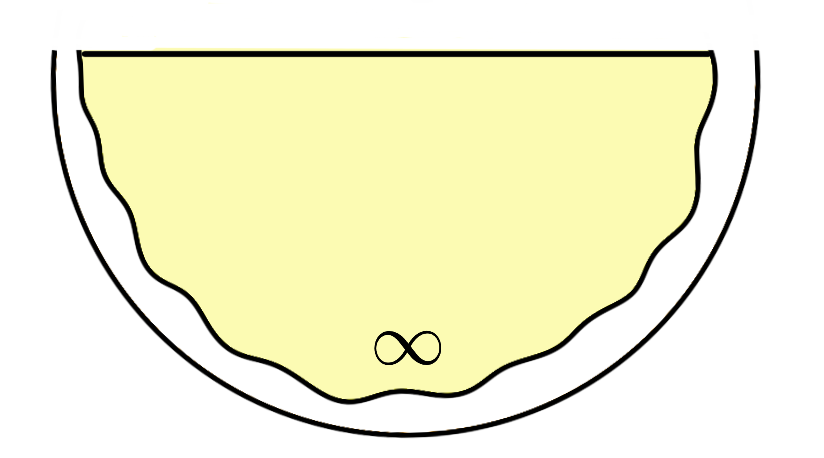}\,,\quad\quad O_i\ket{HH}=\includegraphics[valign=c,width=0.2\textwidth]{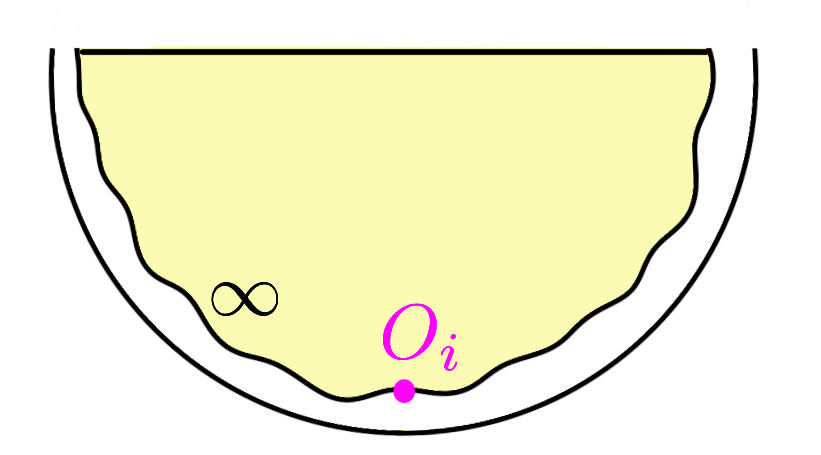}\,.
\label{eq:state-qi}
\ee
where $O_i$ is some neutral and not necessarily BPS operator created from fields from the supergravity effective field theory.\footnote{Alternatively, denoting 
\be 
K_i = \lim_{\beta \to \infty} e^{-\beta H} O_i e^{-\beta H},
\ee
we can equivalently act with $K_i$ either on the left or with its transpose on the right side (making $K_i$ or $K_i^T$ a one-sided operator) to obtain the same state as \eqref{eq:state-qi}. } To evaluate inner products between such states, we consider the boundary conditions for the bra and ket states and glue the two preparations to obtain a spacetime that has a single asymptotic boundary whose overall proper length is, once again, infinite. Thus, evaluating inner products amount to evaluating matter correlation functions on the black hole background with $\beta \to \infty$. At the level of the gravitational EFT we can use a variety of fields in such a way that the correlation function obtained from the inner product of two distinct states vanishes. Consider, 
 $K$ prepared states $\{\ket{q_1},\ldots,\ket{q_K}\}$ with
\be
\ket{q_i}=O_i\ket{HH}\,.
\ee
in such a way that $\braket{q_i}{q_j} = 0$, for all $i\neq j$.  Because all the states involve an infinite amount of Euclidean evolution, from the boundary perspective, we expect all the constructed states $\ket{q_i} \in \mathcal H_\text{BPS} \otimes \mathcal H_\text{BPS}$. All such states have the exact same energy and charge; since they are orthogonal, they are nevertheless distinct states.
The number of states $K$ that we can construct solely depends on the details of the gravitational EFT that we are using and has nothing to do with the actual dimension of the BPS black hole Hilbert space $(d_{BPS})^2 = \mathcal H_\text{BPS} \otimes \mathcal H_\text{BPS}$ of two sided states. In particular, since the number of orthogonal states that we can construct by using a QFT in a given gravitational background is infinite,\footnote{Below, we shall take into account the backreaction of the matter insertion on the black hole background. Taking this backreaction into account does not change the conclusion that the number of available orthogonal excitations is infinite. Additionally, since we always use an infinite period of boundary Euclidean evolution, we can always smear the inserted excitation over a very large region such that the energy density associated to that excitation is arbitrarily small. } we can prepare $K$ orthogonal states with $K > (d_{BPS})^2$.  Since the number of orthogonal states in a Hilbert space cannot exceed its dimension, this is a contradiction and can be viewed as the supersymmetric version of the black hole information paradox.

\subsection{Resolution to the paradox and a way to construct all BPS states}

\label{subsec:resolution-of-paradoxes-summary}

The resolution to the paradox comes from the idea that the overlaps between the $K$ states receive non-perturbatively small corrections given by geometries with a different topology than the black hole geometry \cite{Stanford:2020wkf}. To see this, let us look at the simplest example $|\bra{q_i}\ket{q_j}|^2$ with $i\neq j$. 
To compute this squared overlap, we will sum over all geometries whose boundary conditions correspond to the preparation of each bra and ket. The gravitational path integral thus receives contributions from various geometries, each weighted by the Einstein-Hilbert action, including geometries that can connect the boundary preparing the bra and ket in one inner product to the one preparing the conjugated inner product: 
\be
|\braket{q_i}{q_j}|^2 = \includegraphics[valign=c,width=0.23\textwidth]{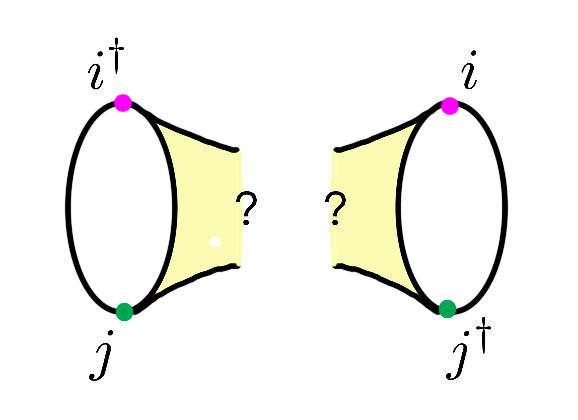}=\underbrace{\includegraphics[valign=c,width=0.25\textwidth]{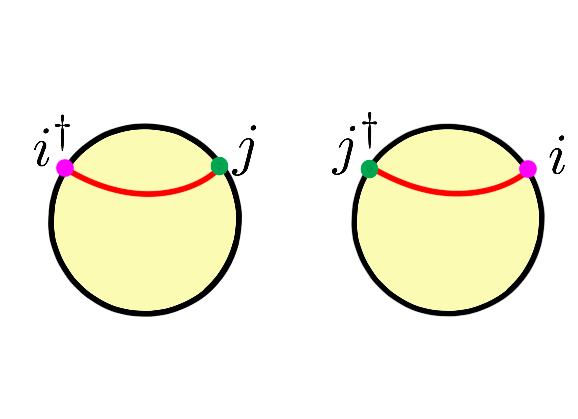}}_{0 \, \text{ for } i\neq j}+\includegraphics[valign=c,width=0.23\textwidth]{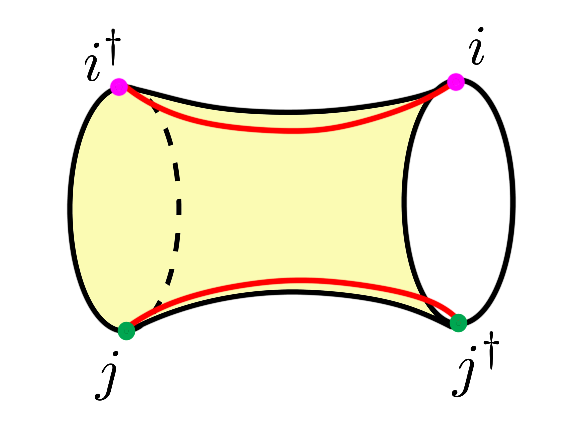}+\cdots\,,
\label{eq:geometries-in-inner-product-squared}
\ee
where in the middle term, we indicated what kind of boundary conditions our geometries need to satisfy but do not specify the geometries that we sum over (therefore, the question marks), and in the last term, we show the connected wormhole  geometries contributing to leading order in the $1/G_N$ expansion. While the wormhole without matter insertions yields a vanishing contribution to the supergravity path integral in the limit $\beta \to \infty$ (as in \eqref{eq:vanishing-wormhole-contribution}), once matter excitations are added, the wormhole contribution no longer vanishes.  The states $\ket{q_i}$ are each normalized by, 
\be
\bra{q_i}\ket{q_i}=\includegraphics[valign=c,width=0.12\textwidth]{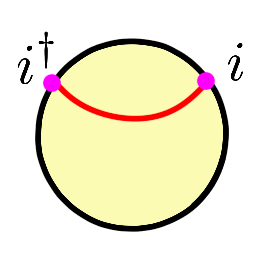} + \dots\,.
\ee
Since the wormhole geometry in \eqref{eq:geometries-in-inner-product-squared} has a larger Einstein-Hilbert action, we thus see that the product squared is non-perturbatively suppressed in $1/G_N$. If, instead of computing its square, we computed $\braket{q_i}{q_j}$ directly, we would not find such a non-perturbatively small correction, and the inner product that the gravitational path integral would capture would still vanish. We can interpret this seeming inconsistency by conjecturing that the gravitational path integral instead captures the statistics of these products \cite{Stanford:2020wkf}. $\braket{q_i}{q_j}$ has arbitrarily oscillating phases even when considering the product of BPS states, and the supergravity path integral indicates a zero average. $|\braket{q_i}{q_j}|^2$, however has a definite sign. Its average cannot vanish and is instead given by the wormhole contribution in \eqref{eq:geometries-in-inner-product-squared}.

Nevertheless, seeing this non-zero square overlap does not tell us how much of the Hilbert space the $K$ states span. How do we harness the fact that these overlaps are non-zero in order to make sure that we constructed a complete basis for the entire Hilbert space? We can actually explicitly calculate the dimension of the black hole BPS Hilbert space by calculating the rank of the density matrix
\be 
\rho_\text{mixed BPS} = \sum_{i=1}^K \ket{q_i} \bra{q_i}\,,
\ee
as a function of $K$. By computing higher moments for the statistics of $\braket{q_i}{q_j}$ we can compute the eigenvalues of $\rho_\text{mixed BPS}$ and determine its rank. We will find that the rank of this density matrix is given by
\be
:\mathrm{rank}(\rho_\text{mixed BPS}):=\begin{cases}K&K<(Z_{BH}^{\beta \to \infty})^2\\(Z_{BH}^{\beta \to \infty})^2 &K>(Z_{BH}^{\beta \to \infty})^2\end{cases}\,,
\label{eq:rank-summary}
\ee
where $Z_{BH}^{\beta \to \infty}$ is the black hole degeneracy in a sector of fixed charge (from now on, we shall suppress the charge indices) obtained from the Gibbons-Hawking prescription as in \eqref{eq:BH-degeneracy-Gibbons-Hawking} and includes the sum over the leading black hole geometry and the subleading orbifolded geometries. The notation $:\dots:$ indicates that we are summing over all geometries (including those that can connect distinct boundaries) associated to a product of inner products or to its analytic continuation.

The dependence of this rank with $K$ can be viewed as a version of the ``Page curve'' for BPS black holes. The rank saturates at the correct dimension of the Hilbert space as derived from the gravitational path integral computation described in section \ref{subsec:Gibbons-Hawking-presc} that uses the Gibbons-Hawking prescription. A similar calculation done in section \ref{sec:reconstructing-an-arbitrary-state} will confirm that any arbitrary state $\ket{\psi}$ can be explicitly reconstructed using the states $\ket{q_i}$. 
Moreover, while we only manage to compute the spectrum of $\rho_\text{mixed BPS}$ on average, and the moments of this spectrum are non-vanishing, we will find that its rank has only vanishing moments. Thus, the rank in \eqref{eq:rank-summary} is not averaged and is instead an exact quantity (in contrast to wormhole computations of the Page curve \cite{Penington:2019kki, Almheiri:2019qdq} or to recent state counts that make use of similar wormhole geometries \cite{Penington:2019kki, Hsin:2020mfa, Balasubramanian:2022gmo, Balasubramanian:2022lnw}).

\section{Counting BPS black hole microstates}

In this section, we calculate the dimension of BPS black hole Hilbert space by calculating the rank of the density matrix
\be 
\rho_\text{mixed BPS} = \sum_{i}^K \ket{q_i} \bra{q_i}\,,
\ee
which naively seems maximally mixed. 
To do this, we first calculate the trace of $n$ such matrices multiplied together and then take the limit $n\rightarrow0$
\be
\mathrm{rank}(\rho_\text{mixed BPS})= \lim_{n \rightarrow 0}\tr \left(\rho_\text{mixed BPS}\right)^n
.
\ee
We use properties of Hartle-Hawking wavefunction to calculate the RHS. Before doing that, we establish the tools in subsections \ref{subsec:basic-ing}, \ref{subsec:basics_correlators}. Then in subsection \ref{subsec:leadingorder}, we present the leading order result of the rank, i.e. the ensemble average of the rank $:\mathrm{rank}(\rho_\text{mixed BPS}):$. In subsection \ref{subsec:non-perturbative}, we calculated non-perturbative corrections.
In subsection \ref{subsec:Interpretation-Haar-random-states}, we identify the Gaussian ensemble model dual to the ground state sector coupled to probe matter. We explain how this model reproduces all the nonzero contributions to the gravitational path integral computing statistics of the products of inner products.
In subsection \ref{sec:vanishing-standard-deviation}, we argue that the standard deviation of the rank completely vanishes thus showing that the calculation of the rank can in principle be made exact. Nevertheless, for the purposes of this paper, we do neglect
the contribution of certain (non-planar) geometries and the full contributions of matter fields present in the EFT; we discuss the role of these sources of error in section \ref{subsec:error}.

\subsection{Review of $\mathcal{N}=2$ JT}
\label{subsec:basic-ing}

The main theory we work with in this paper is $\mathcal{N}=2$ super JT \cite{Stanford:2017thb,Mertens:2017mtv,LongPaper,ShortPaper,Turiaci:2023jfa} which can be used as the effective theory to describe the physics of near-BPS higher dimensional black holes. Of particular interest for us in this paper is the fact that this effective field theory is able to isolate the contribution of BPS states to the gravitational path integral of higher dimensional black holes when taking the limit of zero temperature. As we shall see, this can be traced back to the fact that the density of states has a large number of ground states, separated from finite energy states by an energy gap. As mentioned above, an important example where this effective field theory is applicable is in black holes that are 1/16-BPS black holes in AdS$_5\cross S^5$ \cite{Gutowski:2004ez,Gutowski:2004yv, Chong:2005hr,Cabo-Bizet:2018ehj,Choi:2018hmj,Benini:2018ywd}, but it is more widely applicable for all higher dimensional black holes with SU$(1,1|1)$ near-horizon isometry. Below, we shall use  $\mathcal{N}=2$ super JT not only to make predictions about the spectrum of the theory but also to compute correlation functions in these theories. This will allow us to study the overlap of the BPS states whose preparation we described above. In this paper, we shall work with $\mathcal{N}=2$ super-JT rather than $\mathcal{N}=4$ super-JT, the effective field theory to describe the physics of near-BPS states for black holes in flatspace supergravity, because in the former the Hartle-Hawking wavefunctions and associated correlators are easier to compute. Nevertheless, if one overcomes this computational obstacle, we expect that our conclusions that the states $\ket{q_i}$ can form a complete basis of states will also be valid in the case of BPS black holes in flatspace. 

$\mathcal{N}=2$ super-JT can be thought of as a supersymmetric extension of two dimensional theory of dilaton gravity called JT gravity \cite{Teitelboim, Jackiw, AlmheiriPolchinski,Yang:2018gdb,Maldacena:2016upp,Mertens:2022irh}. Let us first briefly review some aspects of the nonsupersymmetric case. The action of standard JT gravity is given by 
\be 
I_{JT}=-S_0 \chi(\mathcal{M})  -\frac{1}{2}\left(\int_\mathcal{M} \phi(R+2)+2\int_{\partial \mathcal{M}}\phi_b (K-1)\right) ,
\ee
where $g_{\mu \nu}$ is the 2d metric and $\phi$ denotes the dilaton. The first term in the above action is a topological invariant $\chi(\mathcal{M}) = 2- 2g - n$, it depends only on the number of boundaries $n$ and the genus $g$ of the 2d surface $\mathcal{M}$. The second term couples linearly the curvature to the dilaton; integrating out the dilaton in the path integral sets $R=-2$ and so the full path integral of this theory includes only hyperbolic surfaces of negative curvature. The last term is the Gibbons-Hawking term for the boundary, required for studying the partition function with asymptotic AdS boundary conditions.

Non-supersymmetric JT gravity is 
an effective field theory describing fluctuations of the near horizon geometry of higher dimensional black holes which contain a near horizon AdS$_2$ throat (more precisely, black holes with AdS$_2$ throats and near horizon SL$(2,\mathrm{R})$ isometry group). From this perspective, the two dimensional metric captures the $(\tau,r)$ components of the total metric, and the dilaton $\phi$ describes the fluctuations in the area of the transverse closed surface. Performing the dimensional reduction for higher dimensional theory instructs us about the boundary conditions of interest at the disk level. In the usual case, these will be boundary conditions where we take the proper length of the boundary $L = \beta/\epsilon$ and the boundary value of the dilaton $\phi_b = \Phi_r/\epsilon$ to infinity $\epsilon \rightarrow 0$ in such a way that their ratio stays finite.

Imposing the above boundary conditions the total action reduces to just the boundary Schwarzian action
\be
I_{JT, \, g=0}=-S_0-\Phi_r \int d\tau\,\Schw(f(\tau), \tau) , 
\ee
with the function $f(\tau + \beta) = f(\tau)$ parametrizing large diffeomorphisms. The above action has an SL$(2,\mathrm{R})$ symmetry. Because from the higher dimensional perspective configurations related by SL$(2,\mathrm{R})$ transformation are equivalent (because of the near horizon isometry group), this symmetry should be gauged. The total configuration space of the theory should be, therefore, Diff$(S^1)/$SL$(2,\mathrm{R})$. The resulting path integral turns out to be one-loop exact and can therefore be computed exactly. This allows one to compute the corrections from quantum fluctuations of the near horizon geometry for higher dimensional black holes. In recent years this led to a flurry of progress in understanding different aspects of near-extremal black holes \cite{Iliesiu:2020qvm,Iliesiu:2021are,Heydeman:2020hhw,Boruch:2022tno,LongPaper,ShortPaper,Iliesiu:2022kny}. In particular, it gave a resolution of the black hole thermodynamics breakdown puzzle and provided the last necessary ingredient required for precise black hole microstates count from the perspective of gravitational path integral (for certain supersymmetric black holes in four-dimensional $\mathcal{N}=8$ supergravity).

Let us now get back to the supersymmetric case of interest. The generalization will come with some technical complications. However, morally, the story stays the same. So far, we have described everything in the second-order formulation of gravity in terms of the metric. It turns out that for supersymmetric generalizations of the theory, it is more convenient to work with the first-order formulation written in terms of vielbeins and spin connection. The price we pay for this is having to introduce a Lagrange multiplier which ensures vanishing torsion and makes both formulations equivalent. The convenience of the first-order formulations comes from the fact that the total $\mathcal{N}=2$ JT action can now be defined as a topological BF theory with SU$(1,1|1)$ gauge group together with the standard topological term $-S_0 \chi(\mathcal{M})$ and the Gibbons-Hawking boundary term. Imposing the vanishing torsion takes us back now to the second-order formulation. The field content of the resulting theory consists of the 2d metric with complex gravitino, the dilaton with complex dilatino, and the gauge field of the $U(1)$ R-symmetry with a zero-form Lagrange multiplier.

Proceeding now as before, we integrate out the dilaton, dilatino, and the zero-form Lagrange multiplier and impose the standard boundary conditions (focusing now on the disk). This reduces the action to the $\mathcal{N}=2$ super-Schwarzian action 
\be 
I_{\mathcal{N}=2 \, JT,\, g=0}  =- S_0 + \Phi_r \int d\tau\left[- \Schw(f, \tau) + 2(\partial_\tau \sigma)^2 + \text{ (fermions)}\right]\,,
\ee
described in terms of ($f(\tau)$, $\sigma(\tau)$, $\eta(\tau)$, $\bar{\eta}(\tau)$), which parametrize the large superdiffeomorphisms that consist of transformations of the metric, complex gravitino and the U$(1)$ gauge field. This action has SU$(1,1|1)$ symmetry which we'll gauge because we have in mind higher dimensional black holes with SU$(1,1|1)$ isometry group. The resulting path integral is one-loop exact and can be computed exactly. At the disk level the answer is given by 
\cite{Stanford:2017thb,Mertens:2017mtv}
\be
Z(\beta)=e^{S_0}\left(\sum_{|j|<\frac{1}{2}}\cos\pi j+2\sum_j\int_0^\infty ds\,\frac{s\,\sinh2\pi s}{\pi E_{s,j}}e^{-\beta E_{s,j}}\right)\quad\quad E_{s,j}=s^2+\frac{1}{4}\left(j-\frac{1}{2}\right)^2
,
\label{qe:partition-function}
\ee
where we included a sum over sectors of different R-charge $j$, which is quantized in the units of $1/\hat{q}$. As indicated above, \eqref{qe:partition-function} indicates that the spectrum of the theory has a large degeneracy at extremality, given by the first term in the parenthesis. This is followed by a gap above which there is a continuum of states whose density can be identified from the second term in the parenthesis.

\subsection{Multiboundary correlators}
\label{subsec:basics_correlators}

By canonically quantizing the $\mathcal{N}=2$ super-Schwarzian theory, one finds that it desribes a supersymmetric quantum mechanical system with a Liouville-type Hamiltonian \cite{LongPaper,ShortPaper}
\be 
H = - \partial_\ell^2 - \frac{1}{4}\partial_a^2 + i [\bar{\psi}_l \psi_r e^{-\ell/2 - ia} + \psi_l \bar{\psi}_r 
e^{-\ell/2 +ia} + e^{-\ell}
]
,
\ee
where $\ell$ is the bulk geodesic length between the two boundaries of a half disk, $\psi_l,$ and $\psi_r$ are the superpartners of length variable with respect to left and right supercharge respectively, and $e^{i a}$ is the U$(1)_R$ Wilson line between the boundaries. The bulk wavefunctions will then be functions of the above variables. 
The main bulk state we will be working with is the Hartle-Hawking state, defined through a Euclidean path integral over half disk with asymptotic AdS boundary conditions and renormalized proper length of the boundary set to $\beta/2$.
To project it onto the ground state sector we then take $\beta \rightarrow \infty$. 
The resulting state is a ground state of the bulk Hamiltonian $H \ket{\Psi} = 0$. 
Its wavefunction has been found in \cite{LongPaper} and takes the form
\be  
\Psi_{12}^j(\ell_{12}, a_{12}) = e^{\ii j a_{12}} \frac{2 \cos (\pi j)}{\pi} e^{-\ell_{12}/2} 
\left[ 
\xi_{12} \, e^{- \frac{\ii a_{12}}{2}} K_{\frac{1}{2}+j} (2 e^{-\ell_{12}/2})
+
\eta_{12} \, e^{\frac{\ii a_{12}}{2}} K_{\frac{1}{2}-j} (2 e^{-\ell_{12}/2}) 
\right]
,
\label{eq:Hartle-Hawking-wavefunction-N=2}
\ee
where $\eta_{12} = \frac{-i}{\sqrt{2}} \bar{\psi}_r $ and $\xi_{12} = - \frac{1}{\sqrt{2}} \bar{\psi}_l $.
Note that the above expression distinguishes the boundaries.
The wavefunction with opposite orientation is then given by 
\be  
\Psi_{21}^j(\ell_{12}, a_{12}) =  e^{-\ii j a_{12}} \frac{2 \cos (\pi j)}{\pi} e^{-\ell_{12}/2} 
\left[ 
\eta_{12} \, e^{ \frac{\ii a_{12}}{2}} K_{\frac{1}{2}+j} (2 e^{-\ell_{12}/2})
-
\xi_{12} \, e^{\frac{-\ii a_{12}}{2}} K_{\frac{1}{2}-j} (2 e^{-\ell_{12}/2}) 
\right]
,
\ee
which can be also viewed as the wavefunction of the bra state. From the perspective of holographic duality the Hartle-Hawking state can be viewed as the state dual to thermofield double state $\ket{\text{TFD}_{\beta/2}}$. Taking then the zero temperature limit lands us on a maximally mixed state of the ground states.

With the help of the zero temperature Hartle-Hawking state, the disk partition function in a fixed $j$ sector can be computed now by gluing together two Hartle-Hawking wavefunctions along their common geodesic 
\begin{align}
    Z_j(\beta \rightarrow \infty)&=
    \includegraphics[valign=c,width=0.2\textwidth]{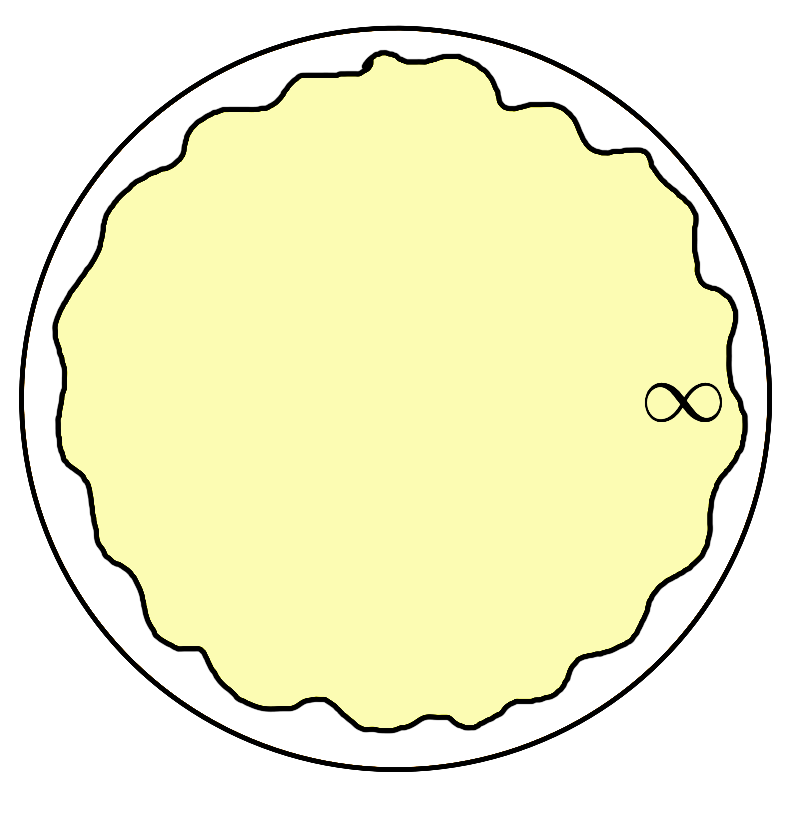}
    =\int d\mu_{12}\includegraphics[valign=c,width=0.2\textwidth]{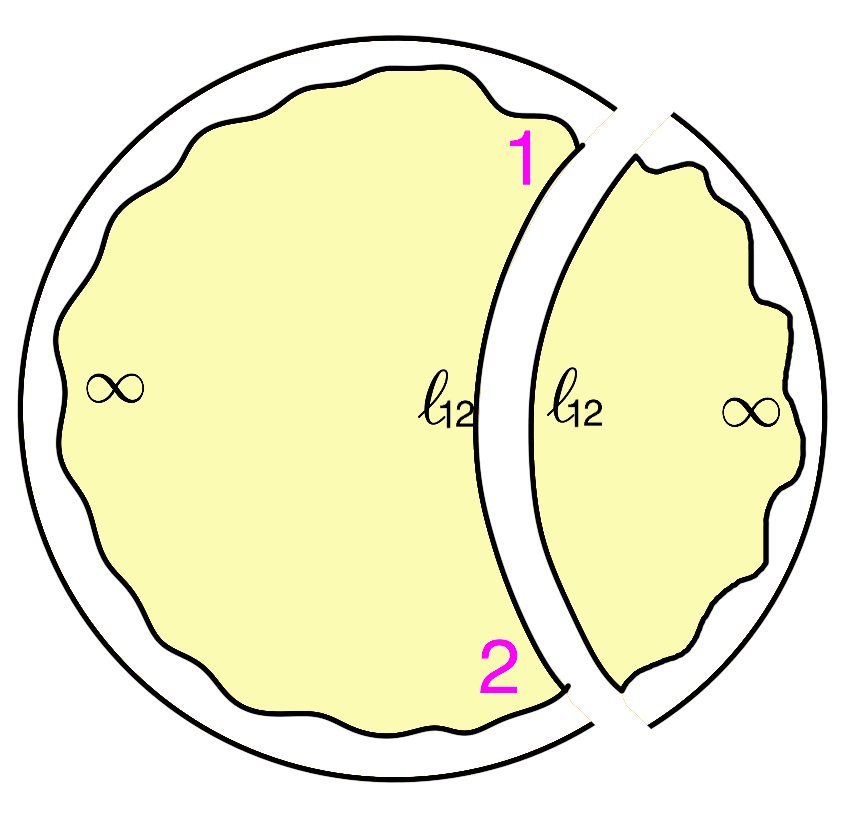}\\
    &=
    e^{S_0}\int d\mu_{12}\,\Psi_{12}^j(\ell_{12}, a_{12})\Psi_{21}^j(\ell_{12}, a_{12}) ,
\end{align}
with the measure
\be  
\int d \mu =
\frac{1}{2 \hat{q}} \int_{-\infty}^{\infty} d\ell \, \int_{0}^{2\pi \hat{q}} da \, \int d\eta \, \int  d\xi ,
\ee
Evaluating the above integral explicitly we verify that 
\begin{align}
Z_j &=  \frac{e^{S_0}}{2\hat{q}}   \int d\ell_{12} \, da_{12} \, d\eta_{12} \, d\xi_{12} \, \Psi^j_{12}   \Psi^j_{21} 
\\
&= e^{S_0} \frac{4 \cos^2(\pi j)}{\pi} 
 \int d\ell_{12} \, 
e^{-\ell_{12}} \left[ 
K_{\frac{1}{2}+j} (2 e^{-\ell_{12}/2})^2 + 
K_{\frac{1}{2}-j} (2 e^{-\ell_{12}/2})^2
\right] 
\\
&= e^{S_0} \cos (\pi j) ,
\end{align}
where we used the relation 
\be 
\int  d\ell \, e^{-\Delta \ell} K_{2\ii s_1} (2 e^{-\ell/2}) K_{2\ii s_2} (2 e^{-\ell/2}) = 
\frac{\Gamma(\Delta \pm \ii s_1 \pm \ii s_2)}{4 \, \Gamma(2\Delta)} 
,\label{BesselRel2}
\ee
with $\Delta=1$. 
Through a similar procedure, we can calculate two-point functions of two boundary insertions of matter fields with conformal weight $\Delta$ by adding a multiplicative factor $e^{-\Delta\ell_{12}}$ to the integral
\begin{align}
    \langle 2\text{pt} \rangle_{j, \, \text{disk}}&
=
\includegraphics[valign=c,width=0.2\textwidth]{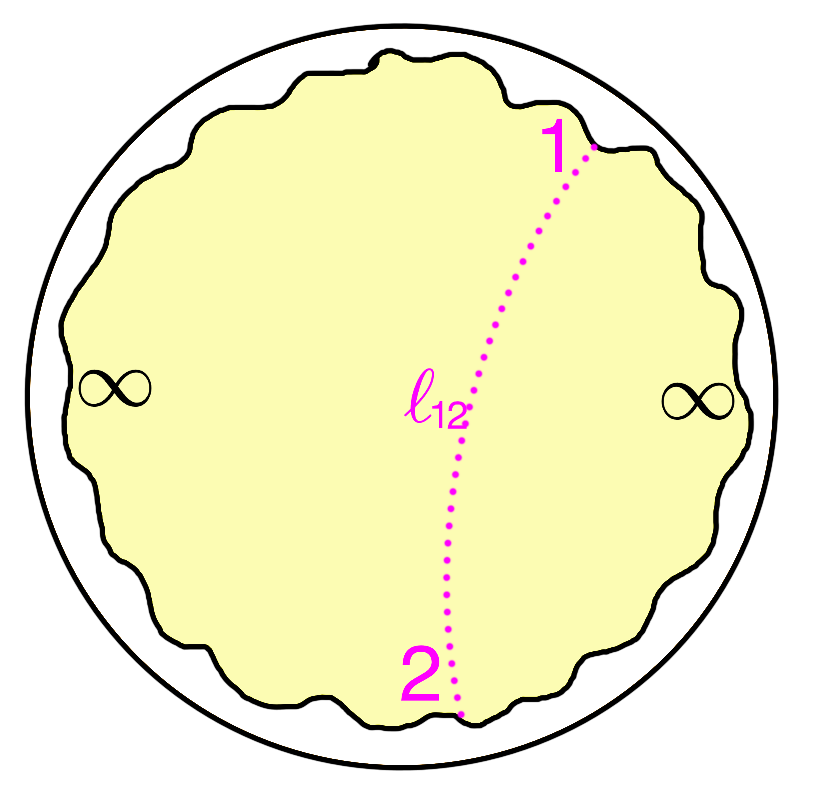}=\int d\mu_{12}\includegraphics[valign=c,width=0.2\textwidth]{split.png}e^{-\Delta\ell_{12}}\\
&
=e^{S_0}\int d\mu_{12}\,\Psi_{12}^j(\ell_{12}, a_{12})\Psi_{21}^j(\ell_{12}, a_{12})e^{-\Delta\ell_{12}} ,
\end{align}
which after using (\ref{BesselRel2}) gives
\begin{align}  
\langle 2\text{pt} \rangle_{j, \, \text{disk}} &= 
e^{S_0}\frac{\cos^2(\pi j)}{2\pi} \frac{\Delta \Gamma(\Delta)^2}{\Gamma(2\Delta)} \Gamma\left(\Delta+\frac{1}{2}\pm j\right) .
\end{align}

\begin{figure}
    \centering
    \includegraphics[width=0.5\textwidth]{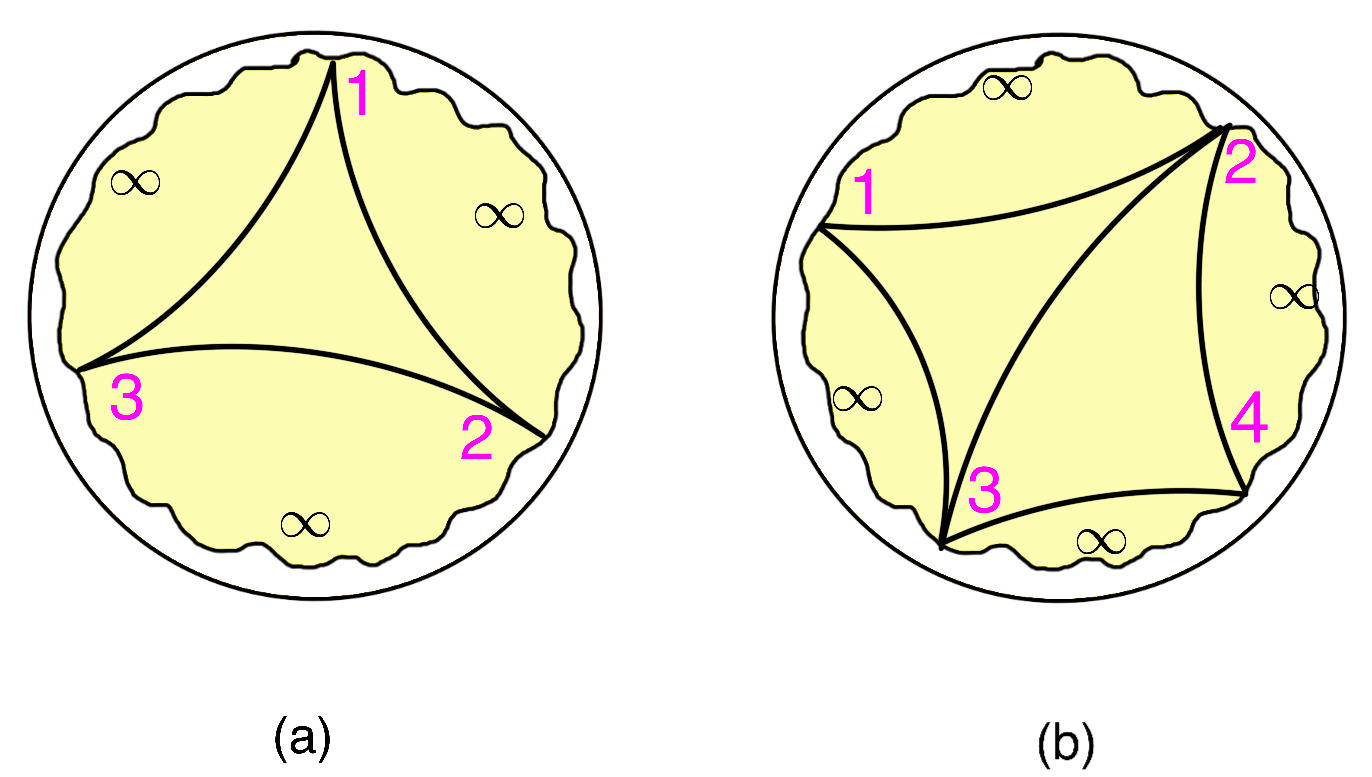}
    \caption{ Consistency conditions for triangle and quadrilateral partition functions: (a) triangle glued to three Hartle-Hawking wavefunctions recovers the disk partition function. (b) Two triangles are glued together to obtain a quadrilateral which we glue to four Hartle-Hawking wavefunctions to recover the disk partition function.}
    \label{fig:triangle}
\end{figure}

The partition function can also be computed by splitting the boundary into more than two Hartle-Hawking states. This introduces the important notion of a "polygon", which is very useful for evaluating correlation functions on multiboundary geometries. We start by computing the partition function of a triangle and we'll glue multiple copies of this geometry to obtain the partition function of any polygon. Consider the function $I(3,2,1)$ (note the opposite orientation of the Hartle-Hawking wavefunctions below)
\be 
I(3,2,1)  \equiv \sum_{j'} \cos(\pi {j'})  \Tilde{\Psi}^{j'}_{32} \Tilde{\Psi}^{j'}_{21} 
\Tilde{\Psi}^{j'}_{13}, \qquad 
\Tilde{\Psi}^j_{12} \equiv 
\frac{\Psi^j_{12}}{\cos(\pi j)} ,
\qquad
\int d\mu_{12} \, \Psi^j_{12} \Tilde{\Psi}^{j'}_{21} = \delta_{j\, j'}\,.
\ee
We can evaluate the disk partition function as 
\be  
Z_j = e^{S_0} \int d\mu_{12} \int d\mu_{23} \int d\mu_{31} \,
I(3,2,1) \Psi^j_{12} \Psi^j_{23} \Psi^j_{31} 
= e^{S_0} \cos(\pi j)
.
\ee
Thus, $I(3,2,1)$ satisfies the correct self-consistency conditions for the partition function of a triangle and is uniquely determined by these conditions \cite{Yang:2018gdb}. 
Two triangles can now be glued along a common geodesic boundary to produce a polygon with four geodesic boundaries, as shown in figure \ref{fig:triangle}. The partition function of the quadrilateral can therefore be written as
\begin{align}
\int d\mu_{23} \, I(3,2,1) I(2,3,4) &= \sum_{j, \, j'} \cos(\pi j) \cos(\pi j') 
\Tilde{\Psi}^j_{21} \Tilde{\Psi}^{j}_{13}
\Tilde{\Psi}^{j'}_{34} \Tilde{\Psi}^{j'}_{42} 
\int d\mu_{23} \, \Tilde{\Psi}^{j'}_{23} 
\Tilde{\Psi}^{j}_{32} , 
\\
&= \sum_j \cos(\pi j) \Tilde{\Psi}^{j}_{21} \Tilde{\Psi}^{j}_{13}
\Tilde{\Psi}^{j}_{34} \Tilde{\Psi}^{j}_{42} 
\equiv I(2,1,3,4) .
\end{align}
This leads us to introduce a polygon geometry with $n$ geodesic boundaries. Its partition function is found to be   
\be  
I (i_1,i_2,\dots ,i_n) = \sum_{j'} \cos(\pi {j'}) \,
\Tilde{\Psi}^{j'}_{i_1 \, i_2} \Tilde{\Psi}^{j'}_{i_2 \, i_3} 
\dots \Tilde{\Psi}^{j'}_{i_n \, i_1} ,
\ee
where one can easily see that it will also reproduce the partition function $Z_j$ after being glued to $n$ boundary ground state Hartle-Hawking wavefunctions of opposite orientation.

To evaluate now a two point function on a two boundary cylinder geometry we consider a polygon with four geodesic boundaries. Two of those boundaries get glued to asymptotic Hartle-Hawking wavefunctions, while the other two get glued together with a factor of $e^{-\Delta \ell}$. We therefore have
\begin{align}
\langle 2 \text{pt} \rangle_{j, \text{cyl}} 
&=\includegraphics[valign=c,width=0.2\textwidth]{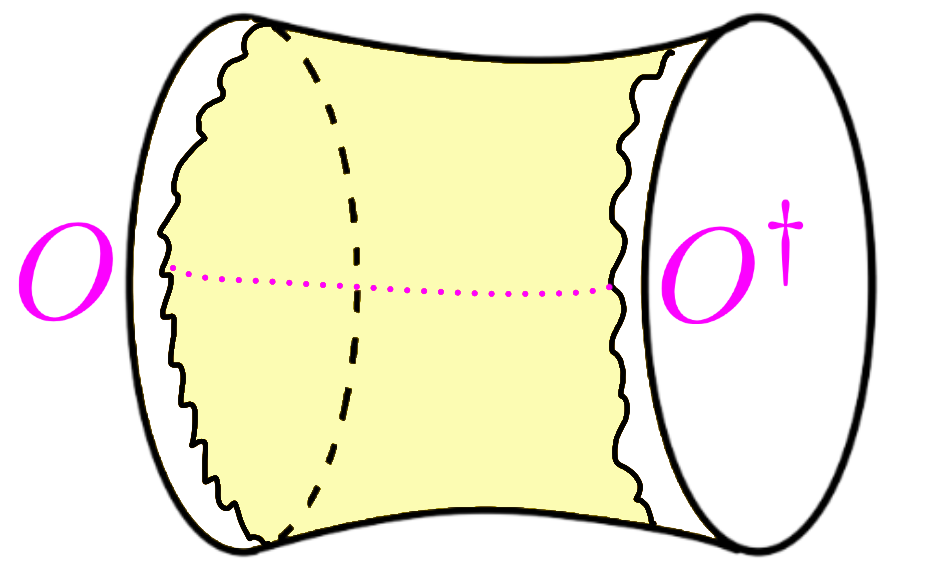}
=\int d\mu_{43}\includegraphics[valign=c,width=0.2\textwidth]{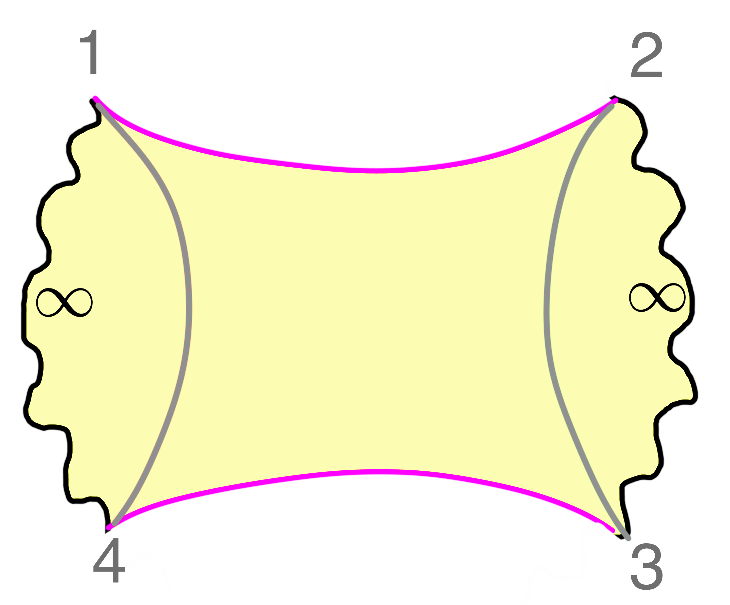}e^{-\Delta \ell_{34}} \\
&= 
\int d\mu_{41} \, \Psi^j_{41} \int d\mu_{23}\,  \Psi^j_{23} \int d\mu_{43}  \, I(\mu_{14},\mu_{43},\mu_{32} ,\mu_{34} ) e^{-\Delta \ell_{34}} 
\\ 
&= 
\sum_{j'} \cos (\pi j') \int d\mu_{41} \tPsi^{j'}_{14} \, \Psi^j_{41} \int d\mu_{23}\,  \Psi^j_{23} \tPsi^{j'}_{32} \int d\mu_{43}  \,  \tPsi^{j'}_{43}  \tPsi^{j'}_{34}  e^{-\Delta \ell_{34}} 
\\ 
&= 
 \cos (\pi j) \int d\mu_{43}  \,  \tPsi^j_{43}  \tPsi^j_{34}  e^{-\Delta \ell_{34}} 
 \\
 &=
\frac{e^{-S_0}}{\cos(\pi j)} \langle 2\text{pt} \rangle_{j, \, \text{disk}} .
\label{eq:wormhole-gluing}
\end{align}
Note that this result precisely matches the equation (72) of \cite{LongPaper} obtained in a different manner.

Finally let us evaluate the $n$-boundary pinwheel geometry with two operator insertions on each boundary, which is the main object we need for the subsequent sections. To compute it we first glue together two polygons of opposite orientation of size $2n$ along every second geodesic. Every geodesic we glue over will also a factor of $e^{-\Delta \ell_{i\, i+1}}$. Each from the rest of $2n$ geodesics are going to be glued to the boundary HH wavefunctions $\Psi^j_{i-1 \,i}$, which will give two Kronecker deltas $\delta_{j\,j'} \delta_{j\,j"}$. We will thus be left with $n$ integrals over $\tPsi^{j}_{i \, i+1} \tPsi^{j}_{i+1 \, i} e^{-\Delta \ell_{i\, i+1}} $ times a factor of $\cos^2(\pi j)$ coming from polygon's definition. Including the topological weighting for $n$ boundaries, we are thus left with 
\begin{align} 
\label{eq:pinwheel-gluing}
Z_{j, \, n}^{\mathcal{O}} 
&=
\includegraphics[valign=c,width=0.2\textwidth]{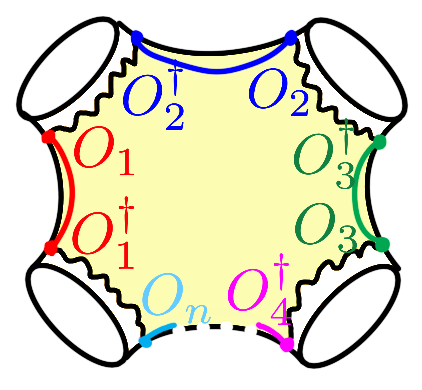}
=\int \prod_{i=\text{even}}^{2n} d\mu_{i,i+1}\,e^{-\Delta \ell_{i\, i+1}} \includegraphics[valign=c,width=0.2\textwidth]{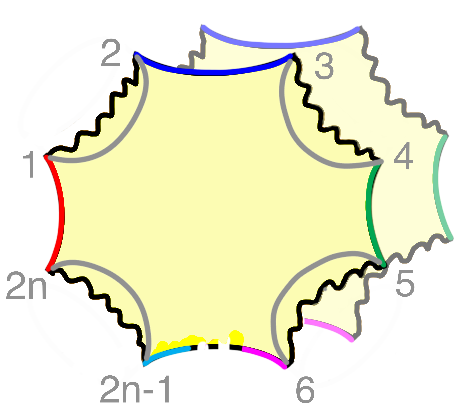}\\
&= e^{S_0 (2-n)} \cos^2(\pi j) \prod_{i=\text{even}}^{2n}\left( 
\int d\mu_{i \, i+1} \, \tPsi^{j}_{i \, i+1} \tPsi^{j}_{i+1 \, i} \, e^{-\Delta \ell_{i\, i+1}} 
\right)
\\
&
= e^{S_0 (2-n)} \cos^2(\pi j) 
\left(
\frac{\Delta \Gamma(\Delta)^2 \,\Gamma\left(\Delta+\frac{1}{2}\pm j\right)}{2\pi \Gamma(2\Delta)}
\right)^n 
.
\end{align}

When computing the boundary-to-boundary propagators on a spacetime with non-trivial topology, one should sum over all matter worldlines propagating between the boundary points, including worldlines that wind around some of the non-contractible cycles of the spacetime. Nevertheless, in the supergravity path integral, one should additionally quotient the integral over all super-geometries by the set of large super-diffeomorphisms that are disconnected from the identity. The quotient between all such super-diffeomorphisms and the super-diffeomorphisms connected to the identity leads to a generalization of the mapping class group of the manifold. Such large transformations can relate geodesics that are in different homotopy classes to each other and, therefore, when computing propagators in the gravitational path integral, one should be careful to not overcount geometries that are, in fact, related by large super-diffeomorphisms. As explained in \cite{Saad:2019pqd, Blommaert:2020seb, Iliesiu:2021ari}, in the case of JT gravity it turns out that the sum over all possible (non-intersecting) geodesics traveling between two boundary points precisely cancels out the effect of the large diffeomorphisms that are generated by twists along the closed geodesic cycles that intersect the boundary-to-boundary geodesic. A similar conclusion holds in the case of supergravity. Thus, the formulas \eqref{eq:wormhole-gluing} and \eqref{eq:pinwheel-gluing} actually correctly take into account the quotient of large super-diffeomorphisms as well as the sum over all propagators connecting the boundary points between an operator insertion and its Hermitian conjugate. 

We should also note that in the above result one cannot simply take the $\Delta \rightarrow 0$ limit to reproduce the cylinder or $n$-boundary wormhole partition function. As discussed in the appendix, both such partition functions should vanish when taking $\beta \to \infty$.  This is because when we insert $e^{-\Delta \ell}$ factor along a geodesic, geometries with different numbers of windings of that geodesic lead to inequivalent contributions to the correlator (see appendix A of \cite{Yan:2023rjh}). This is in contrast with the computation without matter insertion along the geodesic, where geodesics with different windings can be identified through the action of the mapping class group.

\subsection{Leading order result}
\label{subsec:leadingorder}

With the tools described in the previous subsection, we are now ready to use the gravitational path integral to evaluate the rank of the density matrix $\rho_\text{mixed BPS}$. The rank will allow us to probe the dimension of the Hilbert space, by verifying at what value of $K$ it saturates. By definition, the rank can be computed by the following expression
\be
:\mathrm{rank}(\rho_\text{mixed BPS}):= \lim_{n \rightarrow 0} :\tr \left(\rho_\text{mixed BPS}\right)^n: 
.
\ee
The idea is to first compute the trace of the density matrix to some integer power $n$, and then analytically continue it to $n\rightarrow 0$. This will give us the rank of the matrix because only the nonzero eigenvalues will contribute to the resulting sum after analytical continuation. In the above expression, we have used the notation $:\dots:$ to indicate at what point we take the disorder average, if the dual quantum system would correspond to something like SYK. Note that this means that our result will only give us an average rank of $\rho_\text{mixed BPS}$ for the whole ensemble, which in principle can deviate from the average as one moves between different members of the ensemble. These deviations will be analyzed in more detail in section \ref{subsec:error}.
\\ 
\\ 
Let us now discuss the evaluation of 
\be 
:\tr \left(\rho_\text{mixed BPS}\right)^n: = :\sum_{i_1,\dots,i_n=1}^K \bra{q_{i_n}}\ket{q_{i_1}}\bra{q_{i_1}}\ket{q_{i_2}}\cdots\bra{q_{i_{n-1}}}\ket{q_{i_n}}: 
,
\ee
via the gravitational path integral. As in section \ref{subsec:resolution-of-paradoxes-summary}, we will impose the boundary conditions on the asymptotic boundaries of the spacetime that are indicated by preparation of each bra and ket, and then sum over all possible geometries that are consistent with chosen boundary conditions (while being careful not to overcount the geometries). In our case the boundary conditions are specified by the expression above. Each of the above overlaps corresponds to a single boundary thermal circle, with two boundary operator insertions $\mathcal{O}_{q_{i_k}},\mathcal{O}^\dag_{q_{i_{k+1}}}$. This means that in total we will have $n$ boundaries with $2n$ operator insertions. A single bulk contribution will be now specified by deciding how the different boundaries are connected among themselves, as well as how we choose to connect the operators via bulk geodesics. For example, for the two boundary case 
\be  
:\tr (\rho_\text{mixed BPS})^2:  
= :\sum_{i,j=1}^K \bra{q_j}\ket{q_i}\bra{q_i}\ket{q_j}
:,
\ee
we can have, among others, the following contributions 
\begin{align}
\sum_\text{geometries}\sum_{i,j=1}^K\includegraphics[valign=c,width=0.2\textwidth]{2disks0.png}&=\includegraphics[valign=c,width=0.2\textwidth]{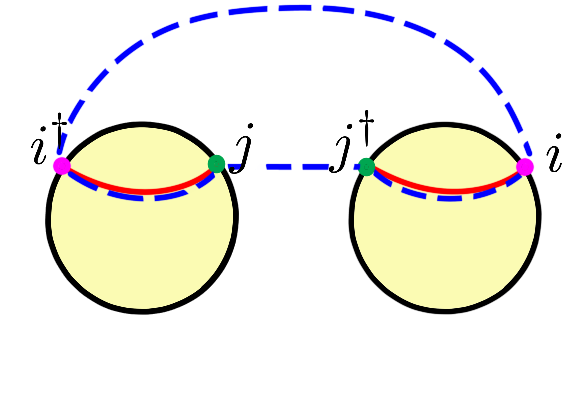} \quad Ke^{2S_0}\nonumber\\
&+\includegraphics[valign=c,width=0.15\textwidth]{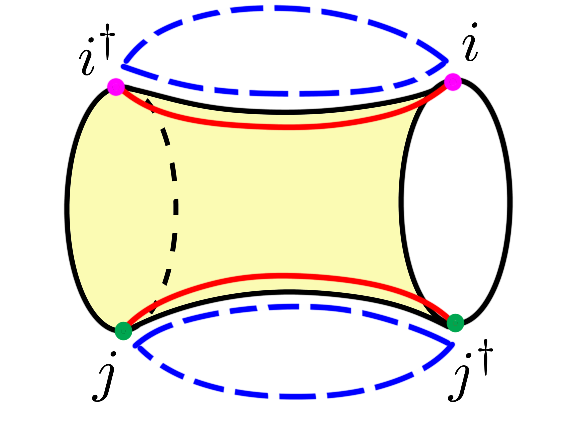} \quad\quad\quad  K^2\nonumber\\
&+\includegraphics[valign=c,width=0.15\textwidth]{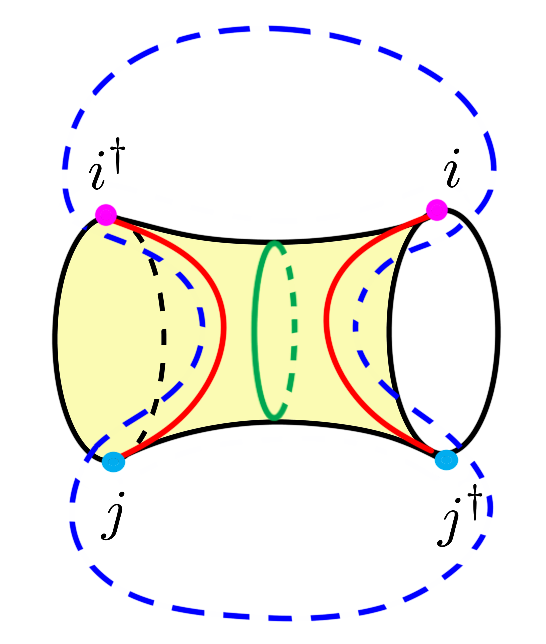} \quad\quad\quad  K.
\label{eq:sum-over-geometries}
\end{align}
Above, the blue dotted loop are loops for operators that share the same index. Thus, in the first line and the third line, we only get contributions from $i=j$, while in the second line on the wormhole geometry, we can have $i \neq j$. Each index loop contributes a factor of $K$. This needs to be multiplied by $e^{S_0 \sum_{\text{Disconnected } \mathcal M}\chi_{\mathcal M}}$ to obtain the overall scaling of each geometry with $K$ and $e^{S_0}$. As we scale $K \sim e^{2S_0}$, the non-perturbative corrections given by the connected wormholes appear at the same order as the disconnected disk contribution in \eqref{eq:sum-over-geometries}. The last line gives an example of a subleading geometry that contains a trumpet, which in the ground state sector is zero as we explain in \eqref{eq:multi-geodesic-trumpet}.

As has been by now very well understood in the literature \cite{Penington:2019kki,Saad:2019pqd,Stanford:2020wkf, Hsin:2020mfa}, it is the wormhole contributions that capture the nonzero variance of the inner products, and they are the key ingredient that allows us to probe the dimension of the Hilbert space.

To resum the geometries for the case with $n$ boundaries we will focus on the leading contributions at large $K$ and $e^{S_0}$. 
These will consist of planar geometries that have leading number of index loops.
This essentially makes the contributing geometries similar to the ones appearing in the West Coast Model \cite{Penington:2019kki}, and allows us to use the gravity analog of resolvent technique to compute the answer.

\subsubsection*{The resolvent}
Let 
\be 
\label{eq:matrix-M-defintion}
M_{ij} \equiv \braket{q_i}{q_j} \,,
\ee 
denote the matrix of overlaps of the basis vectors $\ket{q_i}$. We introduce the resolvent matrix 
\be  
\mathbf{R}_{ij}(\lambda)  =  : \left( 
\frac{1}{\lambda  - M}
\right)_{ij} :
= \frac{\delta_{ij}}{\lambda} + \frac{1}{\lambda} \sum_{n=1}^\infty \frac{:(M^n)_{ij}:}{\lambda^n}
,
\ee
Similar resolvents were considered in \cite{Penington:2019kki, Hsin:2020mfa, Balasubramanian:2022lnw, Balasubramanian:2022gmo} for the case of non-supersymmetric black holes. The geometric expansion of this resolvent can be schematically expressed as 
\begin{align}
\includegraphics[valign=c,width=0.15\textwidth]{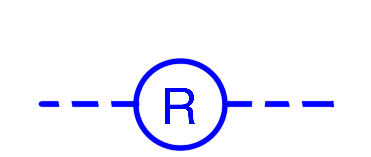}&=\includegraphics[valign=c,width=0.15\textwidth]{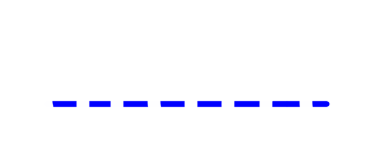} +\includegraphics[valign=c,width=0.13\textwidth]{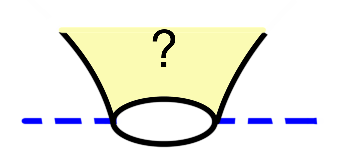} +\includegraphics[valign=c,width=0.2\textwidth]{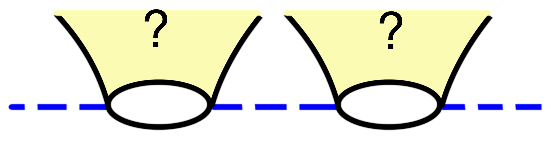}+\cdots\\
&=\includegraphics[valign=c,width=0.15\textwidth]{sd1.png} +\includegraphics[valign=c,width=0.15\textwidth]{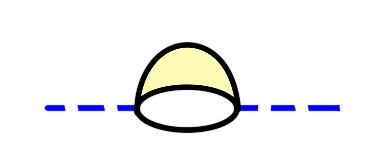}\nonumber\\
&\quad\quad+\includegraphics[valign=c,width=0.25\textwidth]{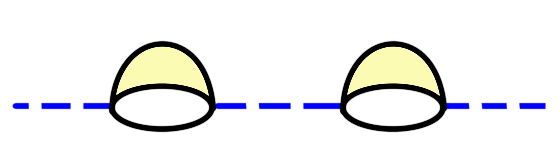}+\includegraphics[valign=c,width=0.25\textwidth]{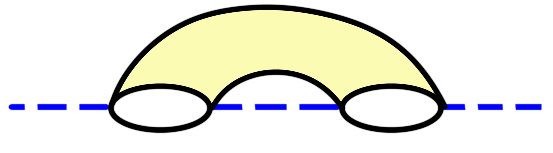}+\cdots
\end{align}

Since the non-zero eigenvalues of the matrix $M$ are the same as the eigenvalues of $\rho_\text{mixed BPS}$, the poles of the trace of the resolvent $R(\lambda) \equiv \Tr \mathbf R(\lambda)$, therefore, give the eigenvalues of the density matrix $\rho_\text{mixed BPS}$. Consequently, we will study the analytic structure of $\mathbf{R}_{ij}(\lambda)$ in order to determine the rank of $\rho_\text{mixed BPS}$. For example, we can use the resolvent to determine the trace of the $n$-th power of the density matrix through 
\be  
:\tr \left(\rho_\text{mixed BPS}\right)^n: = :\tr \left(M\right)^n:  =  \frac{1}{2\pi i}\oint d \lambda\, \lambda^{n} R(\lambda) .
\label{eq:tr_rho_n_power}
\ee
Rewriting things in terms of the resolvent is useful because we can compute it from the Schwinger-Dyson equations \cite{Penington:2019kki}. To write them down, one organizes the infinite expansion in terms of how many boundaries the first boundary is connected to. Then, between each of the connected boundaries one introduces a single power of the resolvent. This will take into account all the planar geometries with leading planar operator contractions contributing to the resolvent.  The resulting consistency equation takes the form 

\begin{multline}
 \includegraphics[valign=c,width=0.15\textwidth]{sd0.png}
=\includegraphics[valign=c,width=0.15\textwidth]{sd1.png} +\includegraphics[valign=c,width=0.2\textwidth]{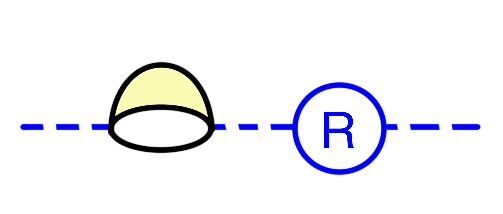}
+\includegraphics[valign=c,width=0.28\textwidth]{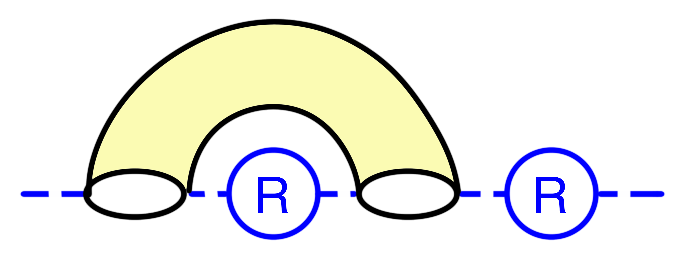}+\cdots   
\nn
\end{multline}

\be
\mathbf{R}_{ij}(\lambda) = \frac{\delta_{ij}}{\lambda} + \frac{1}{\lambda} \sum_{n=1}^\infty Z_n^\mathcal{O} R^{n-1} \mathbf{R}_{ij} (\lambda) ,
\label{SD}
\ee
where  
$Z_n^\mathcal{O}$ is the pinwheel geometry with $n$ boundaries and two operator insertions on each boundary. As explained in the previous subsection, in the ground state sector it has a simple form 
\be  
Z_n^\mathcal{O} =
e^{2\mathbf{S}_j} \, y^n ,
\qquad 
y \equiv e^{-S_0}
\frac{\Delta \Gamma(\Delta)^2 \,\Gamma\left(\Delta+\frac{1}{2}\pm j\right)}{2\pi \Gamma(2\Delta)} ,
\qquad
e^{\mathbf{S}_j} \equiv e^{S_0} \cos(\pi j ) .
\ee
If we take the trace of (\ref{SD}), we further get 
\be 
R(\lambda)=\frac{K}{\lambda} + \frac{1}{\lambda} \sum_{n=1}^\infty Z_n^\mathcal{O} R^{n} ,
\ee
which pictorially we could represent as
\be
R(\lambda)=\frac{K}{\lambda} + \frac{1}{\lambda} \sum_{n=1}^\infty\includegraphics[valign=c,width=0.23\textwidth]{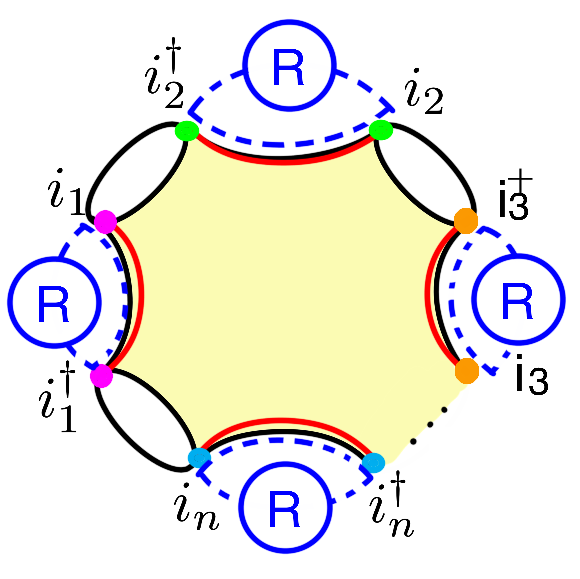}
\ee
Let us explain a bit the above drawing. The main object in the middle which consists of $n$ fully connected boundaries is the pinwheel $Z_n^{\mathcal{O}}$. It consists of $2n$ operators on the boundaries connected in the bulk through the red geodesics. Each of these geodesic connections essentially inserts a Kronecker delta in the operator indices, which is why the indices match at the ends of each geodesic drawn. On the other hand, matrix multiplication $\dots M_{i \, j} M_{j \, k} \dots$ also means that there is a matching index on each two most nearby boundaries. This is why the resolvent matrices $\mathbf{R}_{ij}$ in the drawing are connected to the pinwheel boundaries through a dotted blue index lines. The combined effect of the Kronecker delta from the geodesic and two index lines connecting the resolvent to the ends of the geodesic changes the resolvent insertion into its trace $R(\lambda)$. Because we assumed that all the operators have the same scaling dimensions $\Delta$, we can essentially combine the resolvent traces together into a factor $R^n$. 
Continuing with the computation, after inserting the explicit expression for the pinwheel we get
\begin{align}
R(\lambda) 
=\frac{K}{\lambda} + \frac{e^{2\mathbf{S}_j}}{\lambda} \frac{R y}{1-R y} .
\end{align}
Solving for $R(\lambda)$ in the above equation and choosing the solution that matches the asymptotic $1/\lambda$ behavior, we get
\be  
R(\lambda) = \frac{1}{2y} + \frac{K - e^{2\mathbf{S}_j}}{2\lambda} - \frac{\sqrt{(\lambda-\lambda_+)}\sqrt{(\lambda-\lambda_-)}}{2 \lambda y} , \qquad \lambda_\pm = y (\sqrt{K}\pm e^{\mathbf{S}_j})^2 .\label{Rresult}
\ee

\subsubsection*{Deriving the dimension of the Hilbert space}

With this, we can now evaluate the rank from \eqref{eq:tr_rho_n_power} with $n=0$. The resulting integral expression takes the form
\begin{align}
:\tr \left(\rho_\text{mixed BPS}\right)^n: 
= \frac{1}{2\pi i}\oint d \lambda\, \lambda^{n} R(\lambda)\,.
\label{productresult}
\end{align}
The contour should include all the poles or branch-cuts located along the real line at non-zero values of $\lambda$.  Thus, it should include the branch cut located between $\lambda_-$ and $\lambda_+$. Taking then $n \rightarrow 0$ and introducing the eigenvalue density 
\be 
D(\lambda) = \frac{1}{2\pi i}(R(\lambda - i \epsilon)-R(\lambda + i \epsilon)) ,
\ee
leads to the final expression 
\be
:\mathrm{rank}(\rho_\text{mixed BPS}):= 
\int_{\lambda_-}^{\lambda_+}d\lambda\, D(\lambda)=\begin{cases}K&K<e^{2\mathbf{S}_j}\\e^{2\mathbf{S}_j}&K>e^{2\mathbf{S}_j}\end{cases}\label{dimHbps}
 .
\ee
Even though the properties of operators, such as the scaling dimension, appear in the statistics of the inner product (\ref{productresult}), they don't appear in the final result of the rank. Intuitively this makes sense because the number of BPS states should be independent of the properties of the operator insertions. 

Thus, as we increase the size of the matrix $\rho_\text{mixed BPS}$ above $e^{2\mathbf{S}_j}$ its rank saturates, and all the new states added to the density matrix are linearly dependent on the first $e^{2\mathbf{S}_j}$ states. This identifies the two-sided dimension of the ground state Hilbert space as $\dim (\mathcal{H}_{\text{BPS}} \otimes \mathcal{H}_{\text{BPS}}) =  e^{2\mathbf{S}_j}$. 
As one could expect, the final expression for the rank takes a simple form of the integral of eigenvalue density over the region of nonzero eigenvalues. 

\subsubsection*{Eigenvalues of $M$: from maximally mixed at small $K$, to finding null states at large $K$}

\begin{figure}
    \centering
    \includegraphics[width=0.4\textwidth]{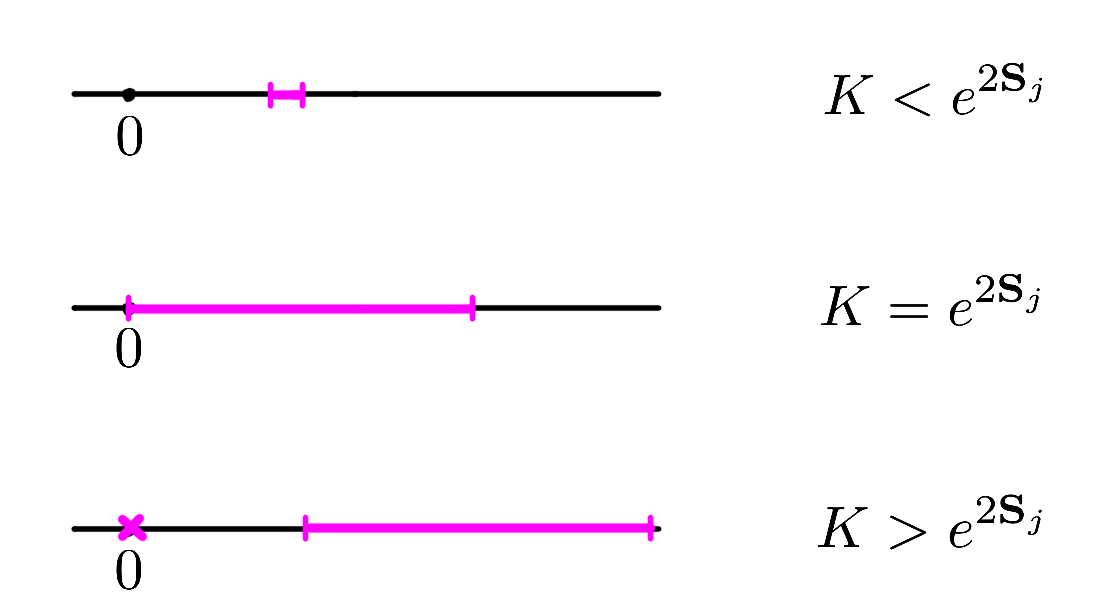}
    \caption{Branch-cut structure in the complex $\lambda$-plane for different values of $K$. This branch-cut gives rise to the density of eigenvalues $D(\lambda)$ shown in \eqref{eq:density-of-eigenvalues} whose behavior we analyze in detail in the paragraph below that equation. }
    \label{fig:brunchcut}
\end{figure}

To see what's happening in more detail, we can write the explicit form of the eigenvalue density of the matrix $M$ defined in \eqref{eq:matrix-M-defintion}. As a reminder, the positive eigenvalues of $M$ are the same as those of $\rho_\text{mixed BPS}$. The eigenvalue density can be extracted from the resolvent by discontinuity along the real axis and is given by:
\be  
\label{eq:density-of-eigenvalues}
D(\lambda) = \frac{\sqrt{(\lambda_+ - \lambda)}\sqrt{(\lambda - \lambda_-)}}{2\pi \lambda y}  + 
\delta(\lambda) (K - e^{2\mathbf{S}_j}) \Theta(K-e^{2\mathbf{S}_j}) 
.
\ee
Depending on the value of $K$, this density has nonzero support only at $\lambda = 0$ and in the region between $\lambda_-$ and $\lambda_+$. We should separate our analysis of the eigenvalue density into three regimes which we also represent in figure \ref{fig:brunchcut}:
\begin{itemize}
    \item $K<e^{2\mathbf{S}_j}$: In this regime the  matrix $M$ only has positive eigenvalues which therefore entirely agree with those of the density matrix $\rho_\text{mixed BPS}$.  When $K\ll e^{2\mathbf{S}_j}$, all eigenvalues are very close to each other (since $(\lambda_+ - \lambda_-)/\lambda_+ \ll 1$) and therefore the matrix $\rho_\text{mixed BPS}\sim \mathbf 1_{K \times K}$ is close to being maximally mixed. This is because the non-perturbative corrections to the inner products play only a small effect when the number of states entering $\rho_\text{mixed BPS}$ is small.
    \item $K=e^{2\mathbf{S}_j}$: In this regime, the eigenvalues range from their minimal possible value $\lambda_- = 0$ (since the matrix $M$ is positive semi-definite) to a maximal eigenvalue that is twice larger than the one encountered in the regime $K \ll e^{2\mathbf{S}_j}$. Consequently, due to the large span of the eigenvalues, $\rho_\text{mixed BPS}$ is very far from being maximally mixed: the small non-perturbative corrections to the inner products consequently cause a large effect due to the size of the matrix $M$. 
    \item $K > e^{2\mathbf{S}_j}$: Once the size of $M$ is larger than $e^{2\mathbf{S}_j}$, the additional $K - e^{2\mathbf{S}_j}$ eigenvalues of $M$ are zero. These are no longer eigenvalues of $\rho_\text{mixed BPS}$. Instead, the existence of such eigenvalues implies that there are linear combinations of the states $\ket{q_i}$ that vanish. Such linear combinations can be straightforwardly obtained from the eigenvectors of $M$ whose eigenvalue vanishes.\footnote{Suppose, $M_{ij} \psi_j= 0$. Then, $\ket{\psi} = \sum_{j=1}^K \psi_j \ket{q_j} = 0$ and obviously $\braket{\psi}{\psi}= 0$. }  Such linear combinations can be understood as null states of the bulk Hilbert space, since their norm is vanishing --- this is highly non-trivial from the bulk perspective: we started with states that were seemingly orthogonal but taking a linear combination of $K > e^{2\mathbf{S}_j}$  of such states can result in a null state.
\end{itemize}

\subsection{Suppressed non-perturbative corrections}
\label{subsec:non-perturbative}

The above computation only takes into account the leading disk geometries to the gravitational path integral. In this subsection, we will take a closer look at the non-perturbative corrections that the resolvent could receive. Potential sources of corrections come from geometries that have closed geodesics not intersecting any of the matter propagators, for instance, higher genus geometries, or from geometries that include defects, such as in the dimensional reduction of the orbifold geometries discussed in section \ref{subsec:Gibbons-Hawking-presc}. Below, we will analyze each such case individually. The result that we shall find is that the only geometries that contribute are those that have a single supersymmetric defect within a patch; all geometries that contain closed geodesics not intersecting any of the matter propagators or geometries with non-supersymmetric or a higher number of supersymmetric defects will all yield a vanishing contribution. 

\subsubsection*{The trumpet geometry }

We will first show that there is no contribution to our count of the BPS sector from geometries containing closed geodesics not intersecting any matter propagator. Before proving this, let us explain which geometries contain such geodesics. Let us describe each connected component $\mathcal M_{g,n}$ (a surface of genus $g$ with $n$ boundaries) of a spacetime by its decomposition into components $\mathcal M_{\Sigma_i}$ that we would obtain if cutting  $\mathcal M_{g,n}$ along the matter propagators connecting different boundary insertion points (with the number of disconnected components labeled by $i=1,\,\dots,\,p$). Each $\mathcal M_{\Sigma_i}$ has boundaries that consists of a union of geodesics and asymptotic boundaries, with each junction characterized by the outside angle between the two. Since there are $2n$-operator insertions on $\mathcal M_{g,n}$ and for each outside angle of $\mathcal M_{\Sigma_i}$ there is an angle in some component $\mathcal M_{\Sigma_k}$ such that the sum between the two angles is $\pi$, one can use the Gauss-Bonnet theorem to relate the Euler characteristic of $\mathcal M_{g,n}$ (denoted by $\chi_{g,n}$) to the characteristics of $\mathcal M_{\Sigma_i}$ 
(denoted by $\chi_{\Sigma_i}$):\footnote{This follows from writing the Gauss-Bonnet theorem in each patch, 
\be 
\frac{1}{2} \int_{\mathcal M_{\Sigma_i}} \sqrt{g} R + \int_{\partial_{\text{asymp.}} \mathcal M_{\Sigma_i}}\sqrt{h} K  + \sum_{j \in \text{Corners of } \mathcal M_{\Sigma_i}} \theta_{j}^{(i)}= 2\pi \chi_{\Sigma_i}\,,
\ee
where the integral over the boundary runs over the asymptotic boundary segments of $\mathcal M_{\Sigma_i}$ where the dilaton is fixed and $\theta_{j}^{(i)}$ is the exterior angle associated to each junction of the patch; along the geodesics, the contribution of boundary term is zero since $K=0$. Writing the Gauss-Bonnet theorem for the entire component $\mathcal M_{g,n}$ of the spacetime, we have 
\be 
\sum_i \left(\frac{1}{2} \int_{\mathcal M_{\Sigma_i}} \sqrt{g} R + \int_{\partial_{\text{asymp.}} \mathcal M_{\Sigma_i}}\sqrt{h} K  \right)= 2\pi \chi_{g,n}\,.
\ee
Using $\sum_{i=1}^p \sum_{\substack{j\in   \text{Corners} \\ \text{of } \mathcal M_{\Sigma_i}}} \theta_{j}^{(i)}  = 2n \pi $, \eqref{eq:Gauss-Bonnet-sum-of-patches} immediately follows.
}
\be 
\label{eq:Gauss-Bonnet-sum-of-patches}
\chi_{g,n}=\sum_{i =1}^p \chi_{\Sigma_i} - n\,.
\ee
Assuming that each component $\mathcal M_{\Sigma_i}$ has genus $g_i$ and $n_i$ number of disconnected boundaries, the above becomes 
\be 
\label{eq:relation-no-of-boundaries-genus}
2-2p = 2g - \sum_{i=1}^p (n_i+2g_i)\,.
\ee
Suppose we are interested in the case where all components  $\mathcal M_{\Sigma_i}$ would not have a closed geodesic on them. They would need to be homotopic to disks and therefore have $n_i =1$ and $g_i = 0$. In such a case, \eqref{eq:relation-no-of-boundaries-genus} becomes $2-p=2g$
which only has one solution (with non-zero $p$) if $g=0$ and $p=2$. Thus, if no component contains a closed geodesic, then each disconnected component of the full spacetime needs to only be separated into two polygon patches -- this is precisely the type of configuration we considered as the starting point in the calculation of the resolvent starting in 
\eqref{eq:pinwheel-gluing}.

As we shall see shortly, all other decompositions, that thus do have at least one component $\mathcal M_{\Sigma_i}$ containing a closed geodesic, yield a vanishing contribution to the gravitational path integral. To prove this, we shall show that first, the trumpet wavefunction is zero in the limit $\beta\rightarrow\infty$, and furthermore, that the result remains zero when gluing this trumpet to any other higher-genus or multi-defect surface. 

In Appendix \ref{app:trumpet}, we show that in $\mathcal{N}=2$ super-JT, the trumpet partition function is zero as $\beta\rightarrow\infty$. However, this is not sufficient for our purposes. We still need to show that after being glued to other geometries, trumpet wavefunction still contributes zero. Before studying the supersymmetric case, let us review why even if the trumpet wavefunction is zero, it is possible that it gives a non-zero contribution after being glued to other geometries. This is what, in fact, happens in non-supersymmetric ($\mathcal N=0$) JT gravity. In this case, the trumpet partition function is given by
\be
\label{eq:N=0-trumpet}
Z_{\text{Trumpet}}^{\mathcal N=0}(\beta,b)=\frac{e^{-\frac{b^2}{2\beta}}}{\sqrt{2\pi\beta}}=\int_0^\infty dE\,\underbrace{\frac{\cos(b\sqrt{2E})}{\pi\sqrt{2E}}}_{\rho(E,b)}e^{-\beta E}
\ee
This gives zero in the limit $\beta\rightarrow\infty$. But the non-supersymmetric partition function has a density of states $\rho(E)$ with support extending all the way to $E\rightarrow0$. When the trumpet is glued to other geometries along the neck $b$, we get an integral over $b$ of the trumpet wavefunction multiplied by a polynomial $P_k(b)$ with degree $k$ over $b$ 
\be
\int db\,Z_{\text{Trumpet}}^{\mathcal N= 0 }(\beta,b)P_k(b)\sim\beta^{k/2}
\ee
so the contribution from the trumpet is no longer zero in the limit $\beta\rightarrow\infty$. Therefore even if in the zero temperature limit (i.e. $\beta\rightarrow\infty$) $Z_{\text{Trumpet}}^{\mathcal N=0}=0$, after glued to other geometries, its contribution may not remain zero.

However, in Appendix \ref{app:trumpet} we showed that in the $\mathcal{N}=2$ super-JT, the density of states of the trumpet does not have support all the way up to $E\rightarrow0$, in contrast to the non-supersymmetric case where from \eqref{eq:N=0-trumpet} one can see that $\rho^{\cN=0}(E,b)$ has support all the way down to $E=0$. In microcanonical ensemble, in $\cN=2$ JT gravity, we find
\be
\rho^{\mathcal N=2}(E,b)=\Theta(E-E_\text{gap})f(E,b)\,.
\ee
In this case, if we glue the trumpet to some other geometries we get the integral
\be
\int db\,\Theta(E-E_\text{gap})f(E,b)P_k(b)=F(E)\Theta(E-E_\text{gap})\, ,
\ee
for some function $f(E)$. Above, $P_k(b)$ could capture the path integral over the $\cN=2$ super-JT moduli space for a bordered Riemann surface of some genus. For instance, in pure $\cN=2$ super-JT gravity  $P_k(b)$ is a polynomial in $b$ \cite{Turiaci:2023jfa}. The integration measure, which could be $b$ dependent can be absorbed into the definition of  $P_k(b)$.
Then in the canonical ensemble
\be
\int dE\,e^{-\beta E}F(E)\Theta(E-E_\text{gap})\sim e^{-\beta E_\text{gap}}f(\beta)\, ,
\ee
so as $\beta\rightarrow\infty$ the above gives zero. Thus, we find that a trumpet whose asymptotic boundary has $\beta \to \infty$ has a vanishing contribution to the path integral. 

\begin{figure}[t!]
    \centering
    \includegraphics[width=0.8\textwidth]{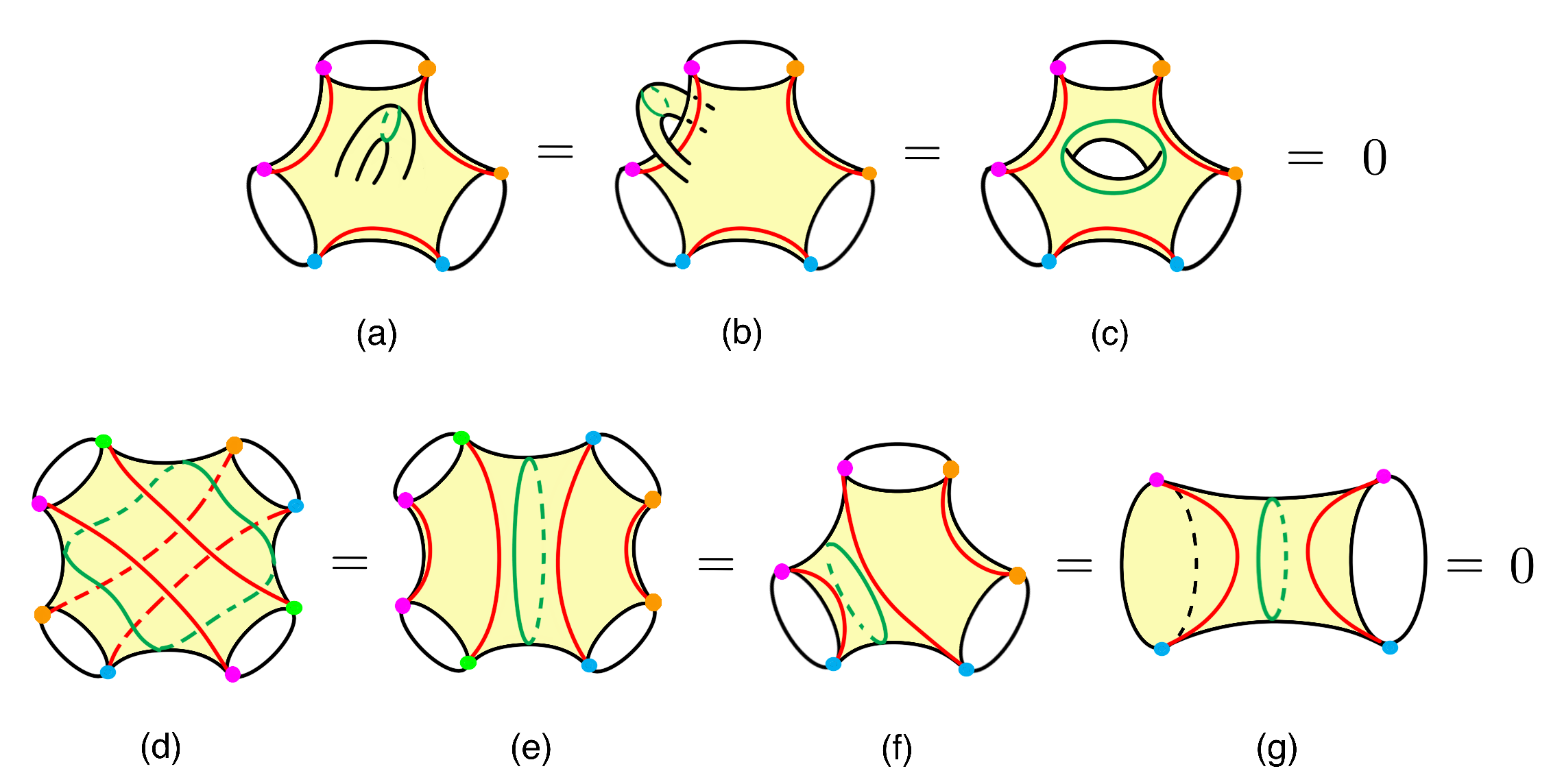}
    \caption{Examples of geometries around which the supergravity path integral yields a vanishing answer. In all cases there is a closed geodesic, shown in green, that does not intersect any of the matter propagators shown in red. The first line (figure a -- c) consists of higher genus geometries. The second line (figures d -- g) contains vanishing geometries that have genus zero but decompose into more than two patches, with at least one patch \textit{not} homotopic to a disk geometry.}
    \label{fig:pinwheelhandles}
\end{figure}

It then follows that the contribution of a geometry that has one closed geodesic (shown in green) and an asymptotic boundary consisting of a union of geodesic segments (shown in red) together with segments along which we fix the proper length and dilaton  (shown in black) vanishes if the total proper length of the latter segments tends to infinity,\footnote{This follows from requiring that when gluing the Hartle-Hawking wavefunctions given in \eqref{eq:Hartle-Hawking-wavefunction-N=2} along each geodesic segment of such a geometry we obtain the trumpet partition function. Since this requirement should hold for Hartle-Hawking wavefunctions with asymptotic boundaries of any possible proper length, this uniquely determines the contribution of the geometries in \eqref{eq:multi-geodesic-trumpet} and implies that when taking $\beta \to \infty$ all geometries containing a patch as shown in \eqref{eq:multi-geodesic-trumpet} should yield a vanishing contribution.     } 
\be 
\includegraphics[valign=c,width=0.3\textwidth]{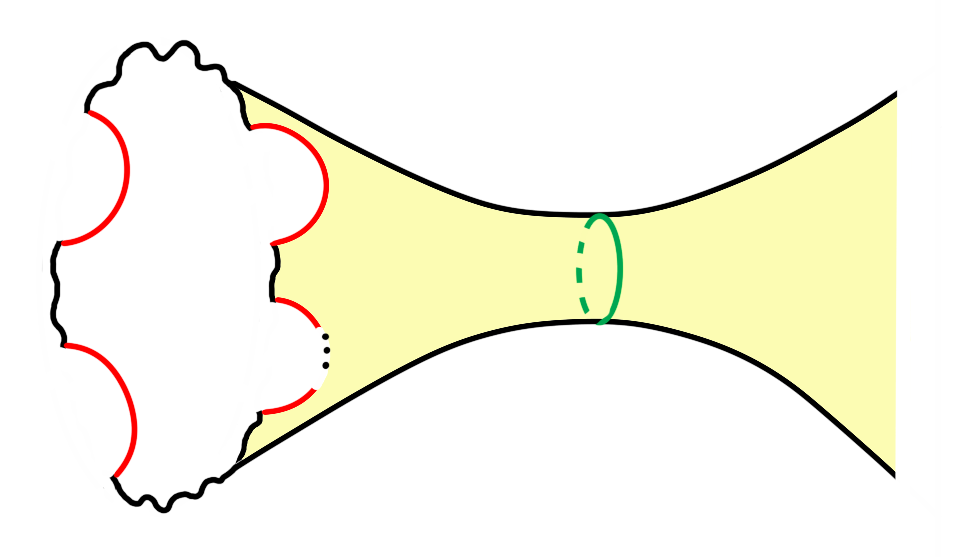} \, = 0\,,
\label{eq:multi-geodesic-trumpet}
\ee
regardless of the geometry that is connected to the right-hand side of the closed geodesic.
Consequently, all geometries that have a closed asymptotic geodesic in each patch vanishes and the only surviving geometry are those where each connected component has $g=0$ and where each connected component decomposes into a sum of two patches that are homotopic to disks. Examples of vanishing geometries are shown in figure \ref{fig:pinwheelhandles}.

\subsubsection*{Non-supersymmetric defect geometries} 

In addition to the geometries of higher topology one could also choose to include geometries which, in the two-dimensional reduction, include defects. Such defects might, in fact, originate from dimensionally reducing smooth higher-dimensional geometries and thus might need to be included in the gravitational path integral. Generically, such defects do not preserve any of the supersymmetries that are present at the disk level. In Appendix \ref{app:defect}, we show that the partition function of non-supersymmetric defects is similar to the trumpet and does not contribute to the BPS sector. For $\theta\neq1-2\phi$
\be
Z_{\text{defect}}=0\quad\quad\text{as}\quad\quad\beta\rightarrow\infty ,
\ee
and because the support has a gap starting above the extremal energy, the partition function of a polygon that includes such a non-supersymmetric defect is also vanishing in the limit in which the proper length of the asymptotic boundary is infinite (similar to how \eqref{fig:pinwheelhandles} yields a vanishing contribution). Thus, such defects do  not correct the computation of the resolvent described in section \ref{subsec:leadingorder}. 

\subsubsection*{The supersymmetric defect geometry}

Before we explain which geometries contribute to our resolvent computation, let us first understand which geometries contribute to the $\beta \to \infty$ partition function in the Gibbons-Hawking prescription. Besides the disk geometry, the only allowed other geometry are defects that preserve a subgroup of the $SU(1,1|1)$ isometry.  Such a defect has its geometric angle related to the $U(1)$ holonomy of the gauge field fixed around the defect (see appendix \ref{app:N=2-super-JT-part-functions}). Such geometries can preserve two out of the four Killing spinors of $SU(1,1|1)$ together with a $U(1) \times U(1)$ bosonic subgroup. For geometries that include more than one such defect no such Killing spinors exist globally and the supergravity path integral around such geometries turns out to yield a vanishing answer in the limit $\beta \to \infty$.

The geometries with a single supersymmetric defect can be viewed as the $2d$ dimensional reductions of the orbifold geometries that appear in the sum over geometries for the higher-dimensional black holes discussed in section \ref{subsec:Gibbons-Hawking-presc}. In those cases, the allowed values of the defect angles are determined by imposing that the higher-dimensional geometry be smooth. In most cases, such as for the orbifolded versions of the $1/16$-BPS black holes in AdS$_5$, this implies that the defect angle is equal to $2\pi/c$ with $c \in \mathbb Z$ with orbifolds having $c>1$. Thus, the sum over all geometries yields a partition function with $\beta \to \infty$, 
\be 
\label{eq:Gibbons-Hawking-with-non-pert-prediction}
Z_{BH}^{\beta \to \infty} = e^{S_0}
\cos{\pi j}
\left(1+\sum_{c} w_c\right) ,
\ee
where $w_c$ denotes the case weight associated to each defect and $S_0$ denotes the extremal entropy. For the higher-dimensional orbifolded geometries discussed above, $w_c$  is non-perturbatively small is $1/G_N$ and, to leading order, is given by $w_c = e^{ \frac{1-c}c S_0}$.

We can now compare \eqref{eq:Gibbons-Hawking-with-non-pert-prediction} to the corrected rank when including the contributions of supersymmetric defects. Following the reasoning above, polygons that include multiple supersymmetric defects yield a vanishing partition function when such polygons are glued to Hartle-Hawking wavefunctions whose boundary has an infinite proper length. Consequently, only one defect is allowed in each polygon. With this, we can very simply incorporate the SUSY conical defects correction into our resolvent analysis: a single SUSY defect of weight $w_i$ can be seen to modify the polygon partition function by multiplying its associated weight factor $w_c$. 
Since we're gluing together two different polygons across a geodesic weighted by $e^{-\Delta \ell}$, each such polygon could either have a defect or not. The total contribution from a wormhole with $n$-boundaries will now be given by a sum over SUSY conical defects with different weights. Thus, \eqref{eq:pinwheel-gluing} is modified to 
\be  
Z_{j,n}^{\mathcal{O}} \rightarrow 
Z_{j,n}^{\mathcal{O},\text{total}} =  
Z_{j,n}^{\mathcal{O}} \left(1+\sum_{c} w_c\right)^2\,.
\ee
This modifies the total resolvent
\begin{align} 
\label{eq:resolvent-including-non-perturbative-corr}
R_{\text{total}}(\lambda) &= \frac{1}{2y} + \frac{K - e^{2\mathbf{S}_j}
\left(1+\sum_{c} w_c \right)^2
}{2\lambda} - \frac{\sqrt{(\lambda-\lambda_+)(\lambda-\lambda_-)}}{2 \lambda y} , 
\\ 
\lambda_\pm &= y \left[\sqrt{K}\pm e^{\mathbf{S}_j}\left(1+\sum_{c} w_c \right)\right]^2 
.
\end{align}
The rank whose maximum captures the dimension of BPS Hilbert space now becomes 
\be
\label{eq:rank-with-SUSY-defects}
\mathrm{rank}(\rho_\text{mixed BPS})= \frac{1}{2\pi i} 
 \oint d\lambda \,  R = \int_{\lambda_-}^{\lambda_+}d\lambda\, D(\lambda)=
\begin{cases}
K,&K<\left(Z_{BH}^{\beta \to \infty}\right)^2
\\
\left(Z_{BH}^{\beta \to \infty}\right)^2, &K>\left(Z_{BH}^{\beta \to \infty}\right)^2
\end{cases} 
.
\ee
Thus, the results obtained from computing the maximal rank and those obtained from computing the degeneracy using the Gibbons-Hawking prescription are in complete agreement and include the same kind of non-perturbative corrections.

\begin{figure}[t!]
    \centering
    \includegraphics[width=0.8\textwidth]{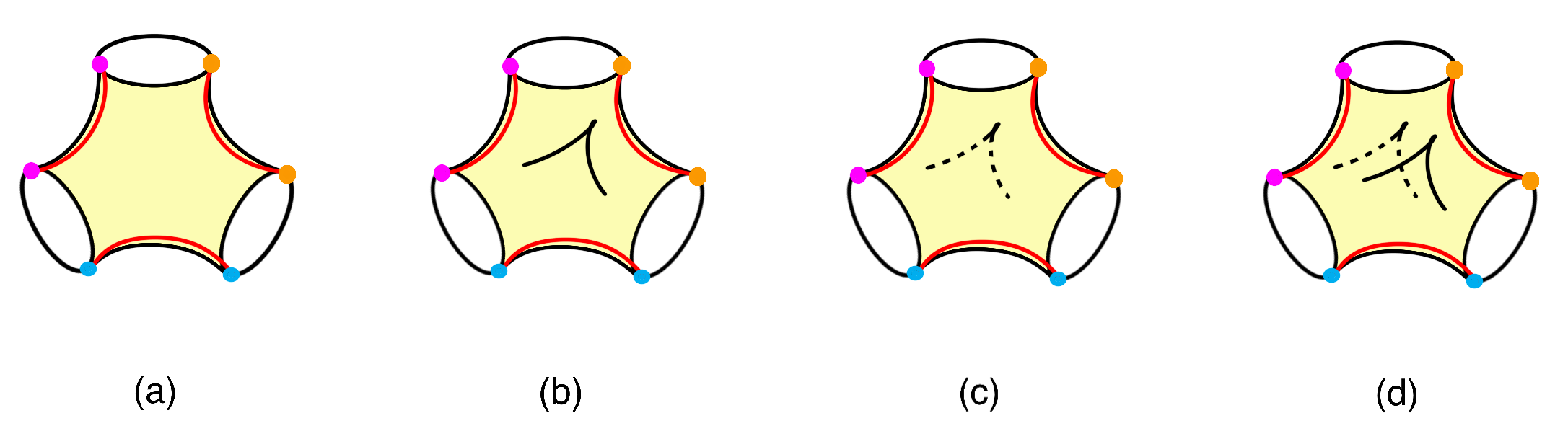}
    \caption{(a) A pinwheel without any defects (b) A pinwheel with one defect on the front polygon (c) A pinwheel with one defect on the back polygon (d) A pinwheel with two defects one on the front and one on the back}
    \label{fig:pinwheeldefects}
\end{figure}

\subsection{Interpretation in terms of boundary Haar-random states}
\label{subsec:Interpretation-Haar-random-states}

The results described above are completely consistent with the identification of the states $\ket{q_i}$ as Haar-random in a Hilbert space that is $\left(Z_{BH}^{\beta \to \infty}\right)^2$-dimensional. Consider the states 
\begin{align}
\label{eq:Haar-random-states}
\ket{q_i} &= \frac{e^{-S_0}}{2\pi} 
\sqrt{\frac{\Delta}{\Gamma(2\Delta)}} \Gamma (\Delta) 
\Gamma(\Delta + \frac{1}{2} +j) 
\sum_{a=1}^{
e^{2\mathbf{S}_j}
} C_{ia} \ket{\text{BPS}_a^{\otimes 2}}
, 
\\
\bra{q_j} &= 
\frac{e^{-S_0}}{2\pi} 
\sqrt{\frac{\Delta}{\Gamma(2\Delta)}} \Gamma (\Delta) 
\Gamma(\Delta + \frac{1}{2} -j) 
\sum_{b=1}^{
e^{2\mathbf{S}_j}
} C_{jb}^* \bra{\text{BPS}_b^{\otimes 2}} 
,
\end{align}
where the states $\ket{\text{BPS}_a^{\otimes 2}}$ (or $\bra{\text{BPS}_a^{\otimes 2}}$)  form a complete orthogonal basis of states in the two-sided Hilbert space $\mathcal H_\text{BPS} \otimes \mathcal H_\text{BPS}$ whose dimension is $(Z_{BH}^{\beta \to \infty})^2 = \left(\dim \mathcal H_\text{BPS}\right)^2$ and the coefficients $C_{ia}$ are independent random numbers taken from a complex Gaussian distribution with\footnote{The width of this Gaussian distribution does not affect the statistics of the inner products when the normalization of such states is taken into account.  }
\be 
\mathbb{E}_C (C_{ia} C^*_{jb}) 
= 
\delta_{ij} \delta_{ab} .
\label{eq:statistic-of-C}
\ee
Just like the states obtained from the gravitational path integral, any two such states are not exactly orthogonal but rather have an inner product that is suppressed by the size of the Hilbert space. The statistics of products of all the inner products between such states precisely take the value determined by our gravitational calculation above  (see also appendix D of \cite{Penington:2019kki} in which similar states were considered).
For example, by using \eqref{eq:statistic-of-C} one can now explicitly verify that we can reproduce gravitational answers as 
\be 
\mathbb{E}_C\left(\bra{q_i} \ket{q_j}\right) = \langle 2\text{pt} \rangle_{j, \text{ disk}} \, \delta_{ij} , 
\qquad 
\mathbb{E}_C\left( \bra{q_i} \ket{q_j}
\bra{q_k} \ket{q_l} \right)
 = \delta_{ij} \delta_{kl} \,
\langle 2\text{pt} \rangle^2_{j, \text{ disk}} + 
\delta_{il} \delta_{jk} \, Z^{\mathcal{O}}_{j,\, 2} ,
\label{eq:statistic-inner-product}
\ee
where for the two-point function squared each term comes from a specific Wick contraction of $CC^*$ in the Gaussian integral.  The above observation generalizes now to any number of boundaries, and the ensemble reproduces all possible nonzero contributions to the gravitational path integral with two operator insertions on each boundary. In particular, one can verify that each Wick contraction appearing in a product of Gaussian random coefficients $C_{ia}$  corresponds to a unique geometric contribution to the path integral. Thus, taking the average over the $C$'s and $C^*$'s exactly reproduces the answer over the pinwheel geometries that we considered above 
\be 
\mathbb{E}_C\left(\bra{q_{i_1}} \ket{q_{i_{2}}} 
\bra{q_{i_3}} \ket{q_{i_{4}}}
\dots 
\bra{q_{i_{2n-1}}} \ket{q_{i_{2n}}} \right)
=  \sum_{\substack{\text{all possible}\\ \text{geometries}}} \includegraphics[width=0.3\textwidth,valign=c]{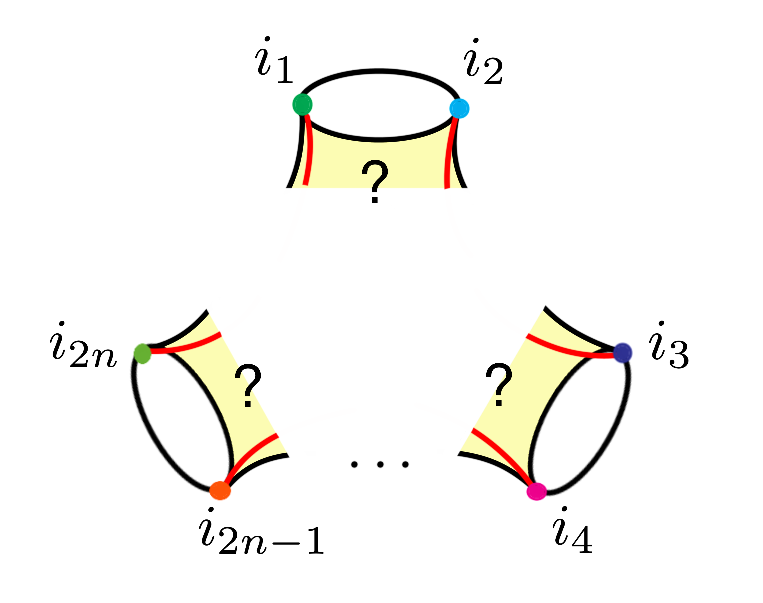}\,,
\label{eq:higher-order-moment-Gaussian}
\ee
where the right-hand side includes a sum over all possible allowed geometries connecting the $n$ boundaries. 
For example, it is easy to see that the leading fully connected contribution coming from pinwheel geometries with $n$-boundaries, $Z_{j, \, n}^\mathcal{O} \left(\delta_{i_2 i_3} \dots \delta_{i_{2n-2} i_{2n-1}} \delta_{i_{2n}i_1} +\text{all fully cyclical contractions} \right)$, is exactly reproduced by taking all Wick contractions of $C$'s and $C^*$'s that are fully cyclical. 
The right-hand side of \eqref{eq:higher-order-moment-Gaussian} includes all the geometries that contribute to the inner product, including the non-planar contributions that we had previously neglected in the computation of the resolvent. 

In matching \eqref{eq:statistic-inner-product} and \eqref{eq:higher-order-moment-Gaussian} to our gravitational computation we have used the fact that we are in the ground state sector and the trumpet is vanishing. For instance, if the trumpet did not vanish there would be additional geometries containing $(ij)$ and $(kl)$ contractions on a connected cylinder geometry, 
\be
\mathbb{E}_C\left(\bra{q_i} \ket{q_j} \bra{q_k} \ket{q_l}\right) \not\supset \underbrace{\includegraphics[valign=c,width=0.2\textwidth]{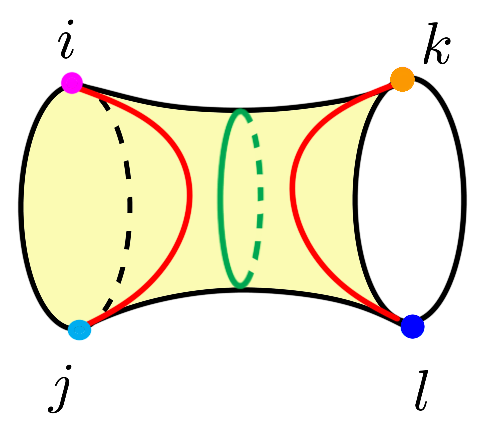}}_{\substack{\text{Vanishing in SUGRA}\\ \text{for } \beta \to \infty}
}
\ee 
which would not be reproduced in the model of Gaussian random states in \eqref{eq:Haar-random-states}. Thus, the vanishing of trumpets is crucial in matching the statistics of inner products in gravity to the statistics of inner products of the Gaussian Haar random states in \eqref{eq:Haar-random-states}.\footnote{In fact, the fact that trumpet geometries vanish in our case also leads to the main difference between the ensemble that we consider and the ensemble considered in \cite{Penington:2019kki}. In \cite{Penington:2019kki}, the Gaussian random ensemble needs to be coupled to the JT matrix integral. In their case, wormhole geometries capture the fluctuation of eigenvalues in the JT matrix integral. In our case, since the number of BPS states is fixed and because the energies of BPS states are not fluctuating, the Gaussian random coefficients $C_{ia}$ are all that is needed in order to specify the states $\ket{q_i}$. We thank D.~Stanford for several discussions about these differences. } 

The only additional non-perturbative corrections that we have encountered in the gravitational path integral in section \ref{subsec:non-perturbative} came from supersymmetric defects that could appear in each polygon that is part of a pinwheel geometry. In the language of Haar random states, the addition of such geometries simply change the overall dimension of the Hilbert space: the index $a$ in \eqref{eq:Haar-random-states} has a range from 1 to $(Z_{BH}^{\beta \to \infty})^2$ when including these non-perturbative corrections instead of going from 1 to $e^{2 \mathbf{S}_j}$ if we only wanted to reproduce the answers from the leading geometries. Including such defects, however, leaves the statistics of the coefficients $C$ completely unaffected. 

 Since all the statistics of the inner products of the states \eqref{eq:Haar-random-states} completely agree to all orders in $1/G_N$ to what we have found from the supergravity path integral, the statistics of the eigenvalues of the density matrix $\rho_\text{mixed BPS}$ will also agree. The result of the rank obtained in gravity in \eqref{eq:rank-with-SUSY-defects}, therefore, also agrees with the corresponding rank of the density matrix constructed from the states in \eqref{eq:Haar-random-states}.

The existence of null states is also clear in the interpretation in terms of Haar random states. Once $K=Z_{BH}^{\beta \to \infty}$, the states span the entire Hilbert space, and if more vectors are added there are always linear combinations of the Haar random states that vanish. 

\subsection{A vanishing standard deviation}
\label{sec:vanishing-standard-deviation}

Above, we computed the dimension of the black hole BPS Hilbert space by calculating $:\mathrm{rank}(\rho_\text{mixed BPS}):$. Nevertheless, since the calculation of this rank in (\ref{dimHbps}) includes wormhole contributions, one might think that the rank itself is an ensemble average rather than an exact count over states. In this subsection, we argue that this is not the case -- even though wormholes contribute to the computation of the rank itself, the standard deviation of this rank vanishes.

It is difficult to compute the standard deviation of the rank from the resolvent method described above -- this would involve finding a Schwinger-Dyson equation for a product of two resolvents, which involves a more intricate geometric expansion. Rather, we will proceed by using the replica trick, writing the standard deviation of the rank as 
\begin{align}
&\sigma_{\mathrm{rank}(\rho_\text{mixed BPS})}  = \sqrt{:\mathrm{rank}(\rho_\text{mixed BPS})^2:-:\mathrm{rank}(\rho_\text{mixed BPS}):^2}
= \nonumber \\  &\qquad=
\lim_{n\rightarrow0}\sqrt{:\tr \left(\rho_\text{mixed BPS}^n\right)\,\tr \left(\rho_\text{mixed BPS}^n\right): - :\tr \left(\rho_\text{mixed BPS}^n\right):^2}
\,.
\end{align}
We would thus like to compute the difference in the latter line for all $n$ and analytically continue the result to $n \to 0$. To get a better sense of $
:\mathrm{tr}(\rho_\text{mixed BPS})^n\mathrm{tr}(\rho_\text{mixed BPS})^n:$, let us first determine what geometries contribute to the variance in the $n=2$ case first
\begin{align}
\label{eq:example-standard-deviation}
&: \mathrm{tr}(\rho_\text{mixed BPS})^2\mathrm{tr}(\rho_\text{mixed BPS})^2: - \left(: \mathrm{tr}(\rho_\text{mixed BPS})^2:\right)^2= \\ &=\sum_{i_L,j_L,i_R,j_R=1}^K :\bra{q_{i_L}}\ket{q_{j_L}}\bra{q_{j_L}}\ket{q_{i_L}}  \bra{q_{i_R}}\ket{q_{j_R}}\bra{q_{j_R}}\ket{q_{i_R}}: - :\bra{q_{i_L}}\ket{q_{j_L}}\bra{q_{j_L}}\ket{q_{i_L}} :: \bra{q_{i_R}}\ket{q_{j_R}}\bra{q_{j_R}}\ket{q_{i_R}}: \nonumber
\end{align}
This difference is graphically represented  as\footnote{The geometries shown below can also be precisely reproduced from the Wick contractions in the Gaussian random model of the boundary states discussed in section \ref{subsec:Interpretation-Haar-random-states}.}
\begin{align}
&\sum_\text{geometries}\sum_{i_L,j_L,i_R,j_R=1}^K \left[\includegraphics[valign=c,width=0.32\textwidth]{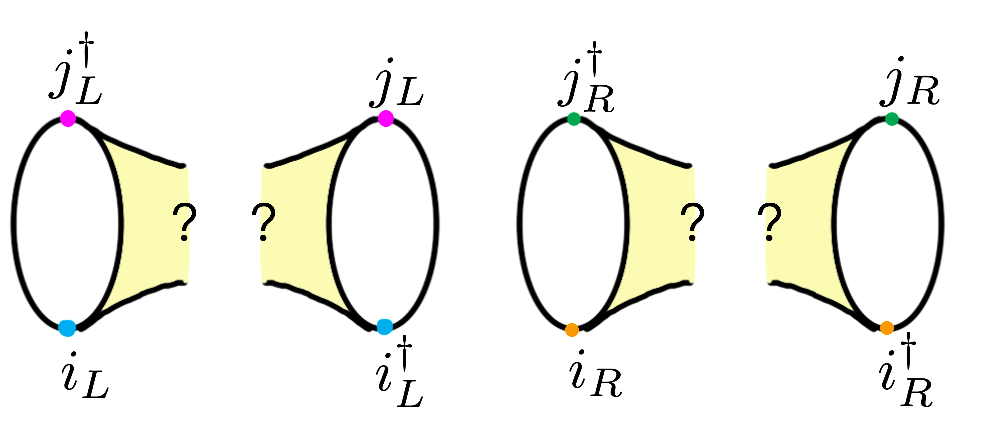}\nonumber -\left(\includegraphics[valign=c,width=0.16\textwidth]{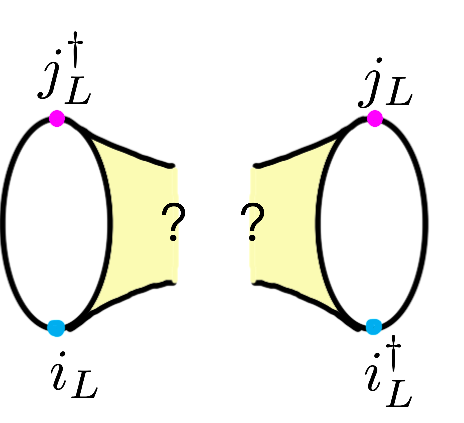}\right)\left(\includegraphics[valign=c,width=0.16\textwidth]{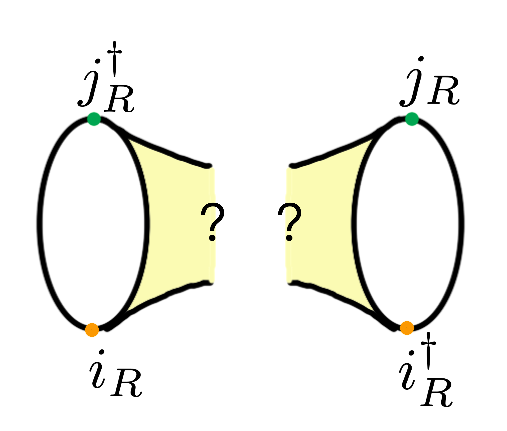}\right) \right]\nonumber\\
=&\,\bigg\{\bigg( \includegraphics[valign=c,width=0.26\textwidth]{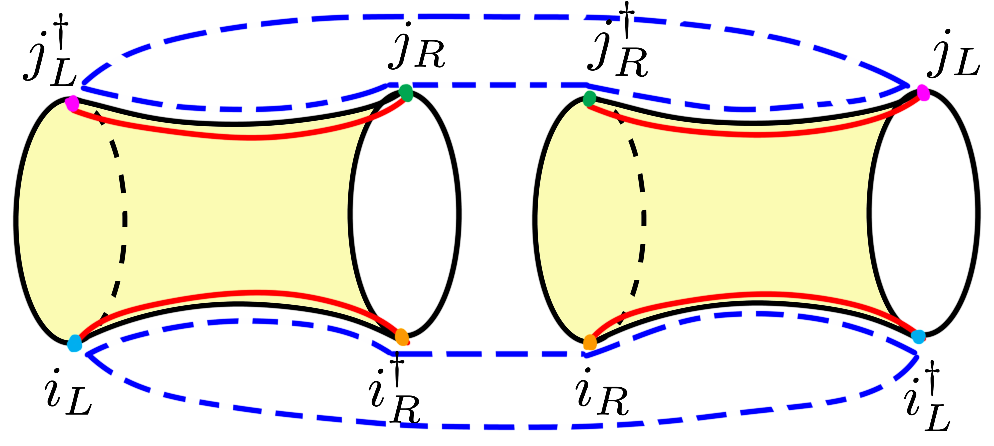}+\text{1 ind.~perm.}
\bigg)+\bigg(\includegraphics[valign=c,width=0.26\textwidth]{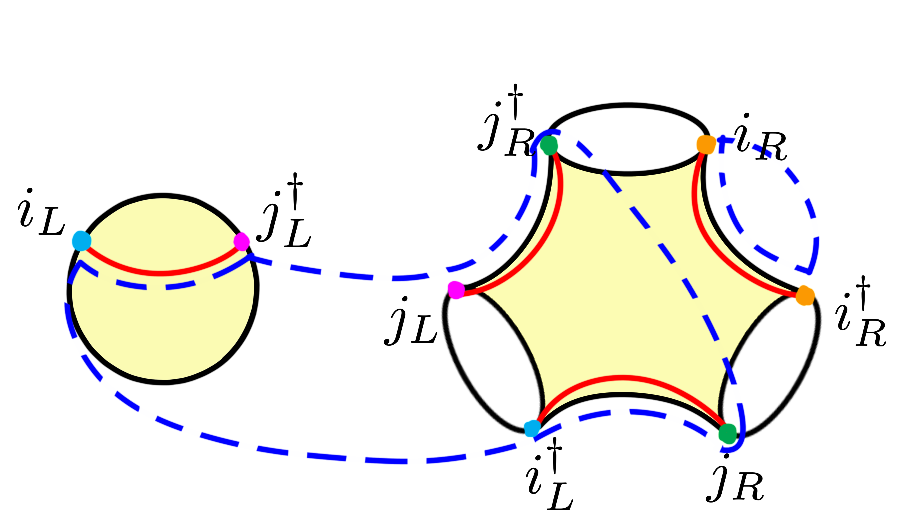}+\text{7 ind.~perm.} \bigg)\bigg\}_{K^2e^0}\nonumber\\
+&\bigg\{\includegraphics[valign=c,width=0.26\textwidth]{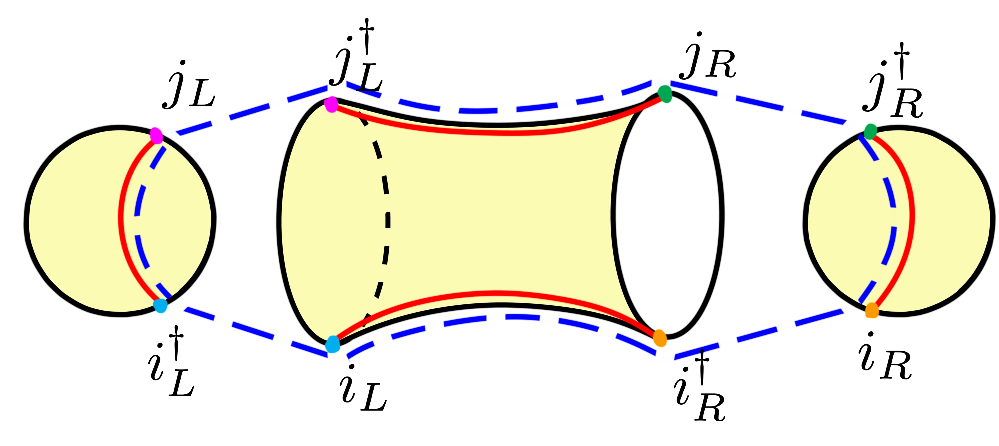}+\text{3 ind.~perm.} \bigg\}_{Ke^{2S_0}}\nonumber
+\bigg\{\includegraphics[valign=c,width=0.2\textwidth]{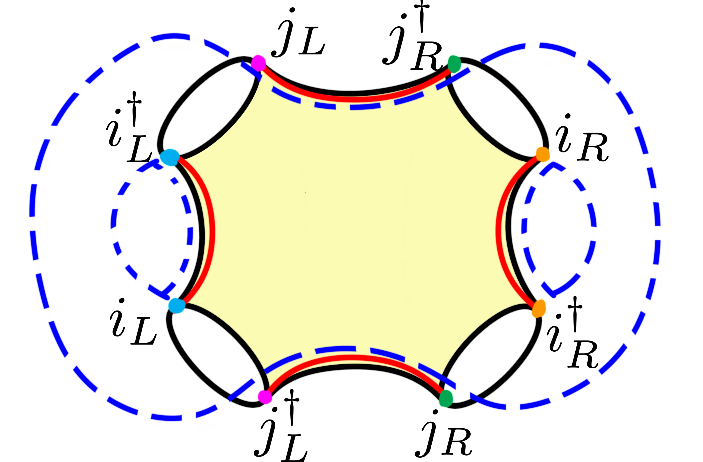} +\text{3 ind. perm.}  \bigg\}_{K^3e^{-2S_0}}\nonumber\\
+& \bigg\{\text{2 subleading geometries of }\bigg\}_{K e^{-2S_{0}}} \nonumber\\
+&\includegraphics[valign=c,width=0.2\textwidth]{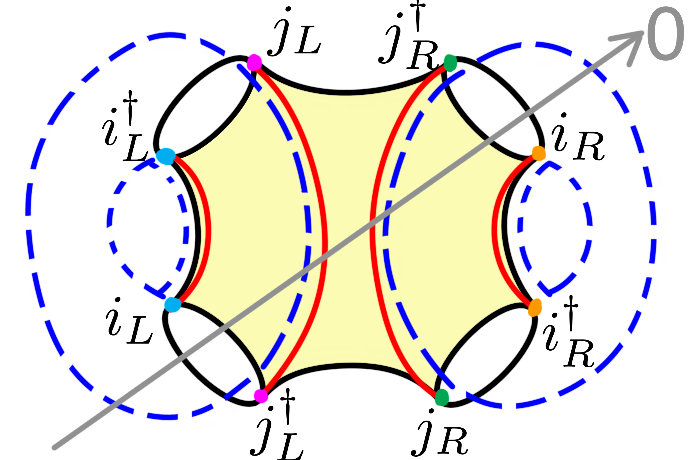}+\includegraphics[valign=c,width=0.3\textwidth]{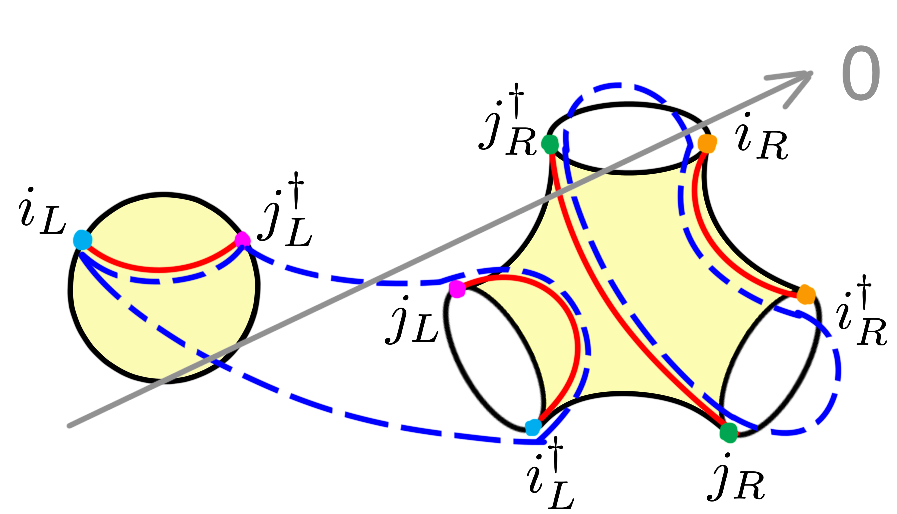} + \text{ higher genus vanishing cont.} \label{sdexample}
\end{align}
where blue dashed lines represent index contractions. As before, each index loop contributes $K$, and each geometry contributes $e^{\chi_{g,n} S_0}$ where a geometry with genus $g$ and $n$ boundaries has Euler character $\chi_{g,n}=2-2g-n$. We have separated the indices between the two copies of the trace by an L or R subscript indicating whether the operator index comes from an inner product in the left trace or from an inner product coming from the right trace.

Above, we only show the leading geometries in that contribute in the limit $K
,\, e^{2S_0} \gg 1$, keeping $K/e^{2S_0} \sim O(1)$. To the right of each curly brackets we indicate the order in $K$ and $e^{S_0}$ at which each geometry contributes with a subscript. 

The geometries in the last line vanish. This is because when taking $\beta \to \infty$ and performing the supergravity path integral, 
the geometry always involves a double-trumpet (whose closed cycles are indicated in the example in \eqref{eq:multi-geodesic-trumpet}), which, as above, yields a vanishing answer. Similarly, higher genus wormholes connecting inner products with left and right indices also give no contributions because they similarly involve a vanishing double trumpet.  The geometries that survive are those that not only connect one of the left inner products in the second line of \eqref{eq:example-standard-deviation} to one of the right inner products through a wormhole (a requirement for a geometry to contribute to the standard deviation of $\tr(\rho_\text{mixed BPS})^2$ and a feature of all geometries in \eqref{sdexample}) but also necessarily include at least one geodesic  (but possibly more geodesics) connecting an operator with left indices to an operator with a right index (this is not the case for the vanishing geometries in the last line of \eqref{sdexample}). For instance, in both the second and third lines of \eqref{sdexample}, there is a geodesic in each geometry connecting the operators with indices $\mathcal O_{j_L}$ and $\mathcal O_{j_R}^\dagger$. 

This requirement holds not only for $n=2$ but for arbitrary $n$ as well. Whenever we have a wormhole geometry that involves both operators with both left and right indices  in which no operator with a left index is connected to an operator with a right index, we can always separate the geometry into two parts -- one with boundaries that only involve left and one with boundaries that only involve right indices -- by a closed curve that does not intersect any of the red matter geodesics, 
\be 
\includegraphics[valign=c,width=0.2\textwidth]{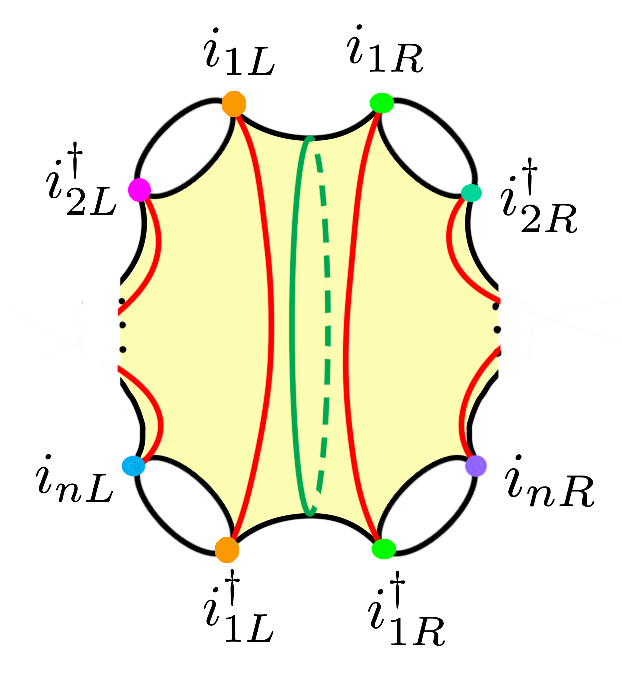} = 0\,.\label{eq:geometry-with-left-and-right-indices-separated}
\ee
Note that while such geometries yield a vanishing contribution in a supergravity path integral where we are taking all proper boundary lengths to be infinite, they would not vanish when computing the standard deviation of similar density matrices in non-supersymmetric theories (or in a supergravity path integral where we are not projecting to the BPS sector). 

Since there are $n$ left inner products and $n$ right inner products, we see that every geometry has a combinatorial factor at least proportional to $n$ coming from choosing a specific operator with a left index to connect through a propagator to any one of the $n$ operators with a right index.\footnote{In most cases, the combinatorial factor is, in fact, proportional to $n^2$ since we can connect any of the $n$ operators with a left index to any of the $n$ operators with a 
right index. However, the second line of \eqref{sdexample}, which includes two cylinders, provides a useful counterexample in which the combinatorial factor is solely proportional to $n$. This is because, in this figure, once we choose a pairing between a specific left operator and any of the $n=2$ right operators, then all other pairings are determined.} Thus, after fixing a connection between some left operator and a specific right operator to obtain an overall proportionality factor of $n$, we  can compute the sum over all remaining possible geometries with all possible index contractions, 
\be 
(\sigma_{\text{Tr}(\rho_\text{mixed BPS}^n)})^2 = n\left(\includegraphics[valign=c,width=0.35\textwidth]{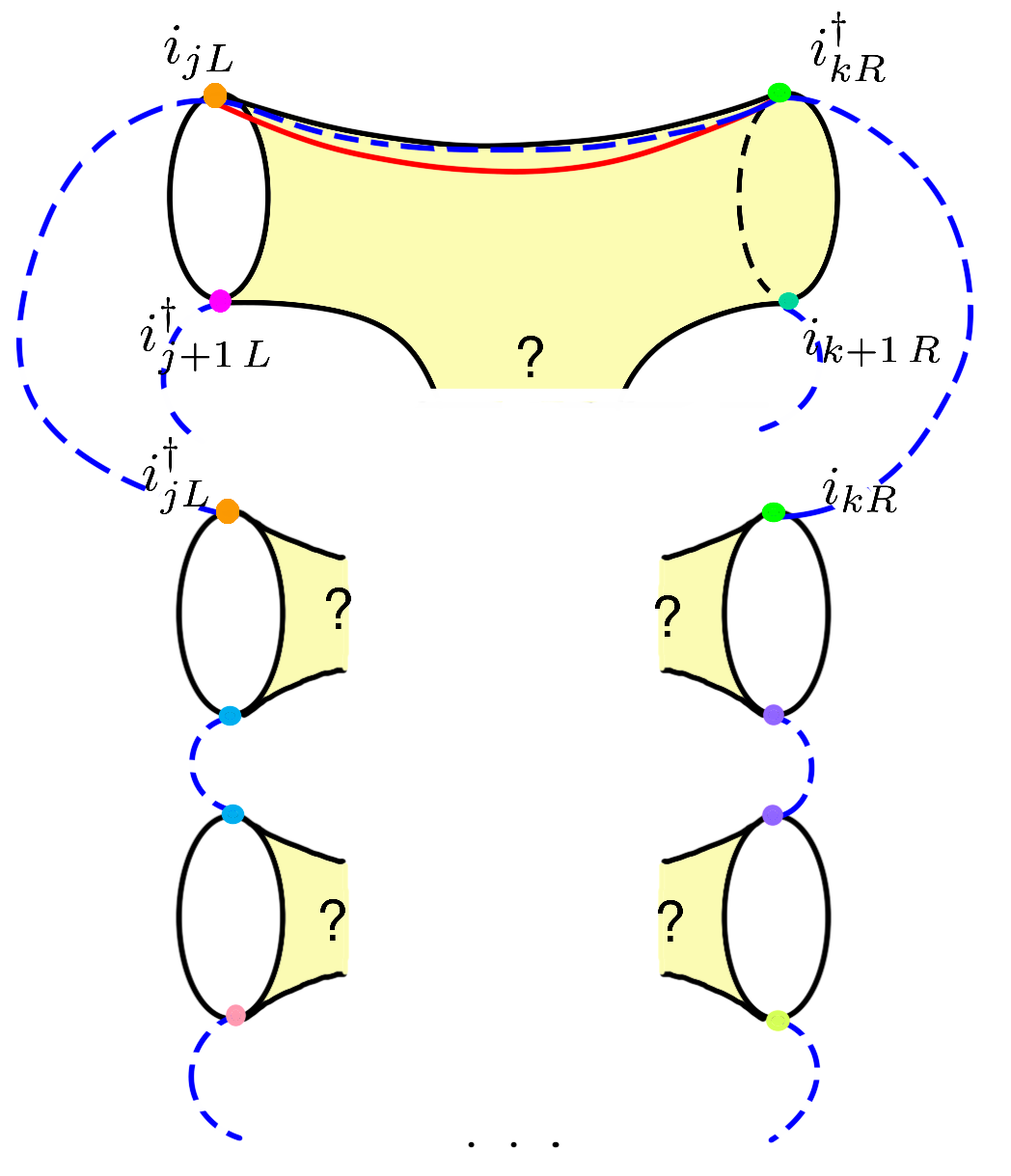}\right)\,,
\ee
where in the figure above, we either fully fix the $j$-index on the L operators or the $k$-index on the R operators. 
 As long as the analytic continuation of this remaining contribution (shown in the parenthesis) does not have a pole as $n \to 0$, the overall proportionality factor of  $n$ would then imply $\sigma_{\mathrm{rank}(\rho_\text{mixed BPS})} = 0$. We have indeed checked the analytic continuation in $n$ for a variety of geometries and never found a pole as $n \to 0$.  For instance, consider the fully connected geometries (such as the one in the third line of \eqref{sdexample}), which provide the seemingly dominant contribution to all standard deviations in the limit $K \gg e^{2S_0}$. The contribution of such geometries is given by 
 \begin{align}
(\sigma_{\text{Tr}(\rho_\text{mixed BPS}^n)}^\text{fully-connected})^2 &= \includegraphics[valign=c,width=0.25\textwidth]{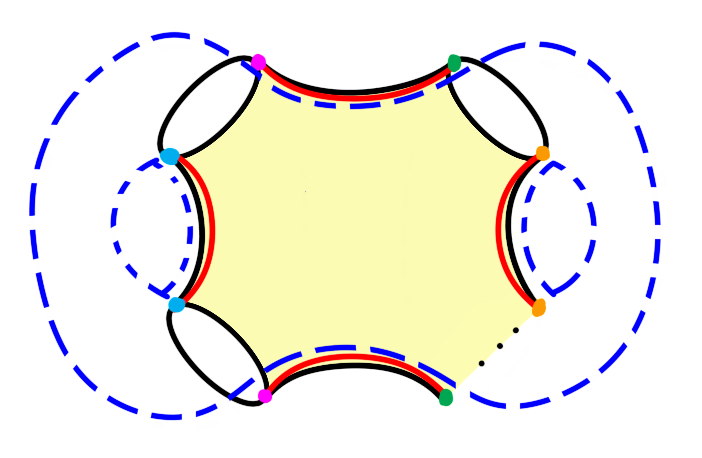}  \nonumber \\ &= n^2 K^{2n-1} e^{S_0(2-2n)} \cos^2(\pi j) 
\left(
\frac{\Delta \Gamma(\Delta)^2 \,\Gamma\left(\Delta+\frac{1}{2}\pm j\right)}{2\pi \Gamma(2\Delta)}
\right)^n 
.
 \end{align}
from which $\sigma_{\text{rank}(\rho_\text{mixed BPS})}^\text{fully-connected} = 0$. We have obtained similar results for the analytic continuation when including numerous other geometries in the standard deviation. To summarize, due to the vanishing contribution of numerous wormhole geometries, we have thus argued that 
\be 
\sigma_{\text{rank}(\rho_\text{mixed BPS})} = 0\,,
\ee
which implies that the rank itself is the same in all members of the ensemble, even though the computation of the rank itself involves numerous connected geometries. Adding the non-perturbative corrections due to supersymmetric defects does not alter our conclusion about the standard deviation vanishing. Including such defects simply rescales the contribution of each polygon involved in the calculation above by a factor of $\left(1+\sum_{c} w_c\right)$ without changing any of the combinatorial factors that result from index contractions. 

The fact that the standard deviation is exactly vanishing is also consistent with the interpretation of our results in terms of Haar-random states. Specifically, one can compute the standard deviation of the rank by explicitly computing the resolvent two-point function in the Gaussian model that specifies the boundary states discussed in 
section \ref{subsec:Interpretation-Haar-random-states}. Taking the double contour integral of this resolvent two-point function one finds the vanishing standard deviation for the rank. Nevertheless, the fact that this standard deviation vanishes has a more intuitive explanation. The density matrix $\rho_\text{mixed BPS}$ is formed from $K$ Haar-random states \eqref{eq:Haar-random-states} which live in a Hilbert space whose dimension is $\left(\dim \mathcal H_\text{BPS}\right)^2$. The only way for the rank of the matrix $\rho_\text{mixed BPS}$ to not become equal to $K$, when $K 
\leq \left(\dim \mathcal H_\text{BPS}\right)^2$, is if the $K$ generated Haar-random states $\ket{q_i}$ are linearly dependent. Since our results for the statistics of the inner products are consistent with the states being chosen from Gaussian independent distributions as in \eqref{eq:Haar-random-states}, the probability of picking a Haar-random state that is exactly a linear combination of the previously chosen $K-1$ states is zero: this would require a randomly chosen vector in a $K$ dimensional space to lie precisely on the $K-1$-dimensional plane spanned by the previously chosen vectors when $K< \left(\dim \mathcal H_\text{BPS}\right)^2 $. Similarly, when $K \geq \left(\dim \mathcal H_\text{BPS}\right)^2$, the Haar random states always span the Hilbert space and once again the standard deviation of the rank vanishes. This thus indicates that the formula for the rank seen in \eqref{eq:rank-with-SUSY-defects} (which also includes non-perturbative corrections) is, in fact, exact even though the states that enter in $\rho_\text{mixed BPS}$ are randomly chosen. The gravitational calculation of the vanishing standard deviation described above is thus another highly non-trivial check of this Haar-random state interpretation.

 Finally, we should again contrast this to how the standard deviation of similar ranks would behave in non-supersymmetric theories or when not projecting down to the BPS sector. As described above, there is no reason for the geometries such as that shown in \eqref{eq:geometry-with-left-and-right-indices-separated} to vanish. Including such geometries does not require us to connect operators with a left index to those with a right index, and therefore the argument that all combinatorial factors are proportional to $n$ does not hold. Specifically, in the fully connected geometry exemplified in \eqref{eq:geometry-with-left-and-right-indices-separated}, all operators are contracted with their conjugates which, consequently, does not lead to additional combinatorial factors proportional to $n$.  Thus the analytic continuation of this result to $n \to 0$ gives a non-vanishing answer resulting in a non-vanishing standard deviation for the rank of non-BPS states. This result is also consistent with the interpretation in terms of Haar-random states which applies equally well to the case of non-supersymmetric states. Nevertheless, in contrast to the BPS case, for non-supersymmetric black holes, the dimension of the Hilbert space within a fixed energy window is fluctuating. Therefore, while the argument about the Haar-random states having a zero probability of becoming linearly dependent as long as $K< \left(\dim \mathcal H_\text{BPS}\right)^2 $ still holds, the standard deviation seen through the contribution of the non-vanishing wormholes should capture the fluctuation in the number of states within the energy window rather than some property of the $K$ Haar-random states.

Our discussion of vanishing standard deviation can be nicely compared with the analysis of \cite{Bousso:2023efc} of the standard deviation of the rank in the context of the PSSY end of the world brane model at fixed energy. There, one has to consider {\it non-tubular} wormholes, analogous in our case to wormholes supported by matter, and {\it tubular} wormholes, analogous to wormholes not supported by matter. The authors find that the non-tubular wormholes capture a random pure state behavior for a system with the fixed Hilbert space dimension, and therefore do not lead to any fluctuations in the rank of the density matrix. On the other hand, the tubular wormholes lead to fluctuations in the rank and capture the fact that the number of states in a given energy window can fluctuate. This is exactly consistent with our results, where we find that the wormholes supported by matter (non-tubular) capture the Gaussian random state behavior and the empty wormholes do not contribute because we're working with exactly degenerate states below the energy gap.\footnote{We thank Masamichi Miyaji for interesting discussions about this.}

\subsection{Sources of error}
\label{subsec:error}

\subsubsection*{Non-planar geometries}
In order to be able to use the resolvent method in the above derivation of the rank, we have focused only on the contributions coming from planar geometries. This means that different connected parts of the geometry never had to cross ``under" each other to connect to other boundaries.
In principle, we should also include non-planar geometries in the full computation of the resolvent.
These, however, will be suppressed in $1/K$ compared to the leading contribution with the same number of boundaries because such geometries lead to a structure of index contractions that contain a smaller number of index loops.
Similarly, we also ignored planar geometries with index structure that has a subleading number of index loops, which are therefore also suppressed in $1/K$. An example would be a pinwheel of \eqref{eq:pinwheel-gluing}  
with boundaries permuted but matter geodesics kept in the same "nearest neighbor" configuration. 
Including both of these contributions would lead to $O(1)$ additive error to the Hilbert space dimension
    \be  
\left( 
\dim \mathcal{H}_{\text{BPS}}
\right)^2
= e^{2S_0} \cos^2(\pi j)\left(1+\sum_{c} w_c\right) (1+ O(e^{-2S_0})) .
\ee
Let us note that while we ignored these contributions in the computation of the resolvent, they are fully taken into account by the Gaussian ensemble described in section \ref{subsec:Interpretation-Haar-random-states}. This allows us to organize the sum over all nonzero geometries in the gravitational path integral and should allow us to include the above contributions in the computation of the resolvent. Nevertheless, the fact that we already obtained the expected answer for the rank without including these geometries suggests that even if these geometries would affect the resolvent at subleading order, their contribution to the rank itself seems to be vanishing.

\subsubsection*{Matter field one-loop determinant }
\label{sec:matter_loop}
\begin{figure}[h]
    \centering
    \includegraphics[width=0.5\textwidth]{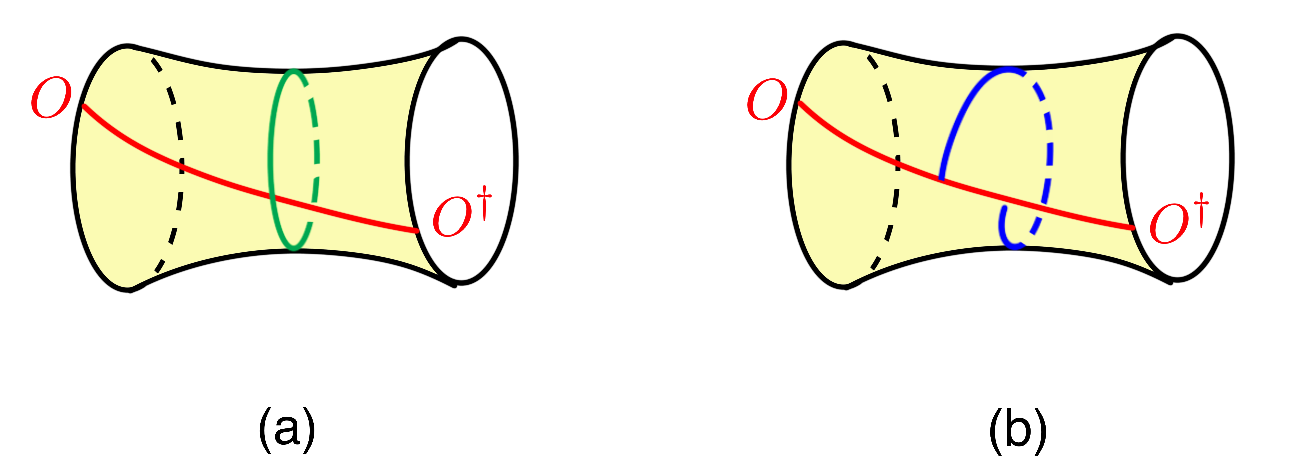}
    \caption{ The pink curve shows a geodesic that connects two particle insertions on the boundary (a) a matter loop that closes (b) the shortest geodesic from the matter geodesic to itself, that wraps the cylinder once but does not necessarily form a closed loop. In the calculation below, we shall use the
    blue geodesic 
    in the right figure to bound the length of the closed geodesic in the left figure.}
    \label{fig:matter_loop}
\end{figure}

In the computation of the higher genus multi-boundary correlators, we have also ignored the contribution coming from the matter one-loop determinant
\cite{Jafferis:2022wez,Saad:2019lba,Saad:2019pqd}.
This is a valid approximation only when the gravitational path integral is dominated by geometries with large throat sizes $b \gg 1$; the matter one-loop determinant contains a tachyonic divergence $\sim e^{\pi^2/6b}$ which becomes important for geometries with small $b$. 
As we explained around \eqref{eq:multi-geodesic-trumpet}, in our case, we do not need to worry about such divergences arising from handles without matter geodesics because such geometries vanish in the BPS sector. Therefore, the only geometries where we need to worry about the matter determinants will be trumpets containing matter geodesics. Below, we provide an estimate for the typical size of the throat for such geometries. We find that the path integral is dominated by geometries with large $b$ as long as the scaling dimension $\Delta$ of boundary operators is large, and therefore, the corrections from matter one-loop determinant are greatly suppressed. 
\\

We'll estimate the size of the throat $b$ on the example of cylinder geometry with two boundary operator insertions connected through bulk matter geodesic (see figure \ref{fig:matter_loop}).
As we show in detail in appendix \ref{app:hyperbolic-geometry}, in terms of the geodesic lengths of the polygon, we can express the size of the throat as 
\be 
\sinh \frac{b}{2} \simeq e^{\frac{1}{4}(\ell_{14}+\ell_{23}-\ell_{12}-\ell_{34})} .
\label{eq:size_throat}
\ee
With this, we can now estimate the size of the throat by inserting the above factor in the computation of the two-point function on the cylinder. We get 
\begin{align}
\biggl< \sinh \frac{b}{2} \biggr>_{j, \text{cyl}}^{\Delta} 
&= 
\int d\mu_{41} \, \Psi^j_{41} e^{\ell_{14}/4} \int d\mu_{23}\,  \Psi^j_{23} 
e^{\ell_{23}/4}
\int d\mu_{43}  \, I(\mu_{14},\mu_{43},\mu_{32} ,\mu_{34} ) e^{-(\Delta+\frac{1}{2}) \ell_{34}} \nonumber
%
\\ 
&= \frac{\cos(\pi j)^3}{4\pi^4} \Gamma\left(\frac{3}{4}\right)^4 
\Gamma\left(\frac{1}{4}\pm j\right)^2 \frac{(2\Delta+1)\Gamma\left(\Delta+\frac{1}{2} \right)^2 \Gamma(\Delta+1 \pm j)}{\Gamma(2\Delta+1)}
.
\end{align}
Denoting
\be 
\langle 2 \text{pt} \rangle^j_{\text{cyl}} 
= 
\frac{\cos(\pi j)}{2\pi} \frac{\Delta \Gamma(\Delta)^2 \Gamma\left(\Delta+\frac{1}{2}\pm j\right)}{2\pi \Gamma(2\Delta)}  ,
\ee
we can write normalized expectation value 
\be 
\frac{\biggl< \sinh \frac{b}{2} \biggr>_{j, \text{cyl}}^{\Delta}}{\langle 2 \text{pt} \rangle^j_{\text{cyl}} 
} = \frac{\cos(\pi j)^2}{\pi^3} \frac{\Gamma\left( \frac{3}{4} \right)^4 \Gamma\left(
\frac{3}{2}+\Delta \right)^2}{(1+2\Delta)\Gamma(1+\Delta)^2} 
\frac{\Gamma\left( \frac{1}{4} \pm j \right)^2 
\Gamma(1\pm j +\Delta)
}{\Gamma\left( 
\frac{1}{2} \pm j +\Delta
\right)} 
,
\ee
which at large $\Delta$ behaves as 
\be 
\frac{\biggl< \sinh \frac{b}{2} \biggr>_{j, \text{cyl}}^{\Delta}}{\langle 2 \text{pt} \rangle^j_{\text{cyl}} 
} \sim \Delta + O\left(\frac{1}{\Delta}\right)\,.
\ee
This then implies that $b$ is large in expectation value.
Similarly, one can also see that at large $\Delta$ we have that, 
\be 
1- \frac{\biggl<\left(\sinh \frac{b}{2}\right)^2 \biggr>_{j, \text{cyl}}^{\Delta}}{\left(\biggl<\sinh \frac{b}{2}\biggr>_{j, \text{cyl}}^{\Delta}\right)^2}  \sim O\left(\frac{1}{\Delta}\right) ,
\ee 
which means that the path integral is dominated by spacetime configurations that have large values of $b$.
Consequently, since for large $\Delta$, the gravitational path integral is completely dominated by geometries with a large 
value of $b$ for which the contribution of the matter one-loop determinant is greatly suppressed, even including such determinants in the gravitational path will not change our computation of the resolvent.

\section{Reconstructing an arbitrary state}
\label{sec:reconstructing-an-arbitrary-state}

In this section, we present a different way to calculate the dimension of the ground state Hilbert space based on the state reconstruction procedure first presented in \cite{Hsin:2020mfa}. We want to ask the following question: given an arbitrary state $\ket{\psi} \in \mathcal{H}_{BPS} \otimes \mathcal{H}_{BPS}$, how much of the state $\ket{\psi}$ can we optimally reconstruct as $\ket{\psi} \simeq \sum_i f_i \ket{q_i}$ with the basis of size $K$. In other words, we will want to optimize the overlap $\sum_i f_i \bra{\psi}\ket{q_i}$ for a given value of $K$.
Reference \cite{Hsin:2020mfa} studied this question in the non-supersymmetric case. This approach has some advantages compared to the approach of just computing the rank. 
It allows to
explicitly derive the coefficients $f_i$, in terms of the state overlaps, required to write $\ket{\psi}$ in the constructed basis. 
It is also more useful for asking more refined questions about state reconstruction in the bulk.\footnote{A detailed analysis of this problem will appear shortly in \cite{ToAppearJan}.}

We start with a short review of the discussion in  \cite{Hsin:2020mfa}. The question of how much of the state we can reconstruct with a given basis of size $K$ can be rephrased in terms of a maximization problem of the normalized overlap 
\be  
: \text{max}_{f_i} \frac{\braket{q_f}{\psi}}{\sqrt{\braket{\psi}\braket{q_f}}} : 
\ee
After maximizing with respect to coefficients $f_i$ subject to the normalization constraint $\braket{q_f}=1$ \cite{Hsin:2020mfa} found that the optimal overlap reduces to 
\be  
: \text{max}_{f_i} \frac{\braket{q_f}{\psi}}{\sqrt{\braket{\psi}\braket{q_f}}} : 
=
\frac{:\sqrt{VM^{-1}V^\dag }:}{\sqrt{\braket{\psi}}} ,
\ee
where $M_{ij} \equiv \braket{q_i}{q_j}$ denotes the matrix of overlaps of the basis vectors $\ket{q_i}$, and $V_i \equiv \braket{\psi}{q_i}$. Taking the square root in the expression above might, in general, be problematic (since the averaging is taken outside of it); however, in our case, we are interested in studying the above overlap close to 1. One can therefore argue that it's enough to study for what $K$ the quantity under the square root is equal to one, which will then imply that the overlap itself is also equal to 1. We are then interested in the following question:
\be  
\text{For what minimal $K$ is    } \frac{:VM^{-1}V^\dag :}{\braket{\psi}} \text{    equal to 1?} 
\ee

To answer this question, we will again turn to the resolvent matrix 
\be  
\mathbf{R}_{ij} = \left( 
\frac{1}{\lambda  - M}
\right)_{ij} 
= \frac{\delta_{ij}}{\lambda} + \frac{1}{\lambda} \sum_{n=1}^\infty \frac{(M^n)_{ij}}{\lambda^n}
,
\ee
in terms of which we can express 
\be  
: V M^n V^\dag: = \frac{1}{2\pi i} \oint d\lambda \, \lambda^n \, :V \mathbf{R} V^\dag: \, \,  .
\ee
The new Schwinger-Dyson equations get slightly modified because in the product $:\braket{\psi}{q_1}\dots \braket{q_n}{\psi}:$ the first and last states are not summed over. The resulting equation takes the form 
\be  
 :V \mathbf{R} V^\dag: = \sum_{n=0}^\infty R^{n+1} Z_{n+2}^{\psi} ,
\ee
which graphically we can represent as
\be
 :V \mathbf{R} V^\dag:=\sum_{n=0}^\infty \includegraphics[valign=c,width=0.23\textwidth]{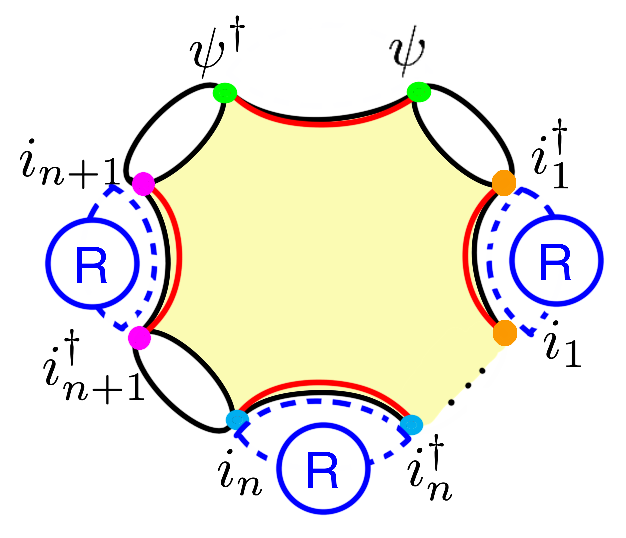}
\ee
As before, $Z_{n+2}^\psi$ is the pinwheel geometry with $n+2$ boundaries with two operator insertions on each boundary; note that the first and the last boundary contain a single $\mathcal{O}_\psi$ operator insertion of the state we want to reconstruct, while the rest of the operators are the basis with which we want to reconstruct $\ket{\psi}$. Working with the case $\Delta_\psi = \Delta$, the pinwheel has the same form as in previous subsection
\be  
Z_n^\psi =Z_n^\mathcal{O} =
e^{2\mathbf{S}_j} \, y^n ,
\qquad 
y \equiv e^{-S_0}
\frac{\Delta \Gamma(\Delta)^2 \,\Gamma\left(\Delta+\frac{1}{2}\pm j\right)}{2\pi \Gamma(2\Delta)} ,
\qquad
e^{\mathbf{S}_j} \equiv e^{S_0} \cos(\pi j ) ,
\ee
where again by $Z_n^\mathcal{O}$ we denoted the pinwheel with only
$\mathcal{O}_{q_i}$ operator insertions. 
With this simple expression, we can resum the series to get 
\be  
 :V \mathbf{R} V^\dag: = 
 e^{2 \mathbf{S}_j} \frac{R y^2}{1-R y},
\ee
Using the results from the last section (\ref{Rresult}), the final expression can now be written as
\be  
: V M^n V^\dag: = \frac{1}{2\pi i} \oint d\lambda \, \lambda^n \, y \, (\lambda R- K) .
\ee
To evaluate this, we deform the integral away from zero towards the branch cut between $\lambda_-$ and $\lambda_+$.
This leads to the reconstruction rate (recall that $\langle \text{2pt} \rangle_{\text{disk}} = e^{2\mathbf{S}_j} y$)
\be  
\frac{: V M^{-1} V^\dag:}{\braket{\psi}} = \frac{y}{e^{2\mathbf{S}_j} y} \int_{\lambda_-}^{\lambda_+} d\lambda \, D(\lambda) = 
\begin{cases}
\, \, K e^{-2 S_0}/[\cos^2(\pi j)] \, ,  \qquad &K<e^{2\mathbf{S}_j} ,
\\ 
\, \, 1   \,  \qquad \qquad \qquad \qquad    &K>e^{2\mathbf{S}_j}.
\end{cases}
\ee
We, therefore, see that for a basis of size $K \geq e^{2S_0} \cos^2(\pi j)$, we can always reconstruct any other excitation in the bulk after appropriately choosing coefficients $f_i$.
The computation was exactly analogous to the fixed energy case in non-supersymmetric JT \cite{Hsin:2020mfa}. This makes sense since by taking an infinite Euclidean time evolution, the operators are projected into the fixed energy ground state sector.

Our result of when we can explicitly reconstruct an arbitrary state $\ket{\psi}$ precisely matches what we obtained from the rank of the density matrix. This is yet another consistency check of the fact that the BPS sector of black holes behaves as a quantum system.

\section{Discussion}
\label{sec:discussion}

In the gravitational path integral we oftentimes encounter two prescriptions:
\begin{itemize}
    \item \textit{The Gibbons-Hawking prescription}  computes the partition function of a gravitational theory by making the Euclidean time circle periodic and integrating over all bulk geometries consistent with these boundary conditions. 
    \item \textit{The inner-product prescription} computes inner-products between gravitational states that are prepared by using the gravitational path integral.
\end{itemize}
In supergravity theories, the Gibbons-Hawking prescription sometimes allows us to determine the exact degeneracy or index of BPS black holes. Nevertheless, even in such cases, why the Gibbons-Hawking prescription yields an actual count over states is unclear. This consequently begs the question: can the second prescription provide an explicit count over states? Yes, it can: in this paper, we have explicitly constructed a complete basis of two-sided BPS black hole microstates by using the supergravity path integral whose dimension precisely agrees with the degeneracy predicted from the Gibbons-Hawking prescription.

 We have presented this construction in the context of $\mathcal{N}=2$ super JT gravity coupled to bulk matter, an effective field theory for the quantum fluctuations relevant for studying the physics of BPS states in a variety of higher dimensional black holes. To prepare our basis of states, we considered a number of states that are orthogonal to leading order and are created by local operator insertions on AdS$_2$ boundary with subsequent infinite Euclidean boundary time evolution. 

Regardless of whether the local operators are BPS or not, the infinite Euclidean time evolution projects them into the BPS sector. By using these states,  one can naively reach a paradox: in principle, it is possible to construct an infinite number of orthogonal states by using matter operator insertions which would contradict the expected finiteness of the BPS Hilbert space predicted by the Gibbons-Hawking prescription. However, by summing over all geometries specified by the boundary conditions used in the preparation of each state, we find tiny nonzero overlaps between constructed states. These tiny overlaps, subleading in $e^{S_0}$, are captured from the bulk perspective by wormhole geometries in the gravitational path integral.\footnote{Our construction showcases how even though wormhole geometries yield vanishing contributions in the zero temperature limit when computing the variance of BPS black hole degeneracies by using the Gibbons-Hawking prescription, such geometries still play an important role in capturing the statistics of inner products between states in the BPS sector.} Taking these wormhole contributions into account results in a maximal number of orthogonal states that agrees with the BPS degeneracy or with the index predicted by the Gibbons-Hawking prescription. In other words, the two prescriptions described above yield the same exact result. This is the case even if the geometries that contribute in the Gibbons-Hawking prescription, which include no wormhole contributions, are completely different than the geometries that appeared in the computation of the Hilbert space dimension using the second prescription, primarily consisting of wormhole geometries with matter sources).

This match is not only successful at leading order in $1/G_N$, but also succeeds in reproducing the kind of non-perturbative corrections that are encountered when computing the ground state degeneracy in a fixed charge sector by using the Gibbons-Hawking prescription. Moreover, while we have used a variety of wormhole geometries to verify that these states form a complete basis, we argue that our count has an exactly vanishing variance. This is because there are no additional wormhole geometries that ``connect'' the several copies of the rank of the BPS density matrix that we considered. Consequently, our count does not suffer from the factorization puzzle that plagues almost all observables in the gravitational path integral. 

Our analysis also shows that there are states prepared by the gravitational path integral, which naively are linear combinations of orthogonal states but are, in fact, null once non-perturbative corrections to the inner products are taken into account. The existence of such states can be seen from the zero eigenvalues of the Gram matrix $M_{ij} = \braket{q_i}{q_j}$ whose eigenvectors can be used to determine which precise linear combination of BPS states is null.  It would be interesting to understand how the preparation of such null states can be used to study the non-isometric code between bulk and boundary states recently discussed in \cite{Akers:2022qdl}.

A different perspective on our results can be gained by studying the classification of the algebra of boundary observables in the BPS sector of the supergravity theory. The importance of this classification in gravitational theories was only understood recently \cite{Penington:2023dql, Leutheusser:2021qhd, Leutheusser:2021qhd, Witten:2021unn, Chandrasekaran:2022cip}. Starting with a quantum field theory on a fixed background, the algebra of observables forms a type III von Neumann algebra. When coupling a matter theory to gravity, the algebra of boundary observables typically changes. For instance, in the case of (non-supersymmetric) JT gravity coupled to matter, including the effect of gravitational fluctuations, leads to a Type II${}_\infty$ algebra of boundary observables \cite{Penington:2023dql}. Above, we have shown that when restricting to the BPS sector and including the effect of wormholes, the two-sided BPS Hilbert space in a model of $\mathcal N=2$ super-JT gravity coupled to matter fields becomes finite-dimensional. This, in turn, suggests that the algebra of boundary observables projected to the BPS sector becomes of type I when including wormholes. This is because the dimension of the two-sided Hilbert space that we generate by acting on the maximally mixed BPS state with one-sided operators that are projected to their BPS sector is equal to the square of the expected Hilbert space dimension associated to each boundary. It would be interesting to understand how one can explicitly construct factorized states (states with zero entanglement entropy between the left and right boundaries) by considering linear combinations of the gravitationally prepared states that we have constructed. This would help resolve the ``factorisation'' puzzle that is typically present when studying two-sided states from the gravitational perspective.

Finally, it would be instructive to discuss our results about the statistics of inner products between BPS states in a concrete boundary dual to the gravitational theory that is not a disordered system. Consider BPS states in $\cN=4$ super-Yang Mills on $\mathbb R \times S^3$ with large enough scaling dimensions that are dual to black hole microstates in a sector with fixed R-charges and angular momenta. As in our construction, we can start with the two-sided maximally entangled BPS state in such a sector. On this state, we can act (on the left or right side) with a set of distinct primary operators (that are not BPS) projected to the BPS sector through an infinite period of Euclidean evolution to obtain a set of two-sided BPS states. Let us denote this set of states by $\mathcal B$. One can consider the various inner products between states in $\mathcal B$ and compute the averages over this set for a variety of (products of) such inner products. The simplest quantity to compute is the squared inner product between distinct states in $\mathcal B$ (exemplified in section \ref{subsec:resolution-of-paradoxes-summary}), but one can also try to compute the more complicated examples of products of inner products that we extensively studied in section \ref{subsec:leadingorder}). The gravitational calculations discussed above predict the average of all such products. Reproducing these results by considering averages of (products of) inner products for states in $\mathcal B$ would provide a boundary meaning to the wormhole contributions discussed throughout this paper without the need to consider an ensemble average of different theories. 

\section*{Acknowledgements}
\vspace{0.1cm}

We thank Henry Lin, Douglas Stanford, Pawel Caputa, Maciej Kolanowski, Masamichi Miyaji, Gustavo-Joaquin Turiaci, and Stephen Shenker for valuable discussions and helpful comments. We would especially like to thank Pawel Caputa, Henry Lin, Juan Maldacena, and Douglas Stanford for comments on the draft. LVI was supported by the Simons Collaboration on Ultra-Quantum Matter, a Simons Foundation Grant with No. 651440. JB is supported by the NCN Sonata Bis 9 2019/34/E/ST2/00123 grant. JB wants to give special thanks to Douglas Stanford and Stanford Institute for Theoretical Physics for generous hospitality and support during his visit at Stanford University, where this work had been initiated.

\appendix 

\section{The $\mathcal N=2$ super-JT partition functions}

\label{app:N=2-super-JT-part-functions}

Below we compute that $\mathcal N=2$ super-JT partition function on the disk, trumpet and defect. The only difference in all these cases, will correspond to the boundary conditions that we impose in the $\mathcal N=2$ super-Schwarzian theory which describes the fluctuation in the boundary shape of the AdS$_2$ spacetime.

\subsection{The disk}

We begin with briefly reviewing the localization calculation of the partition function for $\mathcal{N}=2$ JT supergravity on the disk. This was derived in \cite{Stanford:2017thb}, but it will be useful to revisit this derivation before performing the similar computations for the trumpet and the defect. 

In the $\mathcal{N}=2$ super-Schwarzian theory living on the one dimensional boundary of  an AdS$_2$ disk, we have the following boundary conditions for the reparametrization mode $f(\tau)$, the $U(1)$ mode $\sigma(\tau)$, and the charged fermion $\eta(\tau)$ 
\be 
\label{eq:b.c.-fields-disk}
f(\tau+\beta)=f(\tau)\,, \qquad \sigma(\tau+\beta) = e^{2\pi i \alpha} \sigma(\tau)\,, \qquad \eta(\tau+\beta) = - e^{2\pi i \alpha} \eta(\tau)\,.
\ee 
The action then consists of the Schwarzian action, the particle moving on a $U(1)$ group manifold,  with additional fermionic terms determined by supersymmetry. The  action is thus given by 
\be 
I_{\cN=2 \, \text{Schw}}  = \Phi_r \int d\tau\left[- \Schw(f, \tau) + 2(\partial_\tau \sigma)^2 + \text{ (fermions)}\right]\,,
\ee

We will first quickly go over the bosonic saddle points. For the Schwarzian mode with disk boundary conditions, there is a unique saddle given by $f(\tau) = \tan(\pi \tau/\beta)$. The saddles for the $U(1)$ mode $\sigma(\tau)$ consistent with the boundary conditions are $\sigma(\tau) = \left(n \hat q + \alpha \right) \frac{2\pi \tau}{\beta}$, with each saddle labeled by an integer $n$ that represents the winding of the $U(1)$ mode. The fermions vanish on the saddle point.  The on-shell action is given by 
\be 
-I_{\cN=2 \, \text{Schw}}^\text{on-shell} =  \frac{2\pi^2 \Phi_r}{\beta}\left(1-4(\alpha + \hat{q} n)^2\right)
\ee

We now need to study the one-loop determinant of all these fields around the saddle point. The Schwarzian produces a factor of 
\be
Z_\text{Schw}^\text{1-loop}\sim \left(\frac{\Phi_r}{\beta}\right)^{3/2}\,,
\ee 
while the one-loop determinant of the $U(1)$ mode is given by 
\be 
Z_{U(1)-\text{mode}}^\text{1-loop}\sim \left(\frac{\Phi_r}{\beta}\right)^{1/2}\,,
\ee
where we note that the result is independent of $\alpha$. Finally, the quadratic fluctuations of the fermionic terms are given by the action \cite{Stanford:2017thb}
\be
\label{eq:fermion-quad-expansion-disk}
 -I_{\text{ferm., quad.}} =  \frac{2\pi^2 i \Phi_r}{\beta} \left(1-4m^2\right)(m - n \hat q - \alpha) \eta_{-m + n \hat q +\alpha} \bar \eta_{m - n \hat q -\alpha}\,,
 \ee
 where $m \in \mathbb Z/2$.
 In this expression, the dependence on the inverse-temperature $\beta$ and the chemical potential $\alpha$ appears from the underlying bosonic saddle we perturb around. The modes $m =\pm 1/2$ are gauge modes that correspond to the four $SU(1,1|1)$ fermionic generators. Thus, we should not integrate over them. In total (the $\alpha$ and $\beta$-dependent part of) the one-loop determinant is given by
 \be 
 Z_{\text{fermion modes}}^\text{1-loop}= \left(\frac{\beta}{\Phi_r}\right)^2\prod_{m = \dots,\,-\frac{5}2,\,-\frac{3}2,\,\frac{3}2, \,\frac{5}2,\, \dots} (m-n \hat q - \alpha) = \frac{\cos(\pi (\hat q n+\alpha))}{1-4 (\hat q n+\alpha)^2} \left(\frac{\beta}{\Phi_r}\right)^2\,. 
 \ee
 This gives a total partition function\footnote{Notice that $\alpha\to -\alpha$ together with the redefinition $n \to -n$ leaves the partition function invariant. This guarantees that the density of states is charge conjugation invariant. }
 \be 
 \label{eq:N=2-JT-disk-part-function}
 Z_{\cN=2 \, \text{JT}}^\text{disk} =  e^{S_0}\sum_{n\in \mathbb Z} \frac{\cos(\pi (\hat q n+\alpha))}{1-4 (\hat q n+\alpha)^2}  e^{\frac{2\pi^2 \Phi_r}{\beta}\left(1-4(\alpha + \hat{q} n)^2\right)}
 \ee
We can look at the index which corresponds to setting $\alpha = 1/2$ (periodic fermions in \eqref{eq:b.c.-fields-disk}) and, for simplicity, consider the case $\hat q=1$. In such a case, there is a unique notion of a supersymmetric index instead of having to analyze the refined index discussed in \cite{Fu:2016vas, Boruch:2022tno}. In this limit and for this value of $\hat q$, the index is consequently given by 
\be
 \text{Index}_{\cN=2 \, \text{JT}}^\text{disk}=Z_{\cN=2 \, \text{JT}}^\text{disk}\left(\alpha=\frac{1}2\right) = \frac{\pi}2 e^{S_0}
\ee
which, as expected, is non-zero and temperature independent. 

\begin{figure}
    \centering
    \includegraphics[width=0.75\textwidth]{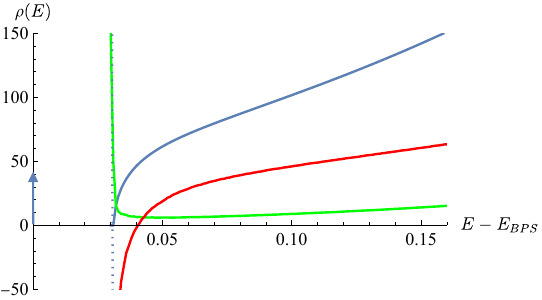}
    \caption{The continuous spectrum in JT gravity for the disk (\textit{in blue}), for the trumpet (\textit{in green}), for the regular defect (\textit{in orange}) and for the  supersymmetry preserving defect (\textit{in red}). Only the disk contribution has (discrete) support at $E-E_\text{BPS} = 0$.  }
    \label{fig:continous-spectrum}
\end{figure}

We can now discuss the BPS and non-BPS densities of states in fixed-charge sectors. First, we note that 
\be 
\label{eq:inv-laplace-transf-disk-part-function}
\int d\beta \,e^{\beta E} \,e^{\frac{2\pi^2 \Phi_r}{\beta}\left(1-4(\alpha + \hat{q} n)^2\right)} = \delta(E) + \sqrt{\frac{2\pi^2 \Phi_r}{E} \left(1-4(\alpha + \hat{q} n)^2\right)} \,\,I_1\left( \sqrt{{8\pi^2 \Phi_r}{E} \left(1-4(\alpha + \hat{q} n)^2\right)}\right)
\ee
Summing over all coefficients multiplying the Dirac-delta functions coming from \eqref{eq:N=2-JT-disk-part-function}, and Fourier transforming with respect to $\alpha$, yields a degeneracy in each charged sector
\be 
d_j = \sqrt{\frac{\pi}{8}} \,e^{S_0}\, \cos\left(\pi j\right) \, \frac{\text{sgn}\left(1-2j\right) + \text{sgn}\left(1+2j\right)}2\,,
\ee
where $j \in \mathbb \mathbb Z/{\hat q}$, and $d_j \neq 0$ only for $|j| \leq \frac{1}2$. 
The sum over all Bessel functions produces a gap in the spectrum. This gap, together with the continuous density of states above the gap, is shown in Figure \ref{fig:continous-spectrum} in blue. 

\subsection{The trumpet}
\label{app:trumpet}

In the $\mathcal{N}=2$ super-Schwarzian theory living on the one dimensional boundary of  an AdS$_2$ trumpet, we have the following boundary conditions for the super-Schwarzian field: 
\be 
\label{eq:b.c.-trumpet}
f(\tau+\beta) = e^{ b} f(\tau)\,, \qquad \sigma(\tau+\beta) = e^{2\pi i(\alpha\pm\phi)} \sigma(\tau)\,, \qquad \eta(\tau+\beta) = -e^{2\pi i \alpha} \eta(\tau)\,.
\ee
The relative sign between $\alpha$ and $\phi$ in the middle equation is fixed by the relative orientation of the holonomy around the trumpet compared to the holonomy around the boundary. Since we want to perform an integral over all metrics we should take into account both possible signs. Below, we shall first compute the on-shell action with a relative $+$-sign and will obtain the final partition function by adding the solution with the $-$-sign.

The saddle point in  this case is given by $f(\tau) = e^{ b \frac{\tau}\beta}$, $\sigma(\tau) = 2\pi(n\hat q + \alpha + \phi) ]\frac{\tau}{\beta}$, with the fermions vanishing once again. The on-shell action is given by 
\be 
-I_{\cN=2 \, \text{Schw}}^\text{Trumpet On-shell} =  \frac{2\pi^2 \Phi_r}{\beta}\left[-\left(\frac{b}{2\pi}\right)^2-4(\alpha + \phi+ \hat{q} n)^2\right]\,.
\ee
The one-loop determinants in this case are given by, 
\be 
\label{eq:1-loop-Schw-and-U(1)-trumpet}
Z_\text{Schw}^\text{1-loop}\sim \left(\frac{\Phi_r}{\beta}\right)^{1/2}\,,\qquad Z_{U(1)-\text{mode}}^\text{1-loop}\sim \left(\frac{\Phi_r}{\beta}\right)^{1/2}
\ee
for the Schwarzian and $U(1)$-modes, while the fermions one-loop determinant is obtained from the quadratic expansion of the action  
\be 
\label{eq:ferm-quad-trumpet}
-I_{\text{ferm., quad.}} =  -\frac{2\pi^2 i \Phi_r}{\beta} \left(\left(\frac{b}{2\pi} \right)^2+4(m+\phi)^2\right)(m - n \hat q - \alpha) \eta_{-m + n \hat q +\alpha} \bar \eta_{m - n \hat q -\alpha}
\ee 
where we have once again decomposed the fermionic fields $\eta(\tau)$ and $\bar \eta(\tau)$ into Fourier modes, once again with $m \in \mathbb Z/2$. Note that there are no longer any modes for which the quadratic expansion vanishes (as the $m=\pm 1/2$ modes in \eqref{eq:fermion-quad-expansion-disk}) -- this is because no trumpet geometry, regardless of the value of $b$ and $\phi$, has globally well defined Killing spinors. Therefore, there are no super-isometries that correspond to modes that we have to quotient out. The fermionic one-loop determinant is thus given by 
\be 
Z_{\text{fermion modes}}^\text{1-loop}= \prod_{m \in \mathbb Z/2} (m-n \hat q - \alpha) ={\cos\left(\pi (\hat q n+\alpha)\right)}
\ee 
The $b$ and $\phi$ dependent factor in \eqref{eq:ferm-quad-trumpet} that would yield a $b$ and $\phi$ dependent one-loop determinant cancels with the terms in the symplectic measure of the $\cN=2$ super-Schwarzian path integral -- the same happens for the $b$-dependent part of the bosonic one-loop determinant coming from the quadratic expansion of the bosonic Schwarzian. The overall partition function is now given by 
\begin{align}
\label{eq:N=2-partition-func-trumpet}
     Z_{\cN=2 \, \text{Schw}}^\text{Trumpet} =  \sum_{n\in \mathbb Z} &\left(\frac{\Phi_r}{\beta} \right) \bigg({\cos(\pi (\hat q n+\alpha))} e^{\frac{2\pi^2 \Phi_r}{\beta}\left[-\left(\frac{b}{2\pi}\right)^2-4(\alpha + \phi+ \hat{q} n)^2\right]} \nonumber \\
 &+ {\cos(\pi (\hat q n-\alpha))} e^{\frac{2\pi^2 \Phi_r}{\beta}\left[-\left(\frac{b}{2\pi}\right)^2-4(\alpha -\phi+ \hat{q} n)^2\right]} 
 \bigg)
\end{align}

Let's again look at the index of the theory with $\hat q=1$. Setting $\alpha = \frac{1}2$ we find that no saddles yield a non-zero contribution due to the fact that $\cos(\pi (\hat q n+\alpha)) =0$ with no other diverging factor in the one-loop determinant as $\alpha \to \frac{1}2$. Thus, we have that 
\be
 \text{Index}_{\cN=2 \, \text{JT}}^\text{Trumpet}=Z_{\cN=2 \, \text{JT}}^\text{Trumpet}\left(\alpha=\frac{1}2\right) =0\,.
\ee

As for the disk, we can also discuss the density of states in a fixed charge sector. Taking the Laplace transform of each term in \eqref{eq:N=2-partition-func-trumpet} gives
\be 
\int d\beta \frac{e^{\beta E}}{\beta} e^{\frac{2\pi^2 \Phi_r}{\beta}\left[-\left(\frac{b}{2\pi}\right)^2-4(\alpha + \phi+ \hat{q} n)^2\right]}  = I_0\left( \sqrt{8\pi^2 E \Phi_r \left[-\left(\frac{b}{2\pi}\right)^2-4(\alpha + \phi+ \hat{q} n)^2\right] } \right)\,.
\ee
Importantly, this Laplace transform does not have a Dirac-delta function which is an indication that the trumpet geometries do not contribute to the extremal degeneracy. Summing over all the Bessel functions in \eqref{eq:N=2-partition-func-trumpet}, once again produces a gap, followed by a continuous density of states, which we show in Figure \ref{fig:continous-spectrum} in green. Thus, because the density of states contains a non-vanishing gap above extremality, all higher genus surfaces that can be obtained through a gluing of the trumpet geometry also yields no contribution to the extremal degeneracy.

\subsection{The defect}
\label{app:defect}

The partition function with a single defect can be obtained through the analytic continuation $b\to 2\pi i \theta$, where $\theta$ parametrizes the deficit angle with $\theta=0$ corresponding to cusp and $\theta =1 $ corresponding to no defect insertion. Without loss of generality we will assume $\theta \in [0,1]$ and $\phi\in [0,1]$ below. The on-shell action (with the boundary condition for $\sigma$ with a relative $+$-sign between $\alpha$ and $\phi$ in \eqref{eq:b.c.-trumpet}) is now given by 
\be 
-I_{\cN=2 \, \text{Schw}}^\text{Defect On-shell} =  \frac{2\pi^2 \Phi_r}{\beta}\left[\theta^2-4(\alpha + \hat{q} n)^2\right]\,.
\ee
The Schwarzian and fermionic one-loop determinants remain the same as in \eqref{eq:1-loop-Schw-and-U(1)-trumpet}. Once again the fermionic one-loop determinant is determined by the quadratic expansion of the action which now becomes
\be 
\label{eq:ferm-quad-defect}
-I_{\text{ferm., quad.}} =  \frac{2\pi^2 i \Phi_r}{\beta} \left(\theta^2-4(m+\phi)^2\right)(m - n \hat q - \alpha) \eta_{-m + n \hat q +\alpha} \bar \eta_{m - n \hat q -\alpha}
\ee 
with the same decomposition into Fourier modes as for the trumpet. As compared to the trumpet, there are now two important cases that we can distinguish.

\subsubsection*{Supersymmetric defect}

Notice that when $\theta = 1-2\phi$ when $\phi\leq 1/2$ (i.e.,~the geometric angle of the defect is related to the holonomy of the $U(1)$
gauge field when going around the defect then the first parenthesis of \eqref{eq:ferm-quad-defect} can be expressed as 
\be 
\theta^2-4(m+\phi)^2 = -(2m+1)(2m+4\phi-1)\,.
\ee
In such a case we again observe a zero-mode when $m=-1/2$. This is because such defects preserve two Killing spinors of the four that existed in $SU(1,1|1)$ when the defect was absent. Thus, once again, these modes have to be quotiented out. The overall one-loop determinant is thus given by 
\be 
 Z_{\text{fermion modes}}^\text{SUSY defect 1-loop}= \left(\frac{\beta}{\Phi_r}\right)\prod_{m = \dots,\,-\frac{5}2,\,-\frac{3}2,\,\frac{1}2, \,\frac{3}2, \, \dots} (m-n \hat q - \alpha) = \frac{\cos(\pi (\hat q n+\alpha))}{1-2 (\hat q n+\alpha)} \left(\frac{\beta}{\Phi_r}\right)\,. 
 \ee
 Thus, the final answer is given by 
 \be 
 \label{eq:part-function-SUSY-defect}
 Z_{\cN=2 \, \text{JT}}^\text{SUSY Defect} =  \lambda e^{S_0}\sum_{n\in \mathbb Z} {\cos(\pi (\hat q n+\alpha))} \left(\frac{e^{\frac{2\pi^2 \Phi_r}{\beta}\left((1-2\phi)^2-4(\alpha + \phi + \hat{q} n)^2\right)}}{{1-2 (\hat q n+\alpha)}  } + \frac{e^{\frac{2\pi^2 \Phi_r}{\beta}\left((1-2\phi)^2-4(-\alpha + \phi + \hat{q} n)^2\right)}}{{1-2 (\hat q n-\alpha)}  } \right)
 \ee
 where $\lambda$ is the weight of the defect insertion in the gravitational path integral. 
 
 Let's again go back to computing the index by taking $\alpha=1/2$ once again in the simplified case when $\hat q=1$. Only the $n=0$ saddle now survives and taking the limit $\alpha \to \frac{1}2$ of \eqref{eq:part-function-SUSY-defect} we find 
 \be 
 \text{Index}_{\cN=2 \, \text{JT}}^\text{SUSY Defect}=Z_{\cN=2 \, \text{JT}}^\text{SUSY Defect}\left(\alpha=\frac{1}2\right) = \frac{\lambda \pi}2 e^{S_0}\,.
 \ee
This is, thus, the only other geometry different than that of the disk that can contribute to the gravitational path integral with the boundary conditions corresponding to the computation of an index. We should again the contribution of this geometry to the density of states. Laplace transforming \eqref{eq:part-function-SUSY-defect} to obtain the density of states (just as in \eqref{eq:inv-laplace-transf-disk-part-function}), we again find a non-trivial contribution to the extremal degeneracy due to the Dirac $\delta$-functions. Summing the overall coefficients in front of the $\delta$-functions to obtain this degeneracy in a fixed charge sector $j$, we find
\be 
d_j = \sqrt{\frac{\pi}{8}} \lambda \,e^{S_0}\, \cos\left(\pi j\right) \, \frac{\text{sgn}\left(1-2j\right) + \text{sgn}\left(1+2j\right)}2\,,
\ee
where once again, the range and values of the charge $j$ are the same as that for the disk. Thus, we find that the supersymmetric defect geometry only changes the extremal degeneracy by an overall proportionality factor, which is $R$-charge independent.

As for all other geometries, the sum over all Bessel functions produces a gap in the spectrum which, together with the continuous density of states above the gap, is shown in Figure \ref{fig:continous-spectrum} in red. 

\subsubsection*{Non-supersymmetric defect}

For $\theta \neq 1-2\phi$ (once again when $\phi \leq 1/2$) we have no supersymmetry enhancement. Therefore, the partition function on this geometry can be obtained in a way completely analogous to that of the trumpet. The resulting partition function is 
\be 
 Z_{\cN=2 \, \text{Schw}}^\text{Non-SUSY defect} =  \lambda e^{S_0}\sum_{n\in \mathbb Z} \left(\frac{\Phi_r}{\beta} \right) {\cos(\pi (\hat q n+\alpha))} \left(e^{\frac{2\pi^2 \Phi_r}{\beta}\left[\theta^2-4(\alpha + \hat{q} n)^2\right]} + e^{\frac{2\pi^2 \Phi_r}{\beta}\left[\theta^2-4(-\alpha + \hat{q} n)^2\right]}\right)\,.
\ee
Due to the temperature and charge dependence, the contribution of the non-supersymmetric defect geometry to the density of states follows a very similar behavior to that of the trumpet.

\section{Length of the supersymmetric Einstein-Rosen bridge}
\label{app:length-of-spatial-wormhole}

Using the result for the disk 2-point function one can compute the expectation value for the length of the bulk spatial wormhole as \cite{LongPaper}
\be 
\langle \ell \rangle_j = - \frac{\partial_\Delta \langle 2\text{pt} \rangle_{j, \, \text{disk}}|_{\Delta=0}}{Z_j} 
= - \psi \left( \frac{1}{2} + j \right) - \psi \left( \frac{1}{2} - j \right) ,
\ee
where $\psi(z) = \Gamma'(z)/\Gamma(z)$ is the digamma function, and 
\begin{align}  
\langle 2\text{pt} \rangle_{j, \, \text{disk}} = e^{S_0}\frac{\cos^2(\pi j)}{2\pi} \frac{\Delta \Gamma(\Delta)^2}{\Gamma(2\Delta)} \Gamma\left(\Delta+\frac{1}{2}\pm j\right) .
\end{align}
In our case, we can also include the nonperturbative corrections to the length coming from supersymmetric defects of weights $w_i$. Because we are inserting a bulk operator $\ell$ between the two boundaries we can have two insertions on the bulk geometry. On the other hand, the partition function by which we normalize can only have a single supersymmetric defect insertion. The total result for the bulk length will therefore have the form 
\be 
\langle \ell \rangle_j^{\text{total}} = - \frac{(1+W)^2\partial_\Delta \langle 2\text{pt} \rangle_{j, \, \text{disk}}|_{\Delta=0}}{(1+W)Z_j} 
= (1+W)\left[- \psi \left( \frac{1}{2} + j \right) - \psi \left( \frac{1}{2} - j \right)  \right] ,
\ee
where again we denoted $W=\sum_i w_i$ for the general case of defects of different flavors in the bulk.

\section{Hyperbolic Geometry}
\label{app:hyperbolic-geometry}

In this appendix we explain how to derive the expression \eqref{eq:size_throat} for the minimal distance between two boundary anchored geodesics on a hyperbolic disk, which we use to estimate the size of the throat in section \ref{sec:matter_loop}.

Hyperbolic disks can be defined as a hyperboloid embedded in Minkowski space 
\be 
ds^2 = - dX_0^2 + dX_1^2 + dX_2^2 , 
\ee
through the embedding 
\be 
-X_0^2 +X_1^2 +X_2^2 = -1. 
\ee
Parametrizing the hyperboloid as 
\be
X=(\cosh\rho,\sinh\rho\cos\theta,\sinh\rho\sin\theta) ,
\ee
leads then to the standard form of the AdS$_2$ metric 
\be 
ds^2 = d\rho^2 + \sinh^2 \rho \,  d\theta^2 .
\ee
Using the embedding coordinates, we can compute the geodesic distance $\ell$ along the hyperboloid between points $X$ and $X'$ through 
\be
\cosh \ell=-X\cdot X' \equiv X_0 X'_0 
-X_1 X'_1 
-X_2 X'_2 
,
\ee
with signature $\mathrm{diag}(-1,1,1)$. Then the distance between 
\begin{align}
    X&=(\cosh\rho_0,\sinh\rho_0\cos\frac{\theta_1}{2},\sinh\rho_0\cos\frac{\theta_1}{2}) , \\
    X'&=(\cosh\rho_0,\sinh\rho_0\cos\frac{\theta_2}{2},\sinh\rho_0\cos\frac{\theta_2}{2}) ,
\end{align}
is given by
\be
\cosh \ell =\cosh^2\rho_0-\sinh^2\rho_0\cos(\theta_1-\theta_2) .
\ee
For large $\rho_0$ we can then approximate this by
\be
\frac{e^d}{2}\approx\frac{e^{2\rho_0}}{2}\frac{1-\cos\theta_0}{2}=\frac{e^{2\rho_0}}{2}\sin^2\frac{\theta_0}{2}
.
\ee

\begin{figure}
    \centering
    \includegraphics[width=0.2\textwidth]{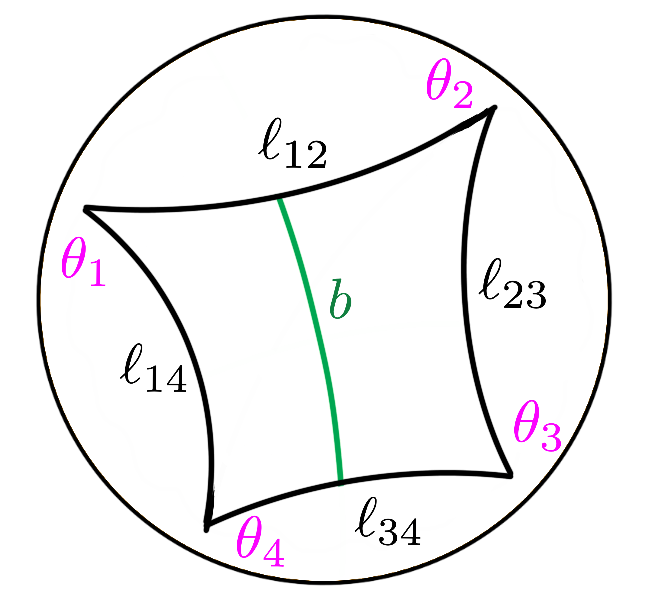}
    \caption{Polygon with four vertices located at $(\rho_0,\theta_1)$, $(\rho_0,\theta_2)$, $(\rho_0,\theta_3)$, $(\rho_0,\theta_4)$}
    \label{fig:polygontheta}
\end{figure}
On the other hand, using the above embedding coordinates, the minimal distance $b$ between two boundary anchored geodesics can be expressed as \cite{Hsin:2020mfa}
\be  
\sinh \frac{b}{2} = \sqrt{\frac{\sin \frac{\theta_{14}}{2} \sin \frac{\theta_{23}}{2} }{\sin \frac{\theta_{12}}{2} \sin \frac{\theta_{34}}{2}}} 
, \qquad
\theta_{ij} \equiv \theta_i - \theta_j ,
\ee
where we choose the cross section $b$ to connect geodesics $(12)$ and $(34)$. Thus, in terms of the geodesic lengths of the polygon we can express the size of the throat as 
\be 
\sinh \frac{b}{2} \simeq e^{\frac{1}{4}(\ell_{14}+\ell_{23}-\ell_{12}-\ell_{34})} .
\ee

\bibliography{biblio.bib}

\end{document}